\documentclass[12pt]{article}

\usepackage[top=80pt,bottom=85pt,left=60pt,right=60pt]{geometry}
\usepackage{amssymb,amsmath,xypic}
\usepackage{slashed}
\usepackage{graphicx,color,xcolor,subfigure}
\usepackage{float}
\usepackage{sectsty}
\usepackage{cite}
\usepackage{dsfont}
\usepackage{esvect}
\usepackage{amsmath}
\usepackage[debug,pageanchor=false]{hyperref}
\definecolor{link}{rgb}{.8,.15,.1}
\hypersetup{colorlinks=true,linkcolor=link,citecolor=link,urlcolor=link,linktocpage}

 \allowdisplaybreaks

\makeatletter
\@addtoreset{equation}{section}

\usepackage[framemethod=tikz]{mdframed}
\usepackage{tikz-3dplot}

\usepackage{tikz}
\usetikzlibrary{positioning}
\usetikzlibrary{arrows}

\usetikzlibrary{shapes.geometric}
\usetikzlibrary{arrows.meta}
\usetikzlibrary{decorations.pathreplacing,calligraphy}

\tikzset{%
  plus/.pic={
   \node[circle,ball color=orange,inner sep=0pt,minimum width=2.5ex]  {+};
   },
  minus/.pic={
   \node[circle,ball color=green,inner sep=0pt,minimum width=2.5ex]  {-};
   },
    zero/.pic={
   \node[circle,ball color=white,inner sep=0pt,minimum width=2.5ex]  {0};
   }
}

\makeatother

\newcommand{\beq}{\begin{equation}}
\newcommand{\eeq}{\end{equation}}
\newcommand{\bea}{\begin{eqnarray}}
\newcommand{\eea}{\end{eqnarray}}
\newcommand{\nn}{\nonumber}

\def\white1{\textcolor[rgb]{0.98,0.98,0.98}}
\usepackage{rotating}

\newcommand{\CP}[1]{\mathbb{CP}^{#1}}

\newcommand{\CY}[1]{\text{CY}_{#1}}

\newcommand{\arctanh}{\text{arctanh}}

\begin{document}

\begin{titlepage}

\begin{center}

\vskip .5in 
\noindent

{\Large \bf{Circle compactifications of Minkowski$_D$ solutions, flux vacua and solitonic branes}}

		\bigskip\medskip
Niall T. Macpherson$^{a}$\footnote{macphersonniall@uniovi.es}, Paul Merrikin$^{b}$\footnote{p.r.g.merrikin.2043506@swansea.ac.uk}, Ricardo Stuardo$^{a}$\footnote{ricardostuardotroncoso@gmail.com}\\

\bigskip\medskip
{\small 
	
	$a$: Department of Physics, University of Oviedo,\\ 
	Avda. Federico Garcia Lorca s/n, 33007 Oviedo
\\
	and
	\\
	 Instituto Universitario de Ciencias y Tecnolog\'ias Espaciales de Asturias (ICTEA),\\ 
	Calle de la Independencia 13, 33004 Oviedo, Spain.\\
~\\
	$b$: Department of Physics, Swansea University, Swansea SA2 8PP, United Kingdom}

	\vskip 3mm

		\vskip 1.5cm 
		\vskip .9cm 
		{\bf Abstract }

		\vskip .1in
\end{center}
    
	\noindent 
 G-structure techniques are used to construct broad classes of circle compactifications of Mink$_{D+1}$ solutions to Mink$_{D}$ embedded into type II supergravity for $D=1,...5$. Under a certain assumptions we show that the conditions that imply supersymmetry for Mink$_{D+1}$ imply those of the Mink$_{D}$ solution, but that Bianchi identities of the fluxes must be modified. This realises an off shell solution generating technique for supersymmetric solutions or a ``supersymmetry generating'' technique. Along the way it is necessary for us to derive G structure conditions for general ${\cal N}=(1,0)$ supersymmetric Mink$_2$ solutions and a restricted class of Mink$_1$ solutions. We apply our results to construct some simple Minkowski flux vacua before turning our attention to ``solitonic branes'' which are generalisations of the AdS soliton. We are able to generalise known examples in two ways: 1) to embed them in terms of generic Sasaki Einstein manifolds. 2) To modify the harmonic factor to include D$p$ brane sources at one  end of the space.

\vskip .1in

\noindent

\noindent

\vfill
\eject

\end{titlepage}

\tableofcontents

\section{Introduction and summary}
Whether ones interests are in the AdS/CFT correspondence, black holes or more traditional flux vacua realising the low energy limits of string theory, there are many reasons to be interested in solutions in type II supergravity. Unfortunately, constructing such solutions can be a difficult task without some sort of guiding principle. 

One method which has born much fruit over the years is to employ consistent truncations to lift solutions of lower dimensional gauged (or ungauged) supergravities to $d=10$. This is particularly useful when one seeks a solution in 10 dimensions that can be separated into some external space realising an effective theory in less dimensions and a compact internal space, because the compact nature of the latter is often guaranteed. A weakness of this approach is that limited consistent truncations are known, the majority of which truncate to either minimal gauged supergravities such as \cite{Gauntlett:2007ma,Passias:2015gya,Larios:2019lxq,Couzens:2022aki} or maximal gauged supergravity from truncations on spheres \cite{deWit:1986oxb,Nastase:1999cb,Nastase:1999kf,Cvetic:2000nc,Guarino:2015vca,Varela:2015ywx}. There have however been recent developments broading the scope of consistent truncations using exceptional field theory \cite{Malek:2018zcz,Malek:2019ucd,Malek:2020jsa,Josse:2021put,Galli:2022idq,Duboeuf:2024tbd}. 

Another method to construct solutions in $d=10$  is to employ the machinery of bilinear and G-structure techniques.  The way this typically works is one makes some assumption about an external space one wants to realise, such as fixing it to be warped Minkowki$_D$ or AdS$_D$, then given this the spinoral conditions for minimal supersymmetry are rephrased in terms of geometric constraints that fix the internal manifold and other physical fields. These constraints are expressed in terms of the forms that span the G-structure of the internal space, a generalisation of group holonomy in which the associated forms (say $(J,\Omega)$ for a Calabi-Yau) are no longer closed, but rather have non trival torsion classes. Such methods have been very successful at classifying and constructing both AdS and Minkowki solutions, for some notable classifications of such solutions see for instance \cite{Grana:2004bg,Gauntlett:2004zh,Grana:2005sn,Gabella:2012rc,Apruzzi:2013yva,Dibitetto:2018ftj,Legramandi:2023fjr}. We should stress though that these methods apply beyond the realm of string vacua, and indeed only the existence of a single supercharge is required  \cite{Gauntlett:2002fz,Gauntlett:2003wb,Tomasiello:2011eb,Legramandi:2018qkr}. This also indicates the weakness of G-structure techniques, they have not historically given access to solutions that break supersymmetry, though there has been some progress in this direction \cite{Legramandi:2019ulq,Menet:2023rnt}.

A third approach for constructing solutions, which has some overlap with the previous methods, is to utilise supergravity solution generating techniques. This usually hinges on string dualites such as T, S and U dualites, or rather there analogues in supergravity as well as non duality commuting coordinate transformations. A notable example in this are T-s-T transformations \cite{Lunin:2005jy} as well as their IIA/$d=11$ analogue, which when performed on solutions with a dual QFT interpretation induce marginal deformations. There are also the sorts of formal U dualities considered in \cite{Maldacena:2009mw} which were used to map (a deformation of \cite{Casero:2006pt}) the wrapped D5 solution of \cite{Maldacena:2000yy}  to the baryonic branch of Klebanov Strassler \cite{Butti:2004pk}. Finally, there exist many works that make use of non-Abelian \cite{delaOssa:1992vci,Lozano:1995jx,Sfetsos:2010uq,Kelekci:2014ima} and Poisson-Lie \cite{Klimcik:1995jn} T-dualities and beyond \cite{Borsato:2021vfy} to construct new solutions from old.\\
~~\\
In this work we we will touch upon all of the above by using G-structure techniques to construct circle compactifications of classes of  $\text{Mink}_{D+1}$ solutions of type II supergravity, where
\beq
ds^2(\text{Mink}_{D+1})\to  ds^2(\text{Mink}_{D})+ D\phi^2,~~~~D\phi=d\phi+ {\cal A},\nn
\eeq
and the $d=10$ fluxes are modified in terms of the field strength ${\cal F}=d{\cal A}$ and G-structure forms supported by the internal space of the Mink$_{D+1}$ parent class. Under certain assumptions we are able to show that the necessary conditions for the compactified Mink$_D$ class to preserve supersymmetry are implied by those of the Mink$_{D+1}$ parent, however the same is not true of the equations of motion. These  only follow for the Mink$_D$ class after making  a modification of the Bianchi identity of the RR fluxes in the parent. As such, this method falls short of being a solution generating technique, but as it does map one solution to the necessary conditions for supersymmetry to another, we find it apt to think of it as a supersymmetry generating technique. 

Our supersymmetry generating technique requires that the field strength ${\cal F}$ is ``primitive'' 2-form\footnote{The analogue of a primitive (1,1)-form on an SU(n)-structure manifold but generalised to more general G-structures}  with respect the G-structure that the internal space of the parent theory supports\footnote{This has some similarity to the constructions in \cite{Kaste:2003dh,Kaste:2002xs}, but is broader}. It works in a wide variety of cases allowing for compactifciations from Mink$_{D+1}\to$ Mink$_D$ for $D=1,...,5$, terminating at this point because for $D+1=7$ one cannot define a primitive 2 form. Proving this requires knowledge of the G-structure conditions for minimally supersymmetric Mink$_{1,..,6}$ and while those of Mink$_{3}$ \cite{Haack:2009jg} and above are known\footnote{Given conditions for Mink$_D$ it is simple to derive them for Mink$_{D+1}$, see appendix \ref{MinkDtoDm1derivation}}, below this they are absent from the literature so our first task is to derive them. More specifically we will derive the supersymmetry conditions for general ${\cal N}=(1,0)$ supersymmetric  Mink$_2$ (see \cite{Rosa:2013lwa} ${\cal N}=(2,0)$ results ) and a subclass of Mink$_1$ solutions that is sufficient for our purposes.\\
~\\
A summary and lay out of the paper is as follows:

We begin in section \ref{sec:Mink2part} by summarising the necessary and sufficient geometric conditions for a Mink$_2$ solution of type II supergravity to preserve  ${\cal N}=(1,0)$ supersymmetry, which are derived in detail in appendix \ref{sec:mink2susyderivation}. We prove that the Bianchi identities of the magnetic NS and RR fluxes for such solutions imply all the other equations of motion when supersymmetry holds in section \ref{eq:integrabilitymink2}. Then in section \ref{eq:Mink2Gstructures} we show that the internal 8-manifold of these solutions supports an SU(4)-structure with a possible enhancement to Spin(7)-structure in type IIB and a G$_2$ structure in type IIA. After the sanity check of the results thus far of section \ref{sec:D1branes}, we consider circle compactification of Mink$_2\to $Mink$_1$ in section \ref{eq:Mink2toMink1} whose results are derived in appendix \ref{sec:compactifyingMink2} which derives geometric conditions for a restricted class of minimally supersymmetric Mink$_1$ solutions\footnote{In general supersymmetric Mink$_1$ solutions have a Killing vector $K$ associated to them that can be either time-like or null. The appendix considers the null case under the assumption that $K$ decomposes in terms of a time-like and space-like Killing vector, that the entire background and spinors it supports are singlets with respect to individually.}. Then in section \ref{sec:D1circle} we derive a simple class of solutions describing D1 branes wrapping a circle that are backreated on a generic manifold of Spin(7) holonomy (a subcase of which is CY$_4$) in the presence of harmonic 3-form flux - we will make use of this later to construct explicit solutions. 

The procedure for compactifying Mink$_2\to $ Mink$_1$ is a little ad hock, where the supersymmetry generating technique really starts to shine is when we consider higher dimensional circle compactifications in section \ref{eq:MinkDp1tominkD}. In section \ref{sec:compact3to2} we consider compactifications of of Mink$_3\to$ Mink$_2$ where we find that all classes of Mink$_3$ solutions can be compactified in this manor. We then use this result to derive a subclass of solutions describing D2 branes wrapping a circle and backreacted on generic G$_2$ manifolds in the presence of harmonic NS flux in section \ref{eq:D2section}. We turn our attention to Mink$_4\to$ Mink$_3$ compactifications in section \ref{sec:compact4to3} finding a more restricted form of the supersymmetry generating technique that works for Mink$_4$ solutions supporting a strict SU(2)-structure in IIA or SU(3)-structure in IIB, we then leverage our results in section \ref{sec:D3circle} to construct a subclass of solutions describing D3 branes compactified on S$^1$ and backreacted on a CY$_3$ with additional harmonic NS flux. We establish in section \ref{sec:compact5to4} that the case of Mink$_5\to$ Mink$_4$ only works for Mink$_5$ solutions supporting a restricted class of SU(2)-structure manifolds that in IIA leads to a circle compactified D4-D8 system backreacted on CY$_2$ with additional NS flux turned on, as explored in \ref{sec:D4D8circle}. Finally only one class of Mink$_6$ can be compactified in this fashion, it describes D5 branes wrapping a circle and backreacted on an arbitary CY$_2$ manifold as we explain in section \ref{sec:D3circle}.

With the scope of our method for constructing circle compactifications established and some simple classes fully elucidated, we explore some explicit solutions we can construct within them in section \ref{eq:explicitsols}. Our focus for applications is two fold considering Minkowski flux vacua in section \ref{sec:minkvac} and generalisations of AdS solitons we will refer to as solitonic branes in \ref{eq:solbranes}.

With respect to flux vacua we begin in section \ref{sec:universalcompactansatz} where, using the classes of sections  \ref{sec:D1circle}, \ref{sec:D3circle} and  \ref{sec:D5circle}, we construct universal Mink$_p$ vacua for $p=1,3,5$. The internal spaces of these solutions decompose as a U(1) fibration over a compact space consisting of a CY cone over any compact Sasaki-Einstein manifold whose radial coordinate is bounded between two Op planes. We show that it is also possible to construct more general boundary behaviour given a specific choice of Sasaki-Einstein in section \ref{eq:Addtionalmink5}. Here we choose the 3-sphere and establish that it is possible to also bound solutions between D5 branes and an O5 plane, or  a regular zero and an O5. Finally we broaden our horizons and consider similar constructions for Mink$_2$ vacua for which the internal space is now a U(1) fibration over a G$_2$ cone in section \ref{sec:D2ex}. Taking S$^3\times$S$^3$ as our choice of Nearly-Kahler base we are able to bond this cone between either D2 branes and an O2 plane, a regular zero and an O2 plane, or two O2 planes. 

We end the main text with section \ref{eq:solbranes} where we turn our attention to embeddings of AdS solitons and  solitonic branes more generally into higher dimensions. In section \ref{sec:solex} we present 3 known solitonic brane solutions of type IIB supergravity that were first constructed in gauged supergravity then lifted to $d=10$ through  consistent truncations on spheres \cite{Kumar:2024pcz}. We show that, despite initial appearances, they in fact lie within  classes of circle compactified D$p$ branes backreacted on generic CY$_{\frac{9-p}{2}}$ manifolds we derived earlier - specifically for CY manifolds that are cones over U($\frac{9-p}{2}$) preserving squashed spheres. This squashing preserves a $\mathbb{CP}^{\frac{7-p}{2}}$ factor which can be replaced with any Kahler Einstein manifold allowing one to easily construct many more solitonic brane solutions as explained in section \ref{sec:solnew} (the solutions are summerised in \eqref{eq: gensolbrane}). Here we are also able to generalise the harmonic function of the solitonic brane solutions allowing us to backreact source D$p$ branes on top of them - see \eqref{eq: gensolbrane2} for a summary. This modification of the harmonic factor, previously found in \cite{Benvenuti:2005qb} for $p=3$, preserves the asymptotics of the original background at long distances, while near the end of the space it behaves as a D$p$-brane source smeared over the Kahler Einstein manifold. Holographically, this corresponds to turning on a vev for a meson operator. Finally, in section \eqref{sec:solAdS4} we show through dualities that one can also use our results to generialise the embedding of the AdS$_4$ soliton \cite{Anabalon:2021tua} into $d=11$  to include generic Kahler Einstein manifolds and source M2 branes. The resulting solution is in \eqref{11DSmeared}.

Although we have chosen to focus on solutions whose internal space are fibrations over conformal holonomy manifolds, let us stress that the scope of the supersymmetry generating technique we have derived is far broader than that. Indeed it should be possible to apply it to a vast swathe of supersymmetric Minkowski solutions in type II supergravity to construct circle compactifications. He hope to report on work in this direction soon and invite those interested to do likewise.

Our work is supplemented by several appendices. There are three that we have not already mentioned explicitly. In appendix \ref{sec:convensions}  our conventions for type II supergravity are spelled out. In appendix \ref{MinkDtoDm1derivation} we derive the G-strucute condtions for Mink$_{D}$ solutions for $D=3,...,6$ starting from those of Mink$_2$ - this is important for our compactification procedure which is a generalisation of this. While appendix \ref{sec:gstructuresapp} gives more details on the G-structures relevant to this work, the forms that span them, and how the bilinears that appear in the supersymmetry conditions can be shown to decompose in terms of these G-structure forms.

\section{Mink\texorpdfstring{$_2$}{2} G-structures and compactification to Mink\texorpdfstring{$_1$}{1}}\label{sec:Mink2part}

In this section we summarise the necessary and sufficient geometric conditions for supersymmetric Mink$_2$ solutions of type II supergravity derived in appendix \ref{sec:mink2susyderivation}. We also find what additional conditions must be imposed to have a solution and parametrise the geometric conditions in terms of G-structures, G$_2$-structure in IIA and SU(4)-structure in IIB.\\
~~\\
A Mink$_2$ solution in type II supergravity is a solution that can be decomposed as
\begin{align}
ds^2&=e^{2A}ds^2(\text{Mink}_2)+ ds^2(\text{M}_8),~~~~H= e^{2A}H_1\wedge \text{vol}(\text{Mink}_2)+ H_3,\nn\\[2mm]
F_{\pm}&= f^{(8)}_{\pm}+ e^{2A}\text{vol}(\text{Mink}_2)\wedge \star_8\lambda(f^{(8)}_{\pm}),\label{eq:Mink2bosonicfields}
\end{align}
where $(e^A,f^{(8)}_{\pm},H_1,H_3)$ and the dilaton only have support on M$_8$ so that the SO(1,1) symmetry of Mink$_2$ is preserved and acting on a $k$-form $C_k$, and  $\lambda(C_k)=(-)^{[\frac{k}{2}]}C_k$.

To have a solution we should solve the type II equations of motion (see appendix \ref{sec:convensions} for the conventions used through this text), which include the Bianchi  identities $dH=0$ and $d_HF_{\pm}=(d-H\wedge)F_{\pm}=0$ - which should hold in regular parts of the space. Given the decomposition \eqref{eq:Mink2bosonicfields} these split into their parts orthogonal and parallel to Mink$_2$ as
\begin{subequations}
\begin{align}
d_{H_3}f^{(8)}_{\pm}&=0,~~~~dH_3=0,\label{eq:magbimink2}\\[2mm]
d_{H_3}(e^{2A}\star_8\lambda(f^{(8)}_{\pm}))&=  e^{2A}H_1\wedge f^{(8)}_{\pm},~~~~ d(e^{2A}H_1)=0,\label{eq:elbimink2}
\end{align}
\end{subequations}
yielding the magnetic and electric contributions to the Bianchi identities.
\\ 
~~\\
A  Mink$_2$ solution preserves ${\cal N}=(1,0)$ supersymmetry when the $d=10$ Majorana-Weyl Killing spinors decompose as
\beq
\epsilon_1=  \zeta^{(2)}_+\otimes  \chi^{(8)}_{1+},~~~~~\epsilon_2=  \zeta^{(2)}_+\otimes  \chi^{(8)}_{2\mp},\label{eq:deq10mink2}
\eeq
where $\zeta^{(2)}_+$ are positive chirality Majorana spinors on Mink$_2$ and $\chi_{i\pm}$ are Majorana-Weyl spinors on the internal manifold M$_8$. The latter can be taken to obey
\beq
|\chi^{(8)}_{1+}|^2= 2e^A c_0 \cos^2\left(\frac{\beta}{2}\right),~~~~|\chi^{(8)}_{2\mp}|^2= 2e^A c_0 \sin^2\left(\frac{\beta}{2}\right),\label{eq:norms8d}
\eeq
without loss of generality for $\beta$, a point dependent angle, and where  $\sin\beta \neq 0$ will be assumed, as otherwise a non trivial RR sector is not possible.
Given this, supersymmetry can be phrased in terms of the bilinear
\beq
\slashed{\Psi}^{(8)}_{\mp}=\frac{1}{e^A}\chi^{(8)}_{1+}\otimes \chi^{(8)\dag}_{2\mp},
\eeq
where $\Psi^{(8)}_{\mp}$ is a Poly-form of odd/even form degree defined in \eqref{eq:psi8poly} which is  ``self dual''  such that
\beq
\star_8\lambda(\Psi^{(8)}_{\pm})=\Psi^{(8)}_{\pm}.
\eeq
The conditions for unbroken supersymmetry are
\begin{subequations}
\begin{align}
&d(e^{2A}\cos\beta)- e^{2A}H_1=0,\label{eq:BPSM81}\\[2mm]
&d_{H_3}(e^{2A-\Phi}\Psi^{(8)}_{\mp})=\pm \frac{e^{2A}c_0}{16}(-\cos\beta f^{(8)}_{\pm}+ \star_8\lambda (f^{(8)}_{\pm})),\label{eq:BPSM82}\\[2mm]
&(\Psi^{(8)}_{\mp},f^{(8)}_{\pm})_7 =\mp \frac{c_0}{8}e^{-\Phi}\star_8( 2 dA- \cos\beta H_1),\label{eq:BPSM83}\\[2mm]
&(\Psi^{(8)}_{\mp},\star_8 \lambda(f^{(8)}_{\pm}))_7 =\pm \frac{c_0}{8}e^{-\Phi}\star_8( 2 \cos\beta dA-  H_1),\label{eq:BPSM84}
\end{align}
\end{subequations}
where $(X,Y)_k=  X\wedge \lambda (Y)\bigg\lvert_k$.

\subsection{Integrability}\label{eq:integrabilitymink2}
Of course \eqref{eq:BPSM81}-\eqref{eq:BPSM84} only ensure consistency with supersymmetry, to have a solution we must also solve the type II equations of motion. We note that if the magnetic Bianchi identities \eqref{eq:magbimink2} are assumed to hold then \eqref{eq:BPSM81} and \eqref{eq:BPSM82} imply their electric counterparts \eqref{eq:elbimink2}. Further as \eqref{eq:deq10mink2} imply a canonical $d=10$  Killing vector $K$ that is null (see appendix \ref{sec:mink2susyderivation}) we can use the integrability results of \cite{Giusto:2013rxa} to address the status of the remaining type II equations of motion. This informs us that  when supersymmetry is preserved \eqref{eq:magbimink2} will imply all but a single component of Einstein’s equations, namely ${\cal E}_{vv}=0$ (where $v$ is a 1-form/vector on Mink$_2$ defined in appendix \ref{sec:mink2susyderivation}). However, for the decomposition \eqref{eq:Mink2bosonicfields} one necessarily has 
\beq
R_{\mu\nu}\propto g^{(2)}_{\mu\nu},~~~\nabla_{\mu}\nabla_{\nu}\Phi\propto g^{(2)}_{\mu\nu},~~~ \iota_{v}H\propto v\wedge H_1,~~~ \iota_vF_{\pm}\propto v\wedge \star\lambda(f^{(8)}_{\pm}),
\eeq 
for $(\mu,\nu)$ coordinates on Mink$_2$ and $g^{(2)}_{\mu\nu}$ its unwarped metric. We thus find that
\beq
{\cal E}_{vv}\propto v^{\mu}v_{\mu},
\eeq
where  ${\cal E}_{MN}$ is defined in \eqref{eq:compactEOM}. But $v$ is null, so ${\cal E}_{vv}=0$ is implied in this case.\\
~\\
We thus conclude that if \eqref{eq:BPSM81}-\eqref{eq:BPSM84} hold, and one additionally imposes \eqref{eq:magbimink2}, one necessarily has an ${\cal N}=(1,0)$ preserving Mink$_2$ solution.

\subsection{Parameterisation in terms of G-structures}\label{eq:Mink2Gstructures}
In this section we will make the form of the bilinears $\Psi^{(8)}_{\mp}$ explicit by decomposing them in terms of the forms that span a G-structure, specifically G$_2$- and SU(4)-structure in respectively IIA and IIB.\\
~\\
It is in general possible to express all the internal spinors appearing in  \eqref{eq:norms8d} in terms of two unit norm Majorana-Weyl spinors $\chi^{(8)}_{\pm}$ and 1-form $U$ that is likewise unit norm as
\begin{align}
\chi^{(8)}_{1+}&= \sqrt{2 c_0}e^{\frac{A}{2}}\cos\left(\frac{\beta}{2}\right)\chi^{(8)}_+\nn,\\[2mm]
\chi^{(8)}_{2-}&= \sqrt{2 c_0}e^{\frac{A}{2}}\sin\left(\frac{\beta}{2}\right)\chi^{(8)}_-,\nn\\[2mm]
\chi^{(8)}_{2+}&= \sqrt{2 c_0}e^{\frac{A}{2}}\sin\left(\frac{\beta}{2}\right)(\cos\alpha\chi^{(8)}_+-\sin\alpha \slashed{U}\chi^{(8)}_-)\label{eq:8dspinordecompIIA}
\end{align}
where the pre-factors are chosen to solve \eqref{eq:deq10mink2}.
Then by substituting these spinors into \eqref{eq:8dformsdef} we find 
\begin{subequations}
\begin{align}
\Psi^{(8)}_-&=\frac{c_0}{16}\sin\beta\bigg(V- \Phi_3+\star_8 \Phi_3-\iota_{V}\text{vol}(\text{M}_8)\bigg)\label{eq:mink2iiA},\\[2mm]
\Psi^{(8)}_+&=\frac{c_0}{16}\sin\beta\text{Re}\bigg(e^{i\alpha} e^{-i J^{(8)}}- e^{-i \alpha}\Omega^{(8)}\bigg),\label{eq:mink2iiB}
\end{align}
\end{subequations}
which is derive in detail in appendix \ref{sec:gstructuresapp}. We have that in  IIA $\Psi^{(8)}_-$ decomposes in terms of the 3-form $\Phi_3$ that defines a G$_2$ -structure in 7 dimensions and a vielbein direction orthogonal to it $V$ such that the metric decomposes as
\beq
ds^2(\text{M}_8)=  ds^2(\text{G}_2)+V^2,
\eeq
with $\Phi_3$ defined in terms of the vielbein components spanning the G$_2$ manifold. In IIB on the other hand, $\Psi^{(8)}_+$ decomposes in terms of SU(4)-structure forms $(J^{(8)},\Omega^{(8)})$ and a function $\alpha$. Fixing $\alpha=0$ in IIB enhances the SU(4)-structure M$_8$ experiences generically to Spin(7) with self dual 4-form given by
\beq
\psi_4=\frac{1}{2}J^{(8)}\wedge J^{(8)}+\text{Re}\Omega^{(8)}.\label{eq:spin74form}
\eeq

\subsection{Recovering D1 branes backreacted on Spin(7)-manifolds}\label{sec:D1branes}
In this section, as a test of our Mink$_2$ supersymmetry conditions we will recover a class of solutions that should be compatible with them, namely D1 branes backreacted on manifolds of Spin(7) holonomy - we will actually be bit more general than this and allow for a non trivial RR 3-from.

We begin by imposing $(\alpha=0,\beta=\frac{\pi}{2})$ in type IIB and assume that the only non trivial RR fluxes are defined by
\beq
f^{(8)}_-=f_5+f_7.
\eeq
This means that \eqref{eq:BPSM81}-\eqref{eq:BPSM84}  reduce to 
\begin{align}
&H_1=0,~~~~d(e^{-2A-\Phi})=0,~~~~ d(e^{2A-\Phi}\psi_4)=0,~~~~H_3\wedge \psi_4=0,\\[2mm]
& e^{2A}\star_8 \lambda (f^{(8)}_-)=- d(e^{2A-\Phi})+e^{2A-\Phi}H_3,\label{eq:susyD1case}
\end{align}
where $\psi_4$ is the 4-form associated to a Spin(7) structure defined in \eqref{eq:spin74form}. We see then that the M$_8$ is conformally a manifold of Spin(7) holonomy, a subset of which, through the decomposition \eqref{eq:spin74form}, are clearly Calabi-Yau 4-folds, \textit{i.e.}
\beq
e^{4A}\psi_4=  \psi_{\text{Spin(7)}}~~~\Rightarrow~~~ d\psi_{\text{Spin(7)}}=0,
\eeq
so that $\psi_{\text{Spin(7)}}$ is the expected tensionless self dual 4-form. Fixing the Mink$_2$ warp factor in terms of another arbitrary function
\beq
e^{-4A}=h_1,
\eeq
The backgrounds compatible with supersymmetry then decompose as
\begin{align}
ds^2&= \frac{1}{\sqrt{h_1}}ds^2(\text{Mink}_2)+ \sqrt{h_1} ds^2(\text{Spin(7)}),~~~~e^{-\Phi}=\frac{e^{-\Phi_0}}{\sqrt{h_1}},~~~~H=H_3,\nn\\[2mm]
F_5&= e^{-\Phi_0}\left(-\star_8 H_3+\frac{1}{h_1}\text{vol}(\text{Mink}_2)\wedge H_3\right),\nn\\[2mm]
F_7&= e^{-\Phi_0}\hat\star_8 d h_1,\label{eq:D1branes}
\end{align}
where $\hat\star_8$ is the hodge dual on the unwarped Spin(7) manifold which $h_1$ has support on and $e^{-\Phi_0}$ is a constant. One has a solution away from possible sources whenever the magnetic Bianchi identities of \eqref{eq:magbimink2} hold, these lead to 
\beq
dH_3=0,~~~d\hat\star_8 H_3=0,~~~\nabla^2_{\text{Spin(7)}}h_1+\hat{H}_3^2=0,
\eeq
through the identities $d\hat\star_8 d h_1=-\nabla^2_{\text{Spin(7)}}h_1\text{vol}(\text{Spin(7)})$ and $\hat\star_8 H_3\wedge H_3=\hat{H}_3^2\text{vol}(\text{Spin(7)})$. We have thus derived a class describing D1 branes backreacted on an arbitrary Spin(7) manifold in the presence of harmonic NS 3-form flux - setting $H_3=0$ gives the expected class of D1 branes backreacted on Spin(7) manifolds.

\subsection{Compactifying  Mink\texorpdfstring{$_2$}{2} to Mink\texorpdfstring{$_1$}{1}}\label{eq:Mink2toMink1}
As we derive in detail in appendix \ref{sec:compactifyingMink2}, it is possible to generalise \eqref{eq:Mink2bosonicfields} by sending 
\beq
ds^2(\text{Mink}_2)\to  -dt^2+D\phi^2,~~~~D\phi= d\phi+{\cal A},~~~~{\cal F}=d{\cal A},
\eeq
for $\partial_{t},\partial_{\phi}$ respectively time- and space-like isometries of the metric and fluxes such that we  fiber Mink$_2$ over the internal space. Of course if we compactify the $\phi$ direction then  Mink$_2\times$M$_8\to$Mink$_1\times$M$_9$ for M$_9$ a circle fibration over M$_8$.  Consistency with supersymmetry imposes that solutions take the form
\begin{align}
ds^2&=e^{2A}(-dt^2+ D\phi^2)+ ds^2(\text{M}_8),\nn\\[2mm]
H&= d\left(e^{2A}\cos\beta  dt\wedge D\phi\right)+ e^{A}H_2\wedge (dt+D\phi)+ H_3,\\[2mm]
F_{\pm}&= f^{(8)}_{\pm}+ e^{2A}dt\wedge D\phi\wedge \star_8\lambda(f^{(8)}_{\pm})+e^A(dt+ D\phi)\wedge g^{(8)}_{\mp}\mp \frac{8}{c_0}e^{2A-\Phi} (dt- D\phi)\wedge{\cal F}\wedge \Psi^{(8)}_{\mp}\nn,
\end{align}
where the only dependence on $(t,\phi)$ is in $(dt,d\phi)$, $\star_8\lambda(g^{(8)}_{\mp})=g^{(8)}_{\mp}$ and ${\cal F}$ is constrained such that in respectively IIA and IIB
\begin{align}
&{\cal F}\wedge\iota_V\star_8\Phi_3=0~~~~\Rightarrow~~~~\star_8 {\cal F}=-{\cal F}\wedge \Phi_3\wedge V,\nn\\[2mm]
\sin\alpha {\cal F}\wedge \text{Im}( e^{-i \alpha}\Omega^{(8)})&=0,~~~~\star_8{\cal F}= - {\cal F}\wedge \left(\frac{1}{2}J^{(8)}\wedge J^{(8)}+\cos\alpha \text{Re}( e^{-i \alpha}\Omega^{(8)})\right),\label{eq:calFMink2toMink1}
\end{align}
which imply that $\star_8\lambda({\cal F}\wedge \Psi^{(8)}_{\mp})=-{\cal F}\wedge \Psi^{(8)}_{\mp}$. These conditions make ${\cal F}$ a ``primitive'' 2-form in the sense that they impose the same constraints on ${\cal F}$ as imposing $\slashed{\cal F}\chi^{(8)}_{1+}=\slashed{\cal F}\chi^{(8)}_{2\mp}=0$ would\footnote{The nomenclature is in reference to the primitive $(1,1)$-forms that can be defined for an SU(n)-structure, indeed if one is in IIB with $\alpha=\frac{\pi}{2}$ then ${\cal F}$ becomes a primitive (1,1)-form.}.  In addition to this, such solutions must still obey the supersymmetry conditions for Mink$_2$, \eqref{eq:BPSM81}-\eqref{eq:BPSM84} as well as some additional constraints on $(H_2,g^{(8)}_{\mp})$, namely
\begin{subequations}
\begin{align}
&(\Psi^{(8)}_{\mp},g^{(8)}_{\mp})_8=0,\label{eq:Mink1bps1}\\[2mm]
&(\Psi^{(8)}_{\mp},g^{(8)}_{\mp})_6=\frac{c_0 e^{-\Phi}}{16}\bigg(\mp \cos\beta\star_8H_2  + X^{\pm}_{4}\wedge H_2\bigg)\label{eq:Mink1bps2},
\end{align}
\end{subequations}
must hold in general for
\begin{align}
X^+_{4}&=\star_8\Phi_3+\cos\beta  V\wedge \Phi_3,\\[2mm]
X^-_{4}&=-\sin\alpha  \text{Im}( e^{-i \alpha}\Omega^{(8)})+\cos\beta(\frac{1}{2}J^{(8)}\wedge J^{(8)}+\cos\alpha \text{Re}( e^{-i \alpha}\Omega^{(8)})),
\end{align}
while in IIB $g_4$ must in general obey
\begin{subequations}
\begin{align}
&\sin\alpha \text{Im}( e^{-i \alpha}\Omega^{(8)}) \wedge g_4=\frac{1}{3} e^{-\Phi}H_2\wedge J^{(8)}\wedge J^{(8)}\wedge J^{(8)}=0,\label{eq:Mink1bps3}\\[2mm]
&\sin\alpha\neq 0~~~~\text{or}~~~~\sin\beta Z_2\wedge J^{(8)}\wedge g_4+\cos\alpha e^{-\Phi}Z_2\wedge\text{Im}( e^{-i \alpha}\Omega^{(8)})\wedge H_2=0,\label{eq:Mink1bps4}
\end{align}
\end{subequations}
where $Z_2$ is defined below \eqref{eq:pairinghard4}.

When supersymmetry holds the Bianchi identities for the RR and NS fluxes are all implied when one imposes
\begin{align}
&dH_3+ e^{A}{\cal F}\wedge H_2=0,\nn\\[2mm]
&d_{H_3}f^{(8)}_{\pm}+{\cal F}\wedge \left(e^{A}g^{(8)}_{\pm}\pm \frac{8}{c_0}e^{2A-\Phi}{\cal F}\wedge \Psi^{(8)}_{\mp}\right)=0,\nn\\[2mm]
&d_{H_3}\left(e^{A}g^{(8)}_{\pm}\pm \frac{8}{c_0}e^{2A-\Phi}{\cal F}\wedge \Psi^{(8)}_{\mp}\right)+e^{A}H_2\wedge f^{(8)}_{\pm}=0,
\end{align}
which of course reduces to the Mink$_2$ result for ${\cal F}=H_2=g^{(8)}_{\mp}=0$. When these conditions hold we know from \cite{Giusto:2013rxa} that all the EOM are implied by supersymmetry except for the component of Einsteins equations ${\cal E}_{vv}$, this is not implied in this case so needs to be additionally imposed - that is unless they happen to be implied for a particular class of solutions, as is the case in the next section.

\subsection{D1 branes wrapping a circle} \label{sec:D1circle}
In this section we will use the Mink$_1$ supersymmetry conditions of section \eqref{eq:Mink2toMink1} to generalise the class of section \ref{sec:D1branes} such that the D1 branes are now compactified on a circle.\\
~\\
We seek a compactification of \eqref{eq:D1branes}, thus we once more assume that  $(\alpha=0,\beta=\frac{\pi}{2})$ and  $f^{(8)}_-=f_5+f_7$ which leads again to \eqref{eq:susyD1case}, while the field strength must obey
\beq
\star_8{\cal F}= -{\cal F}\wedge \psi_4.
\eeq
We now need to choose a $(g^{(8)}_{\mp}, H_2,{\cal F})$ compatible with supersymmetry. We find that a particularly simple way to solve \eqref{eq:Mink1bps1}-\eqref{eq:Mink1bps2} and \eqref{eq:Mink1bps3}-\eqref{eq:Mink1bps4} in general is to fix
\begin{align}
H_2&=0,~~~~  e^{A} g^{(8)}_{-}=  e^{2A-\Phi}p {\cal F}\wedge\lambda(\Psi^{(8)}_{-}),\nn\\[2mm]
e^{A} g^{(8)}_{+}&=  \frac{1}{2}e^{2A-\Phi}p {\cal F}\wedge\left(1+\frac{1}{2}J^{(8)}\wedge J^{(8)}+ \cos\alpha \text{Re}(e^{-\alpha}\Omega^{(8)})\right),
\end{align}
for $p$ an arbitrary function of M$_8$. This permits us to take the ten dimensional flux to be
\beq
F_+= \left(1+\star\lambda \right)\left(\hat\star_8 d h_1-\hat\star_7 H_3+\frac{1}{2}e^{2A-\Phi}D\phi\wedge {\cal F}\wedge\bigg[ p(1+\psi_4)+ \psi_4-1\bigg]\right).
\eeq
If we demand that we only turn on additional components to the RR flux we originally had turned on, we should fix $p=1$ meaning that we now have class of the form
\begin{align}
ds^2&= \frac{1}{\sqrt{h_1}}ds^2(-dt^2+D\phi^2)+ \sqrt{h_1} ds^2(\text{Spin(7)}),~~~~e^{-\Phi}=\frac{e^{-\Phi_0}}{\sqrt{h_1}},~~~H=H_3\nn\\[2mm]
F_5&= e^{-\Phi_0}\left(-\star_8 H_3+\frac{1}{h_1}dt\wedge D\phi\wedge H_3\right),\nn\\[2mm]
F_7&=e^{-\Phi_0}\left(\hat\star_8 d h_1+D\phi\wedge {\cal F}\wedge \psi_{\text{Spin(7)}}\right),
\end{align}
which solves all of the supersymmetry constraints.
The Bianchi gets modified with respect to the non compactified D1 brane as
\beq
\nabla^2_{\text{Spin(7)}}h_1+\hat{H}_3^2+\hat{\cal F}^2=0,
\eeq
where ($\hat{\cal F}^2$, $\hat{H}_3^2$) are defined as in \eqref{eq:sqdefs}, but computed on M$_{\text{Spin}(7)}$ without the warping.
We will return to this class in sections \ref{sec:universalcompactansatz} and \ref{sec:solex} where we construct explicit solutions within it.

\section{Compactifying Mink\texorpdfstring{$_{D+1}$}{(D+1)} to Mink\texorpdfstring{$_{D}$}{D}}\label{eq:MinkDp1tominkD}
Mink$_{D}$ solutions of type II supergravity can be decomposed in the form
\begin{align}
ds^2&=e^{2A}ds^2(\text{Mink}_D)+ds^2(\text{M}_{d}),\nn\\[2mm]
F_{\pm}&=f^{(d)}_{\pm}+ e^{4A}\text{vol}(\text{Mink}_{D})\wedge \star_{d}\lambda (f^{(d)}_{\pm}),\nn\\[2mm]
H&=e^{D A}H_{3-D}\wedge \text{vol}(\text{Mink}_{D})+H_3,
\end{align}
where $(e^{2A}, H_3,f^{(d)}_{\pm},H_{3-D})$ and the dilaton $\Phi$ have support on the $d=10-D$ dimensional internal space $\text{M}_{d}$ so that the isometries of $\text{Mink}_D$ are respected and $H_{3-D}$ is only non trivial for $D=1,2,3$. In this section we derive a procedure to  compactifiy supersymmetric Mink$_{D+1}$ solutions  to Mink$_{D}$ as 
\beq
ds^2(\text{Mink}_{D+1})\to  ds^2(\text{Mink}_{D})+ D\phi^2,~~~~ D\phi=d\phi+{\cal A},~~~{\cal F}=d{\cal A},
\eeq
where ${\cal F}$ has support on M$_{d-1}$ and $\partial_{\phi}$ is an isometry of the full solutions. We have already seen how this works for Mink$_2\to$Mink$_1$ in section \ref{eq:Mink2toMink1}, so our focus here will be on $D=2,3,4,5$. The procedure fixes ${\cal F}$ to be a ``primitive'' 2 form in the sense that $\slashed{\cal F}$ annihilates the spinors that the internal space supports as we show in appendix \ref{sec:gstructuresapp}. Such forms can only exist when  $\text{M}_{d-1}$ supports at least an SU(2)-structure, making Minkowksi $D+1=6$ the largest external space for which this is possible.  

\subsection{Circle compactifications of Mink\texorpdfstring{$_3$}{3}}\label{sec:compact3to2}
We can derive circle compactifications of Mink$_3$ to Mink$_2$ in a similar way to how Mink$_3$ supersymmetry conditions are derived from those of Mink$_2$  in appendix \ref{sec:Mink3fromMink2}.

We begin with a Mink$_2$ solution like \eqref{eq:Mink2bosonicfields} subject to the constraints \eqref{eq:BPSM81}-\eqref{eq:BPSM84}. The first step is to decompose the the $d=8$ bi-linears as
\beq
\Psi^{(8)}_{\mp}=\frac{1}{2}(\Psi^{(7)}_{\mp}+\Psi^{(7)}_{\pm}\wedge W),
\eeq
where the $d=7$ bilinears decompose in terms of $d=7$ SU(3)-structure forms $(V,J^{(6)},\Omega^{(6)})$ and a function $\alpha$ as
\begin{align}
\Psi^{(7)}_+&= \frac{c_0}{8}\sin\beta \text{Re}\bigg[e^{i \alpha}e^{-i J^{(6)}}-e^{-i\alpha}\Omega^{(6)}\wedge V\bigg],\nn\\[2mm]
\Psi^{(7)}_-&= \frac{c_0}{8}\sin\beta \text{Im}\bigg[e^{i \alpha}e^{-i J^{(6)}}\wedge V+e^{-i\alpha}\Omega^{(6)}\bigg], \label{eq:d7biliears}
\end{align}
and experience an enhancement to G$_2$-structure when $\alpha=0$. This step is just as we did to derive Mink$_3$ conditions, only now we identity
\beq
W= e^{A} D\phi,
\eeq
allowing the isometry $\partial_{\phi}$ to be a non trivial fibration over M$_8$,  we assume $\Psi^{(8)}_{\pm}$ are not charged under $\partial_{\phi}$. Decomposing $(f^{(8)}_{\pm},H_3)$ into their parts orthogonal and parallel to $W$ as
\beq
f^{(8)}_{\pm}= f^{(7)}_{\pm}+ W\wedge f^{(7)}_{\mp},~~~\star_8\lambda(f^{(8)}_{\pm})=-W\wedge \star_7\lambda( f^{(7)}_{\pm})+\star_7\lambda(f^{(7)}_{\mp}),~~~H_3=\tilde{H}_3+W\wedge H_2,
\eeq
and inserting our ansatz into \eqref{eq:BPSM81}-\eqref{eq:BPSM82} we find 
\begin{align}
&d(e^{2A}\cos\beta)= e^{2A}H_1,~~~~\iota_WH_1=0,\nn\\[2mm]
&d_{\tilde{H}_3}\left(e^{2A-\Phi}\Psi^{(7)}_{\mp}\right)=\mp e^{2A}\left(e^{A-\Phi}{\cal F}\wedge \Psi^{(7)}_{\pm}-\frac{c_0}{8}(-\cos\beta f^{(7)}_{\pm}+\star_7\lambda(f^{(7)}_{\mp})\right),\nn\\[2mm]
&d_{\tilde{H}_3}(e^{3A-\Phi}\Psi^{(7)}_{\pm})\pm e^{3A-\Phi} H_2 \wedge \Psi^{(7)}_{\mp}=\frac{c_0}{8}e^{3A}\left(\cos\beta f^{(7)}_{\mp}+\star_7 \lambda (f^{(7)}_{\pm})\right).
\end{align}
In general these have little to do with the supersymmetry conditions for Mink$_3$, but with the ansatz 
\beq
H_1=0,~~~~\star_7\lambda(f^{(7)}_{\mp})=\frac{8}{c_0}e^{A-\Phi}{\cal F}\wedge \Psi^{(7)}_{\pm},~~~~H_2=0,~~~~\cos\beta =\frac{\pi}{2},
\eeq
they reduce to
\begin{align}
&d_{H_3}\left(e^{2A-\Phi}\Psi^{(7)}_{\mp}\right)=0,\nn\\[2mm]
&d_{H_3}(e^{3A-\Phi}\Psi^{(7)}_{\pm})=\frac{c_0}{8}e^{3A}\star_7 \lambda (f^{(7)}_{\pm}),\label{eq:mink3differential}
\end{align}
reproducing the differential constrains for Mink$_3$ in the $2\beta=\pi$ limit precisely.  Now, under these assumptions we find that \eqref{eq:BPSM83}-\eqref{eq:BPSM84} reproduce the pairing constraint for Mink$_3$
\beq
(\Psi^{(7)}_{\mp},f^{(7)}_{\pm})_7=0,\label{eq:mink3pairing}
\eeq
as well as some additional conditions involving ${\cal F}$, namely
\begin{subequations}
\begin{align}
(\Psi^{(7)}_{\pm}, {\cal F}\wedge \Psi^{(7)}_{\pm})_6&=0,\label{eq:primativecondmink31s}\\[2mm]
(\Psi^{(7)}_{\pm}, \star_7\lambda ({\cal F}\wedge \Psi^{(7)}_{\mp}))_6&=0.\label{eq:primativecondmink32s}
\end{align}
\end{subequations}
It is possible to show that these constraints on ${\cal F}$ are equivalent to
\beq
{\cal F}\wedge(\frac{1}{2}\cos\alpha J^{(6)}\wedge J^{(6)}+\text{Re}(e^{-i\alpha}\Omega^{(6)})\wedge V)=0,~~~~\sin\alpha{\cal F}\wedge \text{Im}(e^{-i\alpha}\Omega^{(6)})=0,\label{eq:primativecondmink33}
\eeq
and imply that
\begin{align}
\lambda(\star_8({\cal F}\wedge \Psi^{(7)}_{\pm}))= W\wedge {\cal F}\wedge \Psi^{(7)}_{\mp},
\end{align}
enabling us to take the hodge dual of $\star_8\lambda (f^{(8)}_{\pm})$ such that we find
\beq
f^{(7)}_{\mp}=\pm \frac{8}{c_0}e^{A-\Phi}{\cal F}\wedge \Psi^{(7)}_{\mp}.
\eeq
In summary we have derived a class of solutions that take the form
\begin{align}
ds^2&=e^{2A}(ds^2(\text{Mink}_2)+D\phi^2)+ds^2(\text{M}_7),~~~~H=H_3,\nn\\[2mm]
F_{\pm}&=\left(1+\star\lambda\right)\left(f_{\pm}^{(7)}\pm \frac{8}{c_0}D\phi\wedge {\cal F}\wedge(e^{2A-\Phi}\Psi^{(7)}_{\mp})\right),
\end{align}
that are supersymmetric whenever the $d=7$ bilinears obey \eqref{eq:mink3differential} and \eqref{eq:mink3pairing}, which are simply the supersymmetry conditions for Mink$_3$ solutions, and ${\cal F}$ must solve \eqref{eq:primativecondmink33}. 

Consistency with supersymmetry is not sufficient to have a solution, we must also solve the magnetic Bianchi identities of the fluxes, namely \eqref{eq:magbimink2}. Due to \eqref{eq:mink3differential} and $d{\cal F}=0$ these become
\begin{align}
&dH_3=0,\nn\\[2mm]
&(d-H_3\wedge)f_{\pm}^{(7)}\pm \frac{8}{c_0}{\cal F}\wedge {\cal F}\wedge(e^{2A-\Phi}\Psi^{(7)}_{\mp})=0,
\end{align}
meaning that, given a suitable ${\cal F}$, the necessary conditions for supersymmetry are implied by those of Mink$_3$ but the RR Bianchi identity becomes modified with respect to the Mink$_3$ case. 

\subsubsection{D2 branes wrapping a circle}\label{eq:D2section}
We will now derive a subclass of solutions within our ansatz that describe D2 branes wrapping the S$^1$ direction spanned by $\partial_{\phi}$ that are backreacted on G$_2$ manifolds.

We begin by fixing $\alpha=\beta=\frac{\pi}{2}$ which puts us in the G$_2$-structure limit for which
\beq
\Psi^{(7)}_-=\frac{c}{8}\left(-\Phi_3+\text{vol}(\text{M}_7)\right),~~~~\Psi^{(7)}_+=\frac{c}{8}\left(1-\star_7\Phi_3\right).
\eeq
The conditions \eqref{eq:mink3differential} and \eqref{eq:mink3pairing} in type IIA then reduce to
\begin{subequations}
\begin{align}
d(e^{2A-\Phi}\Phi_3)&=0,~~~H_3\wedge \Phi_3=0,~~~H_3\wedge \star_7\Phi_3=0,\\[2mm]
e^{3A}\star_7\lambda(f^{(7)}_+)&=d(e^{3A-\Phi})-e^{3A-\Phi}H_3-d(e^{3A-\Phi}\star_7\Phi_3).
\end{align}
\end{subequations}
We will make the assumption that the internal 7-manifold is conformally a G$_2$ manifold, \text{i.e}
\beq
ds^2(\text{M}_7)=e^{-2A}ds^2(\text{M}_{\text{G}_2}),~~~~ e^{3A}\Phi_3=\Phi_{\text{G}_2},~~~~e^{4A}\star_7\Phi_3=\hat{\star}_7\Phi_{\text{G}_2},
\eeq
for $d\Phi_{\text{G}_2}=d\hat{\star}_7\Phi_{\text{G}_2}=0$ and $\hat\star_7$ the hodge dual on M$_{\text{G}_2}$. Consistency of this ansatz requires that
\beq
d(e^{-A-\Phi})=0~~~~\Rightarrow~~~~  e^{-\Phi}=e^{-\Phi_0}e^{A},
\eeq
for $\Phi_0$ a constant. Defining $e^{-4A}=h_2$, an arbitrary function of the coordinates on M$_{\text{G}_2}$ we then find that
\beq
f_+^{(7)}=e^{-\Phi_0}\left(\hat\star_7dh_2-\hat\star_7H_3\right).
\eeq
Inserting this into our compactification ansatz we arrive at a class of conformal G$_2$ manifold solutions of the form
\begin{align}
ds^2&= \frac{1}{\sqrt{h_2}}(ds^2(\text{Mink}_2)+D\phi^2)+\sqrt{h_2}ds^2(\text{G}_2),~~~~e^{-\Phi}=e^{-\Phi_0}h_2^{-\frac{1}{4}},~~~H=H_3,\nn\\[2mm]
F_4&=e^{-\Phi_0}\left(-\hat{\star}_7 H_3+\text{vol}(\text{Mink}_2)\wedge d\left(\frac{1}{h_2}D\phi\right)\right),\nn\\[2mm]
F_6&= e^{-\Phi_0}\left(\hat{\star}_7d h_2-D\phi\wedge {\cal F}\wedge\Phi_{\text{G}_2}+\frac{1}{h_2}\text{vol}(\text{Mink}_2)\wedge D\phi\wedge H_3\right),
\end{align}
where $ds^2(\text{G}_2)$ can be the metric on any G$_2$ manifold, meaning it supporting a 3-form $\Phi_{\text{G}_2}$ that is both closed and co-closed\footnote{On unwarped $\text{M}_{\text{G}_2}$.}. Supersymmetry further requires that $(H_3,{\cal F})$ are constrained such that
\beq
H_3\wedge\Phi_{\text{G}_2}=0,~~~~H_3\wedge\hat{\star}_7\Phi_{\text{G}_2}=0,~~~~ {\cal F}\wedge \hat{\star}_7\Phi_{\text{G}_2}=0.
\eeq
One has a solution whenever the magnetic components of the NS and RR Bianchi identities are imposed, which requires away from sources that
\begin{subequations}
\begin{align}
&dH_3=0,~~~~~d\hat\star_7 H_3=0,~~~~\nabla^2_{\text{G}_2}h_2+ \hat {H}_3^2+ \hat{{\cal F}}^2=0,
\end{align}
\end{subequations}
where the hats on $(\hat{{\cal F}}_2^2,\hat {H}_3^2)$ indicate that they are computed on the unwarped manifold M$_{\text{G}_2}$. We thus see that we have a class of solutions with D2 branes wrapping a circle that are backreacted on a G$_2$ manifold in the presence of harmonic NS 3-form flux. We find solutions within this class in section \ref{sec:D2ex}.

\subsection{Circle compactifications of Mink\texorpdfstring{$_4$}{4}}\label{sec:compact4to3}
To derive  circle compactifications of Mink$_4$ to Mink$_3$ we generalise the derivation of the supersymmetry conditions for  Mink$_4$ in appendix \ref{sec:Mink5fromMink4} by taking
\beq
\Psi^{(7)}_{\mp}=-(\text{Re}\Psi^{(6)}_{\mp}+\text{Re}\Psi^{(6)}_{\pm}\wedge W),~~~~~\Psi^{(7)}_{\pm}=(\text{Im}\Psi^{(6)}_{\pm}-\text{Im}\Psi^{(6)}_{\mp}\wedge W),~~~~ W=e^{A}D\phi,
\eeq
for $\partial_{\phi}$ an isometry and where the $d=6$ bilinears, which obey $\iota_{W}\Psi^{(6)}_{\mp}=0$, are spanned by the forms that define a $d=4$ SU(2)-structure $(J^{(4)},\Omega^{(4)})$ and a holomorphic vielbein component $Z$ orthogonal to them as
\begin{align}
\Psi^{(6)}_+&=\frac{c_0}{8}\sin\beta e^{i \alpha}\left(\kappa_{\|} e^{-i J^{(4)}}-\kappa_{\perp}\Omega^{(4)}\right)\wedge e^{\frac{1}{2}Z\wedge \overline{Z}},\nn\\[2mm]
\Psi^{(6)}_-&=\frac{c_0}{8}\sin\beta e^{-i \alpha}\left(\kappa_{\perp} e^{-i J^{(4)}}+\kappa_{\|}\Omega^{(4)}\right)\wedge Z\label{eq:SU(3)XSU(3)bilinears}.
\end{align}
In the above we have that $\kappa_{\|}^2+\kappa_{\perp}^2=1$ and $(\alpha,\beta,\kappa_{\|})$ are functions on the M$_6$ defined by $(J^{(4)},\Omega^{(4)},Z)$. We will again focus on the $2\beta=\pi$ limit meaning that our task now is to insert our ansatz for the $d=7$ bilinears into \eqref{eq:mink3differential} and \eqref{eq:mink3pairing}. Similarly to before we will assume that 
\begin{align}
&\star_7\lambda(f^{(7)}_{\pm})= W\wedge \star_6\lambda(f^{(6)}_{\pm})\pm \frac{8}{c_0} e^{A-\Phi}{\cal F}\wedge \text{Im}\Psi^{(6)}_{\mp},\nn\\[2mm]
&\iota_{W}H_3=0,~~~~\beta=\frac{\pi}{2}~~~~{\cal F}\wedge \text{Re}\Psi^{(6)}_{\pm}=0.\label{eq:assumptionsmink4comp}
\end{align}
Our ansatz then reproduces the supersymmetry conditions for Mink$_4$ solutions in the $2\beta=\pi$ limit, namely
\begin{subequations}
\begin{align}
&d_{H_3}(e^{3A-\Phi}\Psi^{(6)}_{\pm})=0,\label{eq:Mink4susy1}\\[2mm]
&d_{H_3}(e^{2A-\Phi}\text{Re}\Psi^{(6)}_{\mp})=0\label{eq:Mink4susy2} ,\\[2mm]
&d_{H_3}(e^{4A-\Phi}\text{Im}\Psi^{(6)}_{\mp})=\mp \frac{c_0}{8} e^{4A}\star_6\lambda (f^{(6)}_{\pm})\label{eq:Mink4susy3},
\end{align}
\end{subequations}
and imposes the following constraint on ${\cal F}$
\begin{align}
(\text{Im}\Psi^{(6)}_{\mp}, {\cal F}\wedge\text{Im}\Psi^{(6)}_{\mp})_6=0.
\end{align}
Together with our assumption ${\cal F}\wedge \text{Re}\Psi^{(6)}_{\pm}=0$, these conditions are more restrictive than what we have encountered before, in that they not only constrain ${\cal F}$ but also the type of possible G-structure. It possible to show that a non trivial field strength is only possible when
\begin{align}
&\text{IIA}:~~~ \kappa_{\|}=0,~~~~ {\cal F}\wedge J^{(4)}={\cal F}\wedge \Omega^{(4)}=0,\nn\\[2mm]
&\text{IIB}:~~~\kappa_{\perp}=0,~~~~{\cal F}\wedge J^{(6)}\wedge J^{(6)}={\cal F}\wedge \Omega^{(6)}=0.\label{eq:Mink4conds}
\end{align}
This means we are in the strict SU(2)-structure limit in IIA and where we can take
\beq
\Psi^{(6)}_+=-\frac{c_0}{8} \Omega^{(4)}\wedge e^{\frac{i}{2}Z\wedge \overline{Z}},~~~~
\Psi^{(6)}_-=\frac{c_0}{8}  e^{-i J^{(4)}}\wedge Z,
\eeq
without loss of generality and the SU(3)-structure limit in IIB where we can take
\beq
\Psi^{(6)}_{+}=\frac{c_0}{8} e^{i \alpha}  e^{-i J^{(6)}},~~~~\Psi^{(6)}_{-}=\frac{c_0}{8}\Omega^{(6)}\label{eq:SU3structure},
\eeq
also without loss of generality. In either case the conditions imply that ${\cal F}$ is a primitive (1,1)-form on the directions spanning the respective SU(n)-structure. This making it quick to establish that
\beq
f^{(6)}_{\mp}=\mp  e^{2A-\Phi}D\phi\wedge {\cal F}\wedge \text{Re}\Psi^{(6)}_{\mp},
\eeq
where the second of these follows from the form of $f^{(6)}_{\mp}$ and  \eqref{eq:assumptionsmink4comp}.

In summary we have derive a class of solutions of the form
\begin{align}
ds^2&=e^{2A}(ds^2(\text{Mink}_3)+D\phi^2)+ds^2(\text{M}_6),~~~~H=H_3\nn\\[2mm]
F_{\pm}&=\left(1+\star\lambda\right)\left(f_{\pm}^{(6)}\mp \frac{8}{c_0}D\phi\wedge {\cal F}\wedge(e^{2A-\Phi}\text{Re}\Psi^{(6)}_{\mp})\right),
\end{align}
that is supersymmetric when \eqref{eq:Mink4susy1}-\eqref{eq:Mink4susy3} are satisfied and, when ${\cal F}\neq 0$,  M$_6$ supports a strict SU(2)-structure in IIA or SU(3)-structures in IIB which ${\cal F}$ is a primitive (1,1)-form with respect to. When these conditions are met one has a solution in regular parts of the space whenever
\begin{align}
&dH_3=0,\nn\\[2mm]
&(d-H_3\wedge)f_{\pm}^{(6)}\mp \frac{8}{c_0}{\cal F}\wedge {\cal F}\wedge(e^{2A-\Phi}\text{Re}\Psi^{(6)}_{\mp})=0,
\end{align}
are satisfied. Thus we have another class of solutions that share the supersymmetry constrains of Mink solutions in 1 dimension higher, but need to obey modified Bianchi identities.

\subsubsection{D3 branes wrapping a circle}\label{sec:D3circle}
In this  section we will derive a subclass of solutions describing D3 branes wrapping a circle that are backreacted on CY$_3$ manifolds. 

We begin by fixing $\alpha=\beta=\frac{\pi}{2}$ and $\kappa_{\perp}=0$ such that our bilinears take the form
\beq
\Psi_+^{(6)}=\frac{c_0}{8}i e^{-i J^{(6)}},~~~~\Psi_-^{(6)}=\frac{c_0}{8}\Omega^{(6)}.
\eeq
We then find that \eqref{eq:Mink4susy1}-\eqref{eq:Mink4susy3} in type IIB reduce to
\begin{subequations}
\begin{align}
&d(e^{2A-\Phi} J^{(6)})=0,~~~~d(e^{3A-\Phi} \Omega^{(6)})=0,\label{eq:firstfsjk}\\[2mm]
&H_3\wedge J^{(6)}=0,~~~~H_3\wedge \Omega^{(6)}=0,\label{eq:secondfsjk}\\[2mm]
&e^{4A}\star_6\lambda(f^{(6)}_-)=d(e^{4A-\Phi})-e^{4A-\Phi}H_3+\frac{1}{2}e^{4A}d(e^{-\Phi})\wedge J^{(6)}\wedge J^{(6)}.
\end{align}
\end{subequations}
We will make the ansatz that the internal 6-manifold is conformally CY$_3$, namely that
\beq
ds^2(\text{M}_6)=e^{-2A}ds^2(\text{CY}_3),~~~e^{2A} J^{(6)}=J_{\text{CY}_3},~~~~e^{3A} \Omega^{(6)}=\Omega_{\text{CY}_3},
\eeq
where $dJ^{(6)}_{\text{CY}_3}=d\Omega^{(6)}_{\text{CY}_3}=0$. Consistency of this ansatz with \eqref{eq:firstfsjk} means that we must take
\beq
d(e^{-\Phi})=0~~~~\Rightarrow~~~e^{-\Phi}=e^{-\Phi_0},
\eeq
for $\Phi_0$ a constant. Thus if we define $e^{-4A}=h_3$ an arbitrary function on $\text{CY}_3$ we arrive at
\beq
f^{(6)}_-=e^{-\Phi_0}\left( \hat{\star}_6d h_3-\hat{\star}_6H_3\right).
\eeq
Substituting what we have  derived thus far into our compactifications ansatz yields a class of conformal CY$_3$ solution of the form
\begin{align}
ds^2&= \frac{1}{\sqrt{h}_3}(ds^2(\text{Mink}_3)+D\phi^2)+\sqrt{h_3}ds^2(\text{CY}_3),~~~~e^{-\Phi}=e^{-\Phi_0},~~~H=H_3,\nn\\[2mm]
F_3&=-e^{-\Phi_0}\hat\star_6 H_3,\nn\\[2mm]
F_5&=e^{-\Phi_0}\left(\hat\star_6 dh_3+D\phi\wedge {\cal F}\wedge J_{\text{CY}_3}-\text{vol}(\text{Mink}_3)\wedge d\left(\frac{1}{h_3}D\phi\right)\right),
\end{align}
for $ds^2(\text{CY}_3)$ the metric on any CY$_3$ manifold. Supersymmetry additionally demands that
\beq
{\cal F}\wedge J^{(6)}\wedge J^{(6)}={\cal F}\wedge \Omega^{(6)}=0,~~~~H_3\wedge J^{(6)}=H_3\wedge \Omega^{(6)}=0,
\eeq
making ${\cal F}$ a primitive $(1,1)$-form and $H_3$ a primitive $(1,2)$-form. One has a solution whenever 
\beq
dH_3=0,~~~~d\hat{\star}_6 H_3=0,~~~\nabla^2_{\text{CY}_3}h_3+\hat{H}_3^2+\hat{\cal F}^2=0.
\eeq
So we see that we have a class of D3 branes compactified on S$^1$ that are backreacted on CY$_3$ in the presence of a harmonic NS 3-form. We construct explicit solutions within this class in sections \ref{sec:universalcompactansatz} and \ref{sec:solex}.

\subsection{Circle compactifications of Mink\texorpdfstring{$_5$}{5}}\label{sec:compact5to4}
We can construct supersymmetric circle compactifications of Mink$_5\to$Mink$_4$ by generalising our derivation in appendix \ref{sec:Mink5fromMink4}, we decompose
\beq
\Psi^{(6)}_{\mp}=\frac{1}{2}(\Psi^{(5)}_{\mp}\pm i \Psi^{(5)}_{\pm}\wedge W),~~~~\Psi^{(6)}_{\pm}=\frac{1}{2}(\tilde{\Psi}^{(5)}_{\pm}\mp i \tilde{\Psi}^{(5)}_{\mp}\wedge W),~~~~W=e^{A}D\phi,\label{eq:6to5s}
\eeq
where the $d=5$ bilinears are in general
\begin{align}
\Psi^{(5)}&=\Psi^{(5)}_++\Psi^{(5)}_-=\frac{c_0}{4}\sin\beta(1+V)\wedge \left((\overline{a}+c Z)\wedge e^{-i J^{(4)}}+\overline{b}\Omega^{(4)}\right),\nn\\[2mm]
\tilde{\Psi}^{(5)}&=\tilde{\Psi}^{(5)}_++\tilde{\Psi}^{(5)}_-=\frac{c_0}{4}\sin\beta(1+V)\wedge \left(a \Omega^{(4)}-(b-c W)\wedge e^{-i J^{(4)}}\right),\label{eq:5dbilinears}
\end{align}
for $|a|^2+|b|^2+c^2=1$ and where the SU(2)-structure forms decompose as
\beq
J^{(4)}=\frac{i}{2}(W\wedge \overline{W}+Z\wedge \overline{Z}),~~~~\Omega^{(4)}=W\wedge Z,\label{eq:decomp}
\eeq
so span an identity structure unless $c=0$ where this is enhanced to SU(2)-structure. We will again assume that $2\beta=\pi$, in which limit the necessary conditions for Mink$_4$ are  \eqref{eq:Mink4susy1}-\eqref{eq:Mink4susy3}. Using the decomposition of \eqref{eq:6to5s} in these and assuming that 
\begin{subequations}
\begin{align}
&\iota_WH_3=0,~~~~\star_6\lambda(f^{(6)}_{\pm})=-W\wedge \star_5\lambda(f^{(5)}_{\pm})\mp\frac{4}{c_0}e^{A-\Phi}{\cal F}\wedge \text{Re}\Psi^{(5)}_{\pm},\\[2mm]
&{\cal F}\wedge \tilde{\Psi}^{(5)}_{\mp}=0,~~~~{\cal F}\wedge \text{Im}\Psi^{(5)}_{\mp}=0,\label{eq:needtosolve}
\end{align}
\end{subequations}
we find that \eqref{eq:Mink4susy1}-\eqref{eq:Mink4susy3} reduce to the supersymmetry conditions for Mink$_5$ in the $2\beta=\pi$ limit, namely
\begin{subequations}
\begin{align}
&d_{H_3}(e^{3A-\Phi}\tilde{\Psi}^{(5)}_{\pm})=0,\label{eq:mink5bpsmainfirst}\\[2mm]
&d_{H_3}(e^{4A-\Phi}\tilde{\Psi}^{(5)}_{\mp})=0,\\[2mm]
&d_{H_3}(e^{3A-\Phi}\text{Im}\Psi^{(5)}_{\pm})=0,\\[2mm]
&d_{H_3}(e^{4A-\Phi}\text{Im}\Psi^{(5)}_{\mp})=0,\\[2mm]
&d_{H_3}(e^{2A-\Phi}\text{Re}\Psi^{(5)}_{\mp})=0 ,\\[2mm]
&d_{H_3}(e^{5A-\Phi}\text{Im}\Psi^{(5)}_{\pm})=\mp \frac{c_0}{4}\star\lambda(f^{(5)}_{\pm}).\label{eq:mink5bpsmainlast}
\end{align}
\end{subequations}
There are no additional conditions in this case, but we still need to actually solve \eqref{eq:needtosolve}, which again restrict both ${\cal F}$ and the possible G-structure. We find that in both type IIA and IIB we must have
\beq
\text{Im}a=b=c=0,~~~~{\cal F}\wedge J^{(4)}=0,~~~~{\cal F}\wedge \Omega^{(4)}=0.\label{eq:thisexampleofcalFconds}
\eeq
This puts us in a subsector of the SU(2)-structure limit for which we can take
\begin{align}
\Psi^{(5)}&=\frac{c_0}{4}(1+V)\wedge  e^{-i J^{(4)}},~~~~
\tilde{\Psi}^{(5)}=\frac{c_0}{4}(1+V)\wedge \Omega^{(4)},\label{eq:drq5bisp}
\end{align}
without loss of generality, and makes ${\cal F}$ a primitive (1,1)-form on the sub-manifold orthogonal to $V$. From this it is a simple matter to establish that
\beq
f^{(6)}_{\pm}=f^{(5)}_{\pm}\mp \frac{4}{c_0}e^{2A-\Phi}D\phi\wedge {\cal F}\wedge \text{Re}\Psi^{(5)}_{\mp}.
\eeq

In summary we have found  a supersymmetric class of solutions that have the decomposition 
\begin{align}
ds^2=e^{2A}(ds^2(\text{Mink}_4)+D\phi^2)+ds^2(\text{M}_5),~~~H=H_3,\nn\\[2mm]
F_{\pm}=\left(1+\star\lambda\right)\left(f^{(5)}_{\pm}\mp \frac{4}{c_0}e^{2A-\Phi}D\phi\wedge {\cal F}\wedge \text{Re}\Psi^{(5)}_{\mp}\right).
\end{align}
They are governed by the conditions for supersymmetric Mink$_5$ solutions \eqref{eq:mink5bpsmainfirst}-\eqref{eq:mink5bpsmainlast} but specialised to the case of $d=5$ SU(2)-structure bilinears given by \eqref{eq:drq5bisp}. In general the internal manifold may be decomposed as
\beq
ds^2(\text{M}_5)=ds^2(\text{SU(2)})+V^2,
\eeq
were $ds^2(\text{SU(2)})$ supports a $d=4$ SU(2)-structure. When it is non trivial, the field strength ${\cal F}$ must be a primitive (1,1)-form defined $ds^2(\text{SU(2)})$ but is otherwise free.

Under the above assumptions on has a solution whenever the following magnetic Bianchi identities are satisfied
\beq
dH_3=0,~~~~ (d-H_3\wedge )f^{(5)}_{\pm}\mp \frac{4}{c_0}e^{2A-\Phi}{\cal F}\wedge {\cal F}\wedge \text{Re}\Psi^{(5)}_{\mp}=0,
\eeq
generalising the Mink$_5$ result.

\subsubsection{D4-D8 brane system wrapping a circle}\label{sec:D4D8circle}
In this section we derive a class describing D4-D8 branes compactified on a circle and backreated on CY$_2$ and an interval. It is in fact the most general class of soltions consistent with our compactification ansatz in type IIA and contains two branches, only one of which actually included D8 branes.

We begin by fixing $2\beta=\pi$ and $\text{Im}a=b=c=0$ in our $d=5$ bilinears such that they that the form \eqref{eq:drq5bisp}. We then insert this into \eqref{eq:mink5bpsmainfirst}-\eqref{eq:mink5bpsmainlast} in type IIA,  
which reduce to
\begin{subequations}
\begin{align}
& d(e^{A}V)=0,~~~~d(e^{3A-\Phi}J^{(4)})=0,~~~~d(e^{3A-\Phi}\Omega^{(4)})=0,\label{eq:D4D8susy1}\\[2mm]
& d(e^{A-\Phi})\wedge V=0,~~~~H_3\wedge V=0~~~~H_3\wedge J^{(4)}=0,~~~~H_3\wedge \Omega^{(4)}=0,\label{eq:D4D8susy2}\\[2mm]
&e^{5A}\star_5\lambda(f^{(5)}_+)=-d(e^{5A-\Phi})+e^{5A-\Phi}H_3- \frac{1}{2} e^{4A} d(e^{A-\Phi})\wedge J^{(4)}\wedge J^{(4)}.\label{eq:D4D8susy3}
\end{align}
\end{subequations}
The constraints \eqref{eq:D4D8susy1} imply that locally we can introduce a coordinate $\rho$ and closed forms $(J_{\text{CY}_2},~\Omega_{\text{CY}_2})$ such that
\beq
e^{A}V=\rho,~~~~ e^{3A-\Phi}J^{(4)}=J_{\text{CY}_2},~~~~e^{3A-\Phi}\Omega^{(4)}=\Omega_{\text{CY}_2},
\eeq
such that the internal 5-manifold decomposes as a warped product of a CY$_2$ and an interval as
\beq
ds^2(\text{M}_5)= \sqrt{\frac{h_4}{h_8}}ds^2(\text{CY}_2)+\sqrt{h_4 h_8}d\rho^2.
\eeq 
The conditions \eqref{eq:D4D8susy2} can be solved in general by introducing a function of the interval $h_8=h_8(\rho)$ and a primitive (1,1)-form ${\cal G}$  on CY$_2$  such that
\beq
e^{A-\Phi}=h_8,~~~~ H_3=\frac{1}{h_8^2} d\rho\wedge {\cal G}.
\eeq
Then if we further introduce an a function $h_4=h_4(\rho,\text{CY}_2)$ such that $e^{-4A}=h_4 h_8$ we find that \eqref{eq:D4D8susy3} implies that
\beq
f^{(5)}_+=-\partial_{\rho}h_8+\frac{1}{h_8}{\cal G}+h_8d\rho\wedge \hat\star_4 \hat{d}h_4+\partial_{\rho}h_4\text{vol}(\text{CY}_2),
\eeq
where we have decomposed $d=\hat{d}+d\rho\wedge \partial_{\rho}$ and $\hat{\star}_4$ is the hodge dual on unwarped CY$_2$. Our compactification ansatz then leads us to a class of conformally CY$_2$ solutions of the form
\begin{align}
ds^2&= \frac{1}{\sqrt{h_4 h_8}}(ds^2(\text{Mink}_4)+D\phi^2)+ \sqrt{\frac{h_4}{h_8}}ds^2(\text{CY}_2)+\sqrt{h_4 h_8}d\rho^2,\nn\\[2mm]
e^{-\Phi}&=h_4^{\frac{1}{4}} h_8^{\frac{5}{4}},~~~~H=\frac{1}{h_8^2} d\rho\wedge {\cal G},\nn\\[2mm]
F_0&=-\partial_{\rho}h_8,~~~~F_2=\frac{1}{h_8}{\cal G},\nn\\[2mm]
F_4&=h_8d\rho\wedge \hat\star_4 \hat d h_4+\partial_{\rho}h_4 \text{vol}(\text{CY}_2)-h_8 D\phi\wedge d\rho\wedge {\cal F},
\end{align}
where $h_4$ has support on $(\text{CY}_2,\rho)$ but $h_8$ only on $\rho$.
Supersymmetry demands that $({\cal G},{\cal F})$ are primitive (1,1)-forms on $\text{CY}_2$, \textit{i.e}
\beq
{\cal F}\wedge J_{\text{CY}_2}={\cal F}\wedge \Omega_{\text{CY}_2}=0,~~~~{\cal G}\wedge J_{\text{CY}_2}={\cal G}\wedge \Omega_{\text{CY}_2}=0,
\eeq
while one has a solution whenever the magnetic Bianchi identities are imposed, which away from sources is equivalent to
\begin{subequations}
\begin{align}
&d{\cal G}=0,\label{eq:BianchiD4D82}\\[2mm]
&h_8''=0,~~~~h_8\nabla^2_{\text{CY}_2}h_4+\partial_{\rho}^2h_4+\frac{1}{h_8^3}\hat{\cal G}^2+h_8\hat{\cal F}^2=0\label{eq:BianchiD4D81}.
\end{align}
\end{subequations}
The last of these is a generalisation, in terms of the flux terms, of the PDEs that define D4-D8 system backreacted on $(\text{CY}_2,\rho)$. On the other hand \eqref{eq:BianchiD4D82} just imposes that ${\cal G}$ is closed. We thus have a class that describes D4-D8 branes compactified on a circle in the presence of additional NS 3-form flux. When  ${\cal F}=0$ we observe that what we have is similar to a class of AdS$_3\times$S$^2$ solutions found in \cite{Lozano:2019emq} which suggests there is likely an embedding of some $d=4$ non minimal supergravity into type IIA that encompasses both classes. The similarity to \cite{Lozano:2019emq} also leads us to interpret the presence of ${\cal G}$ as adding fractional D4 branes, following the argument in \cite{Couzens:2021veb}.

\subsection{Circle compactifications of Mink\texorpdfstring{$_6$}{6}}\label{sec:D5circle}
Finally we will derive compactifications of Mink$_6$ solution on S$^1$ by generalising the derivation of appendix \ref{sec:Mink6fromMink5}. We assume a warped Mink$_5$ factor and decompose the internal $d=5$ bilinears as
\beq
\Psi^{(5)}_{\mp}+\Psi^{(5)}_{\pm}=\Psi^{(4)}_{\mp}\wedge(W\mp 1) ,~~~~\tilde{\Psi}^{(5)}_{\mp}+\tilde{\Psi}^{(5)}_{\pm}=\tilde{\Psi}^{(4)}_{\mp}\wedge(W\mp 1),~~~~W=e^{A}D\phi,\label{eq:5to4polyformss}
\eeq
in terms of $d=4$ bilinears that can in general be expressed in respectively IIA and IIB as
\begin{align}
\Psi^{(4)}_-&=\frac{c_0}{4}\sin\beta W\wedge e^{\frac{1}{2}Z\wedge \overline{Z}},~~~~\tilde{\Psi}^{(4)}_-=-\frac{c_0}{4}\sin\beta Z\wedge e^{\frac{1}{2}W\wedge \overline{W}},\nn\\[2mm]
\Psi^{(4)}_+&=\frac{c_0}{4}\sin\beta \left(\overline{a}e^{-i J^{(4)}}+\overline{b}\Omega^{(4)}\right),~~~~\tilde{\Psi}^{(4)}_+=\frac{c_0}{4}\sin\beta \left(-b e^{-i J^{(4)}}+a\Omega^{(4)}\right),\label{eq:4dbilinears}
\end{align}
which define an identity-structure in IIA and SU(2)-structure in IIB. We now need to substitute our decomposition of the $d=5$ bilinears into the supersymmetry conditions for Mink$_5$ solution, we will assume that
\begin{subequations}
\begin{align}
&\iota_{W}H_3=0,~~~~\beta=\frac{\pi}{2},~~~~\star_{5}\lambda(f^{(5)}_{\pm})=W\wedge \star_4\lambda(f^{(4)}_{\pm})+\frac{4}{c_0}e^{A-\Phi}{\cal F}\wedge \text{Re}\Psi^{(4)}_{\mp}=0,\\[2mm]
&{\cal F}\wedge \text{Im}\Psi^{(4)}_{\mp}=0,~~~~{\cal F}\wedge \tilde{\Psi}^{(4)}_{\mp}=0.\label{eq:needtosolve2}
\end{align}
\end{subequations}
We then find that \eqref{eq:mink5bpsmainfirst}-\eqref{eq:mink5bpsmainfirst} reduce to
\begin{subequations}
\begin{align}
&d_{H_3}(e^{4A-\Phi}\tilde{\Psi}^{(4)}_{\mp})=0,\label{eq:mink61}\\[2mm]
&d_{H_3}(e^{4A-\Phi}\text{Im}\Psi^{(4)}_{\mp})=0,\label{eq:mink62}\\[2mm]
&d_{H_3}(e^{2A-\Phi}\text{Re}\Psi^{(4)}_{\mp})=0,\label{eq:mink63}\\[2mm]
&d_{H_3}(e^{6A-\Phi}\text{Re}\Psi^{(4)}_{\mp})=- \frac{1}{4}c_0 e^{6A}\star_4 \lambda(f^{(4)}_{\pm}),\label{eq:mink64}
\end{align}
\end{subequations}
which are simply the supersymmetry conditions for Mink$_6$ solutions in the $2\beta=\pi$ limit. The only remaining condition to solve to be consistent with supersymmetry is  \eqref{eq:needtosolve2}, which can in fact only be solved for ${\cal F}\neq 0$ in type IIB where it leads to
\beq
\text{Im}a=b=0,~~~~{\cal F}\wedge J^{(4)}=0,~~~~{\cal F}\wedge \Omega^{(4)}=0,\label{eq:prima}
\eeq
again making ${\cal F}$ a primitive (1,1)-form and constraining
\begin{align}
\Psi^{(4)}_+&=\frac{c_0}{4}  e^{-i J^{(4)}},~~~~
\tilde{\Psi}^{(4)}_+=\frac{c_0}{4}\Omega^{(4)},\label{eq:drq4bisp}
\end{align}
without loss of generality. We can now take the hodge dual of $\star_5\lambda(f^{(5)}_{-})$ arriving at
\beq
f^{(5)}_-=f^{(4)}_-+e^{2A-\Phi}D\phi\wedge {\cal F}.
\eeq
Note however that fixing the $d=4$ bilinears as in \eqref{eq:drq4bisp} makes \eqref{eq:mink61}- \eqref{eq:mink64} particularly simple, they in fact reduces to
\begin{subequations}
\begin{align}
&d(e^{2A-\Phi})=0,~~~~H_3=0~~~~\star_4\lambda(f^{(4)}_-)=e^{4A-\Phi} d\left(e^{-4A}\right),\label{eq:firsfline}\\[2mm]
&d(e^{2A}J^{(4)})=0,~~~~ d(e^{2A}\Omega^{(4)})=0.\label{eq:seconfline}
\end{align}
\end{subequations}
Here \eqref{eq:seconfline} implies that the internal manifold is conformally CY$_2$, i.e. we can define
\beq
ds^2(\text{M}_4)=e^{-2A}ds^2(\text{CY}_2),~~~~e^{2A}J^{(4)}=J_{\text{CY}_2},~~~e^{2A}\Omega^{(4)}=\Omega_{\text{CY}_2},
\eeq
where $(J_{\text{CY}_2},\Omega_{\text{CY}_2})$ are necessarily now closed as they should be. We can solve the first condition in \eqref{eq:firsfline} by taking
\beq
e^{-4A}=h_5,~~~~ e^{-\Phi}=e^{-\Phi_0}\sqrt{h_5},
\eeq
where $h_5$ is an arbitrary function on CY$_2$ and $\Phi_0$ a constant. This leads to
\beq
f^{(4)}_-= e^{-\Phi_0}\hat{\star}_4d h_5,
\eeq
where $\hat{\star}_4$ is the hodge dual on unwarped CY$_2$.

In summary we have derived a class of solutions of the form
\begin{align}
ds^2&= \frac{1}{\sqrt{h_5}}(ds^2(\text{Mink}_5)+D\phi^2)+\sqrt{h_5}ds^2(\text{CY}_2),~~~~e^{-\Phi}=e^{-\Phi_0}\sqrt{h_5},\nn\\[2mm]
F_3&=e^{-\Phi_0}\left(\hat{\star}_4d h_5+ D\phi\wedge{\cal F}\right),
\end{align}
for which $F_3$ is the only non trivial flux and ${\cal F}$ is a primitive (1,1)-form on CY$_2$ and $\hat{\star}_4$ is the unwarped hodge on CY$_2$. The Bianchi identity of the 3-form requires that away from sources we have
\beq
\nabla^2_{\text{CY}_2}h_5+\hat{{\cal F}}^2=0,
\eeq
where again the hat on ${\cal F}^2$ indicates it is computed on the unwarped CY$_2$ manifold. Thus we have derived a class of D5 branes wrapping a circle that are backreacted on  CY$_2$.  We note that the embedding of $d=5$ supergravity coupled to an Abelian vector multiplet in \cite{Lima:2022hji} reproduces this class in its Mink$_5$ limit\footnote{The apparent sign differences with respect to that reference are due to differing conventions for the hodge dual}. We construct solutions within this class in sections \ref{sec:universalcompactansatz}, \ref{eq:Addtionalmink5} and \ref{sec:solex}.

\section{Explicit solutions}\label{eq:explicitsols}
In this section we derive some explicit solution from the classes we derived earlier in the paper. We start by focusing on Mink$_{p}$ vacua (with $p=1,3,5$), with the internal space being spheres and the radial coordinate being bounded by either D$p$ brane source and a O$p$ plane or two O$p$ planes. Then we proceed to perform a similar analysis for Mink$_{2}$ vacua, with the internal space being a G$_{2}$ cone. We finish this section by discussing how to embed "solitonic branes" in our ansatz allowing us to generalise existing solution such that they can have a far broader array of internal spaces and also including D brane sources.

\subsection{Minkowski flux vacua from compactified D branes}\label{sec:minkvac}
In this section we will derive some compact Minkowski vacua, i.e.  solution with a Minkowski factor and bounded internal space. Our starting point will be non compact D brane solutions.

\subsubsection{Universal Mink\texorpdfstring{$_p$}{p} vacua with O\texorpdfstring{$p$}{p} planes for \texorpdfstring{$p=1,3,5$}{p=1,3,5}}\label{sec:universalcompactansatz}
We begin by considering the classes of solution derived in sections \ref{sec:D1circle}, \ref{sec:D3circle} and \ref{sec:D5circle}. We note that contained within these are a series of classes with D$p$ branes backreacted on Calabi-Yau manifolds for $p=1,3,5$. These can be expressed in a unified way as 
\begin{align}
ds^2&= \frac{1}{\sqrt{h_p}}(ds^2(\text{Mink}_{p})+D\phi^2)+ \sqrt{h_p}ds^2(\text{CY}_{\frac{9-p}{2}}),~~~~e^{-\Phi}= e^{-\Phi_0}h^{\frac{p-3}{4}}_p,\nn\\[2mm]
F^{\text{mag}}_{8-p}&=e^{-\Phi_0}\left(\hat\star d h_p+\frac{1}{(\frac{5-p}{2})!}D\phi\wedge {\cal F}\wedge \hat{J}^{\frac{5-p}{2}}\right),\label{eq: DponCY}
\end{align}
where we only write the magnetic parts of the flux in each case and $\hat{J}$ is the Kahler form on the relevant CY manifold. In each case ${\cal F}$ should be a primate (1,1)-form meaning it should obey
\beq
{\cal F}\wedge \hat {J}^{\frac{7-p}{2}}=0,~~~~{\cal F}\wedge \hat \Omega=0,\label{eq:primativecondsexamplesec}
\eeq
where $(\hat J,\hat \Omega)$ define the CY manifold.
Given a choice of CY manifold the equation which governs solutions, away from sources, is
\beq
\hat\nabla^2 h_p+\hat{\cal F}^2=0,\label{eq:defineingCYPDE}
\eeq
where the hats indicate that computations are performed on the unwarped CY manifold\footnote{Note that one can easily compute $\hat{\cal F}^2$ through the identity
\beq
\frac{1}{(\frac{5-p}{2})!}{\cal F}\wedge {\cal F}\wedge \hat{J}^{\frac{5-p}{2}}=-\hat{\cal F}^2\text{vol}(\text{CY}^{\frac{9-p}{2}})\nn.
\eeq}. The most simple solutions to \eqref{eq:defineingCYPDE} with ${\cal F}=0$ are given by D$p$ branes in flat space with spherical symmetry imposed on $h_p$, \textit{i.e}
\beq
ds^2(\text{CY}^{\frac{9-p}{2}})=dr^2+r^2ds^2(\text{S}^{8-p})~~~~\Rightarrow~~~~  h_p\bigg\lvert_{{\cal F}=0}=1+\frac{Q_p}{r^{7-p}}\label{eq: flatspacecase}.
\eeq
However one can equally well place a D$p$ brane at the tip of a CY cone whose base is any Sasaki-Einstein manifold by decomposing
\beq
ds^2(\text{CY}^{\frac{9-p}{2}})=dr^2+r^2ds^2(\text{SE}^{8-p}),
\eeq
and it is in fact more convenient to find solutions with non trivial ${\cal F}$ in this language. Sasaki-Einstein manifold can themselves be expressed as a U(1) fibration over a Kahler-Einstein manifold as
\beq
ds^2(\text{SE}^{8-p})= ds^2(\text{KE}^{7-p})+D\psi^2,~~~~D\psi=d\psi+\eta,
\eeq
where the connection term $\eta$ gives rise to the Kahler form on $\text{KE}^{7-p}$ as
\beq
d\eta= 2J_{\text{KE}},\label{eq:kalherform}
\eeq
while this has associated to it a holomorphic $\frac{7-p}{2}$-form  $\Omega_{\text{KE}}$ such that $(J_{\text{KE}},\Omega_{\text{KE}})$ span an SU($\frac{7-p}{2}$)-structure, and additionally obey\footnote{Note for the $p=5$ case this would make the KE forms span an SU(1)-structure, which doesn't really exist. However one can still define forms with the appropriate behaviours under $d$ and $\wedge$, see for instance B.6 of\cite{Legramandi:2020txf} for the case of the SE-structure on S$^3$.}
\beq
dJ_{\text{KE}}=0,~~~ d\Omega_{\text{KE}}=i \frac{9-p}{2} D\psi\wedge \Omega_{\text{KE}}.\label{eq:KEstructure}
\eeq
These conditions imply that we can define the 2 form and holomorphic $\frac{9-p}{2}$ form on $\text{CY}^{\frac{9-p}{2}}$ to be
\beq
\hat J = dr\wedge (r D\psi)+ r^2 J_{\text{KE}},~~~~\hat\Omega= r^{\frac{7-p}{2}}e^{i\frac{9-p}{2}\psi}\left(dr+i r D\psi\right)\wedge \Omega_{\text{KE}},
\eeq
which are closed by construction. It is not hard to show that
\beq
{\cal A}=\frac{b}{r^{7-p}} D\psi,~~~~db=0~~~~\Rightarrow~~~~{\cal F}^2=(9-p)(7-p)\frac{b^2}{r^{2(9-p)}},\label{eq:choiceofcalF}
\eeq
which is by no means the only choice, leads to ${\cal F}=d{\cal A}$ being a primitive (1,1)-form as required. Assuming we place our D$p$ brane at the tip of the CY cone we have $r^{8-p}\hat\nabla^2h_p=\partial_{r}(r^{8-p}\partial_r h_p)$ and so \eqref{eq:defineingCYPDE} is solved for $p=1,3,5$ by
\beq
h_p=c_1+\frac{c_2}{r^{7-p}}-\frac{(7-p)b^2}{2(8-p)r^{2(8-p)}},\label{eq:hp gen}
\eeq
where $(c_1,c_2)$ are constants and the generalisation of the flat space case in \eqref{eq: flatspacecase} is given by taking $\text{KE}^{7-p}=\mathbb{CP}^{7-p}$. The new term we have added dominates the behaviour as $r\to 0$ so we appear to have spoiled the physical D$p$ brane behaviour one find there when $b=0$. Note however that this term comes with a minus sign, so we never actually reach $r=0$, instead the space terminates at some $r=r_{\text{min}}$ past which $h_p$ becomes negative, and as $h_p$ is an even function we can assume $r_{\text{min}}>0$ without loss of generality. At $r=r_{\text{min}}$ $h_p$ has a zero and the behaviour at this point depends on the order of that zero. In particular an order 1 zero for which $h_p\sim (r-r_{\text{min}})$ still yields a physical singularity, just not that of a D$p$ brane -rather it is that of an Op plane at the tip of a CY cone. It is not hard to show that only an order 1 zero is compatible with a real metric so for physical solutions the space begins with an Op plane at $r=r_{\text{min}}$ - what happens in the rest of the space depends on the tuning of $(c_1,c_2,b^2)$ which can lead to $r$ spanning either a bounded or semi infinite interval. If bounded then we have a viable Mink$_{p}$ flux vacua whenever $\text{KE}^{7-p}$ is compact. This happens whenever $h_p$ has a second zero at $r=r_{\text{max}}>r_{\text{min}}$ and it is not hard to show that tuning $(c_1,c_2)$ such that $h_p$ has zeros (of any order) at $(r_{\text{min}},r_{\text{max}})$ leads to
\begin{align}
h_p&=  \frac{(7-p)b^2}{2(8-p)r_{\text{min}}^2r_{\text{max}}^2}\frac{1}{r^{2(8-p)}}(r^2-r_{\text{min}}^2)(r_{\text{max}}^2-r^2)\tilde{h}_p,\label{eq:hpsol1}\\[2mm]
\tilde{h}_5&=1+\left(\frac{1}{r_{\text{min}}^2}+\frac{1}{r_{\text{max}}^2}\right)r^2,\\[2mm]
\tilde{h}_3&=1+\left(\frac{1}{r_{\text{min}}^2}+\frac{1}{r_{\text{max}}^2}\right)r^2+\left(\frac{1}{r^4_{\text{min}}}+\frac{1}{r^4_{\text{max}}}+\frac{1}{r^2_{\text{min}}r^2_{\text{max}}}\right)\left(1+\frac{r^2}{r^2_{\text{min}}+r^2_{\text{max}}}\right)r^4,\nn\\[2mm]
\tilde{h}_1&=1+\left(\frac{1}{r_{\text{min}}^2}+\frac{1}{r_{\text{max}}^2}\right)r^2+\left(\frac{1}{r^4_{\text{min}}}+\frac{1}{r^4_{\text{max}}}+\frac{1}{r^2_{\text{min}}r^2_{\text{max}}}\right)r^4\nn\\[2mm]
&+\left(\frac{1}{r_{\text{min}}^2}+\frac{1}{r_{\text{max}}^2}\right)\left(\frac{1}{r_{\text{min}}^4}+\frac{1}{r_{\text{max}}^4}\right)\left(1+\frac{r_{\text{min}}^2+r_{\text{max}}^2}{r_{\text{min}}^4+r_{\text{min}}^2r_{\text{max}}^2+r_{\text{max}}^4}\left(1+\frac{r^2}{r_{\text{min}}^2+r_{\text{max}}^2}\right)r^2\right)r^6,\nn
\end{align}
where in each case $\tilde{h}_p$ is necessarily positive. So provided that $r_{\text{min}}\leq r\leq r_{\text{max}}$, we have a physical solution that is bounded between two Op plane singularities  -  clearly the RHS of \eqref{eq:defineingCYPDE} should be modified by $\delta$-function sources to accommodate this which the presence of ${\cal F}$ in \eqref{eq:defineingCYPDE} facilitates. We find no issues with flux quantisation.

\subsubsection{Additional Mink\texorpdfstring{$_5$}{5} vacua}\label{eq:Addtionalmink5}
The universal choice we made for ${\cal F}$ in \eqref{eq:choiceofcalF} is not the only option available to us, indeed for specific choices of CY manifold their exist specific connections yielding their own behaviour. To illustrate this point let us consider $\text{CY}_2=\mathbb{R}^4$ with
\beq
ds^2(\text{CY}_2)=dr^2+r^2ds^2(\text{S}^3),
\eeq
and assume $h_5=h_5(r)$. The 3 sphere supports two sets of SU(2) invariant forms $(L_i,R_i)$, namely the left and right invariants, that obey the relations
\beq
dL_i=\frac{1}{2}\epsilon_{ijk}L_j\wedge L_k,~~~~dR_i=-\frac{1}{2}\epsilon_{ijk}R_j\wedge R_k,~~~~ds^2(\text{S}^3)=\frac{1}{4}(L_i)^2=\frac{1}{4}(R_i)^2.\label{eq:leftrightsu2}
\eeq
One can define the forms on CY$_2$ in terms of either of these, in term of $L_i$ they take the form
\beq
\hat{J}=\frac{r}{2}dr\wedge L_3+\frac{r^2}{4} L_2\wedge L_3,~~~~\hat{\Omega}=\frac{r}{2}dr\wedge (L_1+i L_2)+\frac{r^2}{4}(L_2\wedge L_3+i L_3\wedge L_1).
\eeq
Given this choice of CY$_2$ forms ${\cal A}=\frac{b}{2 r^2} L_3$ reproduces \eqref{eq:choiceofcalF} for $p=5$, which we now see must break the SO(4) isometry of the 3-sphere to SU(2)$\times$U(1). But we could equally well take
\beq
{\cal A}= \frac{1}{2r^2}\left(b_1 L_1+b_2 L_2+b_3 L_3\right)~~~~\Rightarrow~~~ {\cal F}^2=8\frac{b^2}{r^{8}},~~~~b^2=b_1^2+b_2^2+b_3^2,
\eeq 
which leads to exactly the same PDE, and so the same warp factor given in \eqref{eq:hp gen} yeilding further solutions bounded between two O5 planes, but now with SO(4) broken to SU(2). This is not all one can do however as one can also define ${\cal A}$ in terms of $R_i$ due to another identity the two sets of SU(2) forms obey, namely
\beq
\epsilon_{ijk}L_j\wedge L_k\wedge R_{\hat{i}}=\epsilon_{\hat{i}\hat{j}\hat{k}}R_{\hat{i}}\wedge R_{\hat{j}}\wedge L_{i}.
\eeq
This means another valid choice of primitive (1,1)-form is given by
\beq
{\cal A}=\frac{r^2}{2}\left(b_1 R_1+b_2 R_2+b_3R_3 \right)~~~~\Rightarrow~~~ {\cal F}^2=8 b^2,~~~~b^2=b_1^2+b_2^2+b_3^2,
\eeq
which also generically breaks SO(4) to SU(2) with an enhancement to SU(2)$\times$U(1) when $b_1=b_2=0$. The solution to \eqref{eq:defineingCYPDE} now becomes 
\beq
h_5=c_1+ \frac{c_2}{r^2}-b^2 r^2,\label{eq:D5O5}
\eeq
which appeared before in a slightly different context in \cite{Lima:2022hji}. While this may look formally similar to \eqref{eq:hp gen} the types of solution it can accommodate are more broad. Indeed now it is possible to bound the $r$ interval between a D5 and an O5,  two O5s or a regular zero and an O5  - the details are given in \cite{Lima:2022hji}. It is also possible to define a primitive (1,1)-form in terms of both sets of SU(2) invariant forms, but this is incompatible with $h_5=h_5(r)$.

\subsubsection{Mink\texorpdfstring{$_2$}{2} vacua with O2 planes and D2 branes}\label{sec:D2ex}
It is also possible to construct compact Minkowski solutions from D2 branes wrapping a circle and backreated on G$_2$ manifolds. We make use of the class derived in  section \ref{eq:D2section} and fix $H_3=0$  which results in a class of the form
\begin{align}
ds^2&=\frac{1}{\sqrt{h}_2}(ds^2(\text{Mink}_2)+D\phi^2)+\sqrt{h_2}ds^2(\text{G}_2),~~~~e^{-\Phi}=e^{-\Phi_0}h_2^{-\frac{1}{4}},\nn\\[2mm]
F_{6}&=e^{-\Phi_0}\left(\hat \star_7d h_2-D\phi\wedge {\cal F}\wedge \Phi_{\text{G}_2}\right),
\end{align}
where the closed and co closed 3-form $\Phi_{\text{G}_2}$ defines the type of G$_2$ manifold the internal space is - as before quantities with hats are computed on the unwarped internal space. The 2-form ${\cal F}=d{\cal A}$ has support on G$_2$ and is constrained by supersymmetry to be primitive, \textit{i.e}
\beq
{\cal F}\wedge \hat{\star}_7\Phi_{\text{G}_2}=0.\label{eq:calfconds}
\eeq
Given a choice of G$_2$ manifold and ${\cal F}$, the equation defining solutions in this class is
\beq
\hat\nabla^2 h_2+\hat{\cal F}^2=0.\label{eq:defineingG2PDE}
\eeq
The most simple solution to this equation with ${\cal F}=0$ is given by the D2 brane in flat space with SO(7) rotational symmetry which yields
\beq
h_2=1+\frac{Q_2}{r^5},
\eeq
but one can actually place such a D2 at the tip of any G$_2$ cone, \text{i.e} cones over nearly Kahler 6-manifolds of the form
\beq
ds^2(\text{G}_2)=dr^2+r^2 ds^2(\text{NK}_6),
\eeq 
and the warp factor will be the same. The closed forms whos closure ensures that this space is a G$_2$ manifold are
\beq
\Phi_{\text{G}_2}= r^2 dr\wedge  J_{\text{NK}}-r^3\text{Im}\Omega_{\text{NK}},~~~~\hat\star_7\Phi_{\text{G}_2}= r^4 J_{\text{NK}}\wedge   J_{\text{NK}}+r^3\text{Re}\Omega_{\text{NK}}\wedge dr,\label{eq:g2coneforms}
\eeq
where $(J_{\text{NK}},\Omega_{\text{NK}})$ are the SU(3)-structure forms on the base of the cone that solve
\beq
dJ_{\text{NK}}=-3\text{Re}\Omega_{\text{NK}},~~~ d\Omega_{\text{NK}}=-2J_{\text{NK}}\wedge J_{\text{NK}},
\eeq
which are the analogue of the analogue of Kahler-Einstein conditions \eqref{eq:KEstructure} in this case. To be able to turn on the connection term in $D\phi$, we need to be able to define a closed 2-form in the G$_2$ cone, but the forms comprising the nearly Kahler structure do not include such a form. This prevents us from making a universal ansatz like in section \ref{sec:universalcompactansatz} and so we need to be specific in our choice of nearly Kahler base. Up to orbifolds, there are 4 examples of compact NK$_6$ manifolds with explicitly known metrics, the 6-sphere, $\mathbb{CP}^3$, the flag manifold $\mathbb{F}^3$ and a fibration of S$^3$ over S$^3$, we will focus on the latter.

The forms describing the nearly Kahler structure on S$^3\times$S$^3$ are
\begin{align}
J_{\text{NK}}&=\frac{1}{\sqrt{3}6}(R_1\wedge \tilde{R}_1+R_2\wedge \tilde{R}_2+R_3\wedge \tilde{R}_3),\nn\\[2mm]
\text{Re}\Omega_{\text{NK}}&=\frac{1}{27}\left(\tilde{R}_1-\frac{1}{2}R_1\right)\wedge \left(\tilde{R}_2-\frac{1}{2}R_2\right)\wedge \left(\tilde{R}_3-\frac{1}{2}R_3\right)-\frac{1}{72}\epsilon_{ijk}\left(\tilde{R}_i-\frac{1}{2}R_i\right)\wedge R_j\wedge R_k,\nn\\[2mm]
\text{Im}\Omega_{\text{NK}}&=\frac{1}{\sqrt{3}36}\epsilon_{ijk}\left(\tilde{R}_i\wedge R_j\wedge R_k-R_i\wedge \tilde{R}_j\wedge \tilde{R}_k\right),\label{eq:S3S3Jomega}
\end{align}
where $(R_i,\tilde{R}_i)$ are two sets of SU(2) right invariant forms on two separate 3-spheres that obey \eqref{eq:leftrightsu2}. The corresponding metric on our internal space is then
\beq
ds^2(\text{G}_2)=dr^2+  \frac{r^2}{12}(R_i)^2+ \frac{r^2}{9} \left(\tilde{R}_i-\frac{1}{2}R_i\right)^2.
\eeq
Inserting \eqref{eq:S3S3Jomega} into \eqref{eq:g2coneforms} it is possible to show that the connection
\beq
{\cal A}= \frac{r^2}{6}\left(b_1(R_1+\tilde{R}_1)+b_2(R_2+\tilde{R}_2)+b_3(R_3+\tilde{R}_3)\right)~~~~\Rightarrow~~~ {\cal F}^2=14b^2,~~~~b^2=b_1^2+b_2^2+b_3^2,
\eeq
is such that ${\cal F}=d{\cal A}$ solves \eqref{eq:calfconds}. We then find that \eqref{eq:defineingG2PDE} is solved in terms of constants $(c_1,c_2,b)$ by
\beq
h_2= c_1+ \frac{c_2}{r^5}- b^2r^2,\label{eq:D2O2warp}
\eeq
which is essentially a D2 brane analogue of the warp factor in \eqref{eq:D5O5}. Indeed for different tunings of $(c_1,c_2,b^2)$ \eqref{eq:D2O2warp} gives rise to 3 distinct ways to bound the interval spanned by $r$ namely between, a regular zero at $r=0$ and an O2 plane at $r=r_{\text{max}}>0$ for which
\beq
h_2= b^2(r_{\text{max}^2}-r^2),
\eeq
D6 branes at $r=0$ and an O2 plane at $r=r_{\text{max}}>0$ for which
\beq
h_2=\frac{(r_{\text{max}}-r)}{r_{\text{max}}^7r^5}\left(c_1r_{\text{max}}^5(r+r_{\text{max}})r^5+c_2\sum_{i=0}^6r^ir_{\text{max}}^{6-i}\right),~~~ c_2>0,~~~b^2=\frac{c_2+c_1r_{\text{max}}^5}{r_{\text{max}}^7}>0,
\eeq
 or two O2 planes with $0<r_{\text{min}}\leq r\leq r_{\text{max}}$ for which
\begin{align}
h_2&=b^2 (r_{\text{max}}-r)(r-r_{\text{min}})\bigg[1+\frac{r_{\text{min}}+r_{\text{max}}}{r}+\frac{r_{\text{min}}^4r_{\text{max}}^4(r_{\text{min}}+r_{\text{max}})}{r^5\sum_{i=0}^4r_{\text{min}}^ir_{\text{max}}^{4-i}}\bigg(1+\nn\\[2mm]
&\left(\frac{1}{r_{\text{min}}}+\frac{1}{r_{\text{max}}}\right)\left(1+\left(\frac{1}{r^2_{\text{min}}}+\frac{1}{r^2_{\text{max}}}\right)r^2\right)r+\left(\frac{1}{r_{\text{min}}^2}+\frac{1}{r_{\text{max}}^2}+\frac{1}{r_{\text{min}}r_{\text{max}}}\right)r^2\bigg)\bigg].
\end{align}
We find no issue with flux quantisation in any of these cases, so they all represent viable Mink$_2$ vacua.

\subsection{Solitonic branes}\label{eq:solbranes}
In this section we will find solitonic brane solutions whose internal spaces are generic squashed Sasaki-Einstein manifolds, we will again make use of the classes of solution derived in sections \ref{sec:D1circle}, \ref{sec:D3circle} and \ref{sec:D5circle}.\\
~\\
An AdS soliton, originating in \cite{Horowitz:1998ha}, is a simple deformation of the Poincare patch of AdS$_{p+2}$ in which one spacial direction of the Mink$_{p+1}$ factor is compactified and the space is deformed in terms of a function $f=f(r)$. The metric of such solutions takes the form 
\beq
ds^2=  \frac{r^2}{\ell^2}\bigg[ds^2(\text{Mink}_p)+f d\phi^2\bigg]+  \frac{\ell^2}{r^2 }\frac{dr^2}{f},\label{eq:AdS soliton}
\eeq
where $f$ could be any function such that $f\to 1$ as $r\to \infty$, which also has an order 1 zero at $r=r_0>0$ and has no zeros or infinities for $r_0<r<\infty$. If we assume $f=f_0(r-r_0)$ at $r=r_0$ then the metric has a regular zero at $r=r_0$ when $\phi$ has period $\frac{4\pi l}{f_0 r_0^2}$ and tends to AdS$_{p+2}$ when $r\to \infty$.   The particular case of  \cite{Horowitz:1998ha} is
\beq
f=1-\left(\frac{r_0}{r}\right)^{3+p},
\eeq
however other more general choices of $f$ have been shown to give rise to AdS$_4$ and AdS$_5$ solitons in 4 and 5d gauged supergravities \cite{Anabalon:2021tua,Bobev:2020pjk}. These can be lifted to $d=11$ and type IIB supergravities respectively and are also compatible with supersymmetry.

In this section we focus on the embedding of a generalisation of \eqref{eq:AdS soliton} into higher dimensions where we now take the metric to be  
\beq
ds^2=  \left(\frac{r}{\ell}\right)^{\frac{7-p}{2}}\bigg[ds^2(\text{Mink}_p)+f d\phi^2\bigg]+  \left(\frac{\ell}{r}\right)^{\frac{7-p}{2}}\frac{dr^2}{f},\label{eq:brane soliton}
\eeq
where $f$ has the same behaviour described earlier. We will refer to such solutions as solitonic branes, because $\frac{\ell^2}{r^2 }$ has been replaced with the near horizon limit of a D$p$ brane warp factor. The field theory dual of such backgrounds correspond to the reduction on a circle of quantum field theories preserving different amounts of supersymmetry. These reductions where previously considered in \cite{Anabalon:2021tua,Bobev:2020pjk}, where it was showed that a supersymmetric (gravitational) soliton is dual to the twisted circle compactification of the dual theory. To preserve SUSY there is a twist between the isometry of the circle and the U(1) subgroup of the R-symmetry: there is a constant background gauge field of the R-symmetry with components only along the circle direction. This line of research lead to generalizations in different contexts \cite{Anabalon:2022aig,Nunez:2023nnl,Nunez:2023xgl,Fatemiabhari:2024aua,Anabalon:2024che,Kumar:2024pcz,Chatzis:2024kdu,Chatzis:2024top}.

\subsubsection{Finding known Solitonic brane solution in our ansatz}\label{sec:solex}
Here, we are interested in backgrounds constructed with only one type of D$p$-branes, which may have different internal spaces. To connect with our reduction ansatz, we note that (see \cite{Kumar:2024pcz}) we can write the supersymmetric soliton background for the D$p$-brane as (here $p=1,3,5$)
\begin{align}
        ds^{2} &= \frac{1}{\sqrt{h_p}}\left( ds^2(\text{Mink}_p) + fd\phi^{2}       \right) + 
        \sqrt{h_p}\left( \frac{dr^{2}}{f} + r^{2}\sum^{\frac{9-p}{2}}_{i=1} (d\mu_{i})^{2} 
             + \mu^{2}_{i}\left(  d\phi_{i} + {\cal B} \right)^{2} \right),\nn\\[2mm]
        F^{\text{el}}_{p+2}  &= e^{-\Phi_0}\text{vol}(\text{Mink}_p) \wedge \left( d\left(\frac{1}{h_p}\right)\wedge d\phi-{\cal B}_{0} \sum^{\frac{9-p}{2}}_{i=1} d(\mu^{2}_{i})\wedge (d\phi_{i}+{\cal B}) \right),\nn\\[2mm]
        e^{-\Phi} &= e^{-\Phi_0}h^{\frac{p-3}{4}}_p,~~\label{SUSYSoliton}
    \end{align}
where we only write the electric portion of the flux\footnote{this is relevant because while $F_{p+2}=F^{\text{el}}_{p+2}$ for $p=1,5$, $F_5= (1+\star )F_5^{\text{el}}$}, $\mu_i$ are embedding coordinates constrained as $\sum_{i}\mu^{2}_{i}=1$,  $(h_p,f,\zeta)$ are functions of $r$ and $\mathcal{B}_0$ is a constant, they take the form
    \begin{equation}
    \begin{aligned}
        &h_p = \left(\frac{\ell}{r}\right)^{7-p}, \quad
        f = 1 - \left(\frac{r_0}{r}\right)^{9-p},\quad
       {\cal B} = {\cal B}_{0} \zeta(r) d\phi, \\
        &\zeta = \frac{1}{r^{2}} - \frac{1}{r^{2}_{0}},\quad
        r_{0} = \left( Q^{3+p}\ell^{12} \right)^{\frac{1}{9-p}}, \quad
        {\cal B}_{0} = \frac{(r_{0})^{\frac{9-p}{2}}}{\ell^{\frac{7-p}{2}}}.\label{eq:solitonicvals}
    \end{aligned}
    \end{equation}
In the above $r_0\leq r\leq \infty$ with the lower bound a regular zero, provided the period of $\phi$ is $\phi\sim \phi + \frac{4\pi}{f'(r_{0})}\left(\frac{\ell}{r_{0}}\right)^{\frac{7-p}{2}}$  and $r$ is interpreted as a holographic coordinate.
At first sight \eqref{SUSYSoliton} does not look very similar to the sort of compactification ansatze we have been considering in this work, but this is just down to the coordinates used to express the deformed $(8-p)$-sphere it contains. Indeed if we express S$^{8-p}$ as a U(1) fibration over $\CP{\frac{7-p}{2}}$ the relation becomes clear, \text{i.e} 
\begin{align}
\sum^{\frac{9-p}{2}}_{i=1} (d\mu_{i})^{2} 
             + \mu^{2}_{i}\left(  d\phi_{i} + {\cal B} \right)^{2}  &= 
      ds^{2}(\CP{\frac{7-p}{2}}) + (d\psi + \eta + {\cal B})^{2},\nn\\[2mm]
\sum^{\frac{9-p}{2}}_{i=1} d(\mu^{2}_{i})\wedge (d\phi_{i}+{\cal B})&=-2J_{\mathbb{CP}},\label{eq:startargument}
\end{align}
where $\eta$ gives rise to the Kahler form on $\CP{\frac{7-p}{2}}$ as in \eqref{eq:kalherform}. With these substitutions we can rewrite the metric and flux in \eqref{SUSYSoliton} as
\begin{align}
ds^{2} &=   \frac{1}{\sqrt{h_p}}\left( ds^2(\text{Mink}_p)+ D\phi^2\right)+ 
        \sqrt{h_p}\left(\frac{dr^{2}}{f} + r^{2}ds^{2}(\CP{\frac{7-p}{2}}) + r^2f\tilde{D}\psi^{2}\right),\nn\\[2mm]
			D\phi&=d\phi+  q\tilde{D}\psi,~~~~\tilde{D}\psi=d\psi+ \eta-\frac{{\cal B}_0}{r_0^2}d\phi\nn\\[2mm]
F^{\text{el}}_{p+2}&= e^{-\Phi_0}\text{vol}(\text{Mink}_p) \wedge d\left( h^{-1}_{p}D\phi \right),  \label{SUSYSoliton2},
\end{align}
with $q= h_p{\cal B}_{0}$. This looks a lot more familiar, indeed locally we can perform a coordinate transformation such that 
\beq
\tilde{D}\psi\to  D\psi=d\psi+\eta,
\eeq
then from this perspective it is clear that the solutions actually lie in the class of circle compactified D$p$ branes on CY manifolds in \eqref{eq: DponCY}, one should identify
\beq
{\cal A}=qD\psi,~~~~ds^2(\text{CY}_{\frac{9-p}{2}})=\frac{dr^{2}}{f} + r^{2}ds^{2}(\CP{\frac{7-p}{2}}) + r^2 f D\psi^{2}.
\eeq
Formally this is rather similar to our universal ansatz in section \ref{sec:universalcompactansatz}, only now an additional squashing function $f$ appears - note that the presence of either $f$ or ${\cal A}$ break the SO(9-$p$) isometry of S$^{8-p}$ to U($\frac{9-p}{2}$). In the next section we will generalise the solutions of \eqref{SUSYSoliton} to arbitary squashed SE manifolds and general $h_p$.

\subsubsection{New solitonic brane solutions in type IIB}\label{sec:solnew}
We will again take the class of \eqref{eq: DponCY} as our starting point, and make the following ansatz for our CY forms and connection
\beq
\hat J = dr\wedge (r D\psi)+ r^2 J_{\text{KE}},~~~~\hat\Omega= r^{\frac{7-p}{2}}e^{i\frac{9-p}{2}\psi}\left(\frac{dr}{f}+i rf D\psi\right)\wedge \Omega_{\text{KE}},~~~~{\cal A}= q D\psi,
\eeq
where the forms $(J_{\text{KE}},\Omega_{\text{KE}})$ obey \eqref{eq:KEstructure} and we take $(f,q)$ to be arbitrary functions of $r$. This choice makes our CY manifold a cone over a squashed Sasaki-Einstein manifold
\beq
ds^2(\text{CY}_{\frac{9-p}{2}})=\frac{dr^{2}}{f} + r^{2}ds^{2}(\text{KE}^{\frac{7-p}{2}}) + r^2 f D\psi^{2}.
\eeq
The conditions $d\hat\Omega=0$ and \eqref{eq:primativecondsexamplesec} demands that we in general fix
\beq
f= 1+  \frac{f_0}{r^{9-p}},~~~~q=\frac{q_0}{r^{7-p}},~~~~\Rightarrow~~~~{\cal F}^2=(9-p)(7-p)\frac{q_0^2}{r^{2(9-p)}},\label{eq:solsimpsol}
\eeq
while $d\hat J=0$ hold automatically. The PDE that defines $h_p$ away from sources in \eqref{eq:defineingCYPDE} then becomes
\begin{align}
\frac{1}{r^{8-p}}\partial_{r}(r^{8-p}f\partial_r( h_p))+ \frac{(9-p)(7-p)q_0^2}{r^{2(9-p)}}=0.\label{eq:solgenode}
\end{align}
A particularly simple solution to this ODE is given by demanding that 
\beq
h_p=\left(\frac{\ell}{r}\right)^{7-p},~~~~f_0=-\frac{q_0^2}{\ell^{7-p}}.
\eeq
If we further tune  $q^2_0=\ell^{7-p}r_0^{9-p}$ we derive a generalisation of \eqref{SUSYSoliton} to arbitrary squashed Sasaki-Einstein manifolds. Performing the coordinate transformation $\psi\to \psi-\frac{{\cal B}_0}{r_0^2}d\phi$ we find these take the form
\begin{align}
ds^2&= \frac{1}{\sqrt{h_p}}\left( ds^2(\text{Mink}_p)+ fd\phi^{2}\right)+ 
        \sqrt{h_p}\left(\frac{dr^{2}}{f} + r^{2}ds^{2}(\text{KE}^{\frac{7-p}{2}}) + r^2(d\psi+ \eta+{\cal B})^2\right),\nn\\[2mm]
F^{\text{el}}_{p+2}&= e^{-\Phi_0}\text{vol}(\text{Mink}_p) \wedge d\left( h^{-1}_{p} \left(d\phi + q \left(D\psi - \frac{\mathcal{B}_{0}}{r^{2}_{0}}d\phi \right) \right) \right)    ,\nn\\[2mm]
e^{-\Phi}&=e^{-\Phi_0}h_p^{\frac{p-3}{4}},\label{eq: gensolbrane}
\end{align}
where $(f,h_p,{\cal B})$ are defined as in \eqref{eq:solitonicvals}. These solutions still interpolate between a regular zero at $r=r_0$ and the same $r$ dependent boundary behavior as $r\to\infty$. The generalised $p=3$ case,  which is asymptotically AdS$_5$, has appeared before in \cite{Chatzis:2024kdu} (although only the case SE$^5=Y^{p,q}$ is studied in detail), which use the Sasaki-Einstein reduction of \cite{Gauntlett:2007ma} to embed the AdS$_5$ soliton into type IIB. The generalised case for $p=1,5$ are totally new, we expect them to exhibit similar physical behaviour to their squashed sphere cousins: these backgrounds preserve four Poincaré supercharges, so that the effective 3D dual field theories are  $\mathcal{N}=2$. Also, a smoothly shrinking circle of the bulk geometry realises confinement in the dual theory. It would be interesting to see if the general SE cases are gapped as the $p=3$ and SE=$S^{5}$ case.\\
~\\
The simple solution of \eqref{eq:solsimpsol} is by no means the only solution to \eqref{eq:solgenode}, indeed the general solution for $f_0\neq 0$ is given by
\beq
h_p=c_1-\frac{q_0^2}{f_0 r^{7-p}}+c_2 r^2 \!~_2F_1\left(1,\frac{2}{9-p},1+\frac{2}{9-p},- \frac{r^{9-p}}{f_0 }\right),\label{eq:gensol}
\eeq
for $\!~_2F_1\left(a,b,c;z\right)$ the standard hypergeometric function and  $(c_1,c_2)$ integration constants which have been set to zero in \eqref{eq:solsimpsol}. This of course contains distinct solutions depending on how the constants $(f_0,q_0,c_1,c_2)$ are tuned, let us focus on those for which $ds^2(\text{CY}_{\frac{9-p}{2}})$ contains a regular zero at $r=r_0$  and for which $h_p\to c+\frac{l^{7-p}}{r^{7-p}}$ as $r\to \infty$, so that if we fix $c=0$ we recover the UV behaviour of \eqref{eq: gensolbrane} and so have a solution which still yields a holographic description of a QFT - this requires that we  fix
\beq
f_0=-r_0^{9-p},~~~~q^2_0=r_0^{9-p}\left(\ell^{7-p}-\frac{(9-p)br_0^{7-p}}{7-p}\right).\label{eq:hpgensolq0}
\eeq
For the specific values of $p$ we are interested in, we find then find that \eqref{eq:gensol} simplifies in terms of less esoteric functions as
\begin{align}
h_p&=c+ \left(\frac{\ell}{r}\right)^{7-p}+ b\tilde{h}_p,\nn\\[2mm]
        \tilde{h}_5 &=  2\arctanh\left( \frac{r_0^{2}}{r^{2}} \right)- \frac{ 2r_0^{2}}{r^{2}},\\[2mm]
        \tilde{h}_3 &=  
            \frac{1}{2}\ln\left( \frac{(r^{4}+r^{2}r_0^{2}+r_0^{4})}{(r^{2}-r_0^{2})^{2}} \right) - \sqrt{3}\arctan\left( \frac{\sqrt{3}r_0^{2}}{2r^{2}+r_0^{2}}\right) -\frac{3r_0^{4}}{2r^{4}},\nn\\[2mm]
        \tilde{h}_1 &=    2\arctanh\left(\frac{r_0^{2}}{r^{2}}\right) -2\arctan\left(\frac{r_0^{2}}{r^{2}}\right)-  \frac{
        4 r_0^{6} }{3r^{6}},\label{eq:hpgensol}
    \end{align}
where $\tilde{h}_p\sim r^{-2(8-p)}$ as $r\to \infty$ and is thus subleading at this boundary. The same is not true close to $r=r_0$ where in each case we find
\beq
h_p\sim |\log(r-r_0)|,\label{eq:hplimit}
\eeq
which is the behaviour of a codimension 2 source, and as we have only electric $p+2$ flux turned on, this can only correspond to a partially localised stack D$p$ smeared over the Kahler-Einstein base of the squashed Sasaki-Einstein manifold.

If we perform the coordinate transformation $\psi\to \psi -\frac{r_0^{7-p}}{q_0}\phi$ we then arrive at a further generalisation of the solitonic brane solutions of \eqref{SUSYSoliton} of the form
\begin{align}
ds^2&= \frac{1}{\sqrt{h_p}}\left( ds^2(\text{Mink}_p)+ \frac{f}{\Delta}d\phi^{2}\right)+ 
        \sqrt{h_p}\left(\frac{dr^{2}}{f} + r^{2}ds^{2}(\text{KE}^{\frac{7-p}{2}}) + r^2\Delta(d\psi+ \eta+{\cal B})^2\right),\nn\\[2mm]
F^{\text{el}}_{p+2}&= e^{-\Phi_0}\text{vol}(\text{Mink}_p) \wedge d\left( h^{-1}_{p} \left(d\phi + q \left(D\psi - \frac{r^{7-p}_{0}}{q_{0}}d\phi \right) \right) \right),\nn\\[2mm]
e^{-\Phi}&=e^{-\Phi_0}h_p^{\frac{p-3}{4}},\label{eq: gensolbrane2}
\end{align}
where
\beq \label{FunctionsSolitonKE}
\Delta=f+\frac{q_0^2 }{r^{2(8-p)}h_p},~~~f=1-\left(\frac{r_0}{r}\right)^{9-p},~~~~{\cal B}=\left(\frac{q_0}{r^{9-p} \Delta h}-\frac{r_0^{7-p}}{q_0}\right)d\phi,
\eeq
where $h_p$ is given by \eqref{eq:hpgensol} and the constant $q_0$ by \eqref{eq:hpgensol}. The coordinate transformation in $\psi$ ensures that ${\cal B}\to 0$ as $r\to r_0$, while on the other hand we find that $\Delta\to \frac{q^2_0}{r_0^{2(8-p)}h_p} $ in this limit with the warp factor itself behaving as \eqref{eq:hplimit}. This ensures that close to $r=r_0$ the metric tends to what one would expect for D$p$ branes extended in $(\text{Mink}_p,\psi)$ that are partially delocalised along the $\text{KE}^{\frac{7-p}{2}}$ directions, if the period of $\phi$ is $\frac{4\pi q_0 r_0^{7-p}}{9-p}$. Taking the $c=0$ limit in \eqref{eq:hpgensol} means that these solutions are asymptotically \eqref{eq: gensolbrane} and as such have dual QFT descriptions. The effect of the additional D brane sources that follow when including a non trivial $b$ in h$_p$ should be to add add a vev to a mesonic operator in the dual QFT that \eqref{eq: gensolbrane} describes as in \cite{Benvenuti:2005qb}. It would be very interesting to study this in more detail, but we leave that for future work.

The brane charge of the background is given by the flux of the magnetic $F^{\text{mag}}_{8-p}$ across the Sasaki-Einstein 
    \begin{equation}\label{BraneCharges}
        N_{D_{p}} = \frac{1}{(2\pi)^{7-p}}\int_{\text{SE}^{8-p}}  F^{\text{mag}}_{8-p} =  \frac{(7-p)\ell^{7-p}}{(2\pi)^{7-p}}\text{Vol}(\text{SE}^{8-p}).
    \end{equation}

\subsubsection{New AdS\texorpdfstring{$_4$}{4} solitons in \texorpdfstring{$d=11$}{d=11}}\label{sec:solAdS4}
The supersymmetric  AdS$_{4}$ soliton was constructed in \cite{Anabalon:2021tua} and uplifted to $d=11$ supergravity where it consists of a  deformation of AdS$_{4}\times $S$^{7}$ of the form\footnote{Here we use a slightly different parametrization when compared to \cite{Anabalon:2021tua}, in order to match our conventions.}
    \begin{equation}
    \begin{aligned}
        ds^{2} &= \frac{r^{4}}{\ell^{4}}\left( 
        ds^{2}(\text{Mink}_{2})    +  f d\phi^{2} \right)
                     + \frac{dr^{2}}{\frac{r^{2}}{4\ell^{2}}f} 
                     + 4\ell^{2} \sum^{4}_{i=1}\left( d\mu_{i}^{2} 
            + \mu^{2}_{i}(d\phi_{i} + \mathcal{B} )^{2}  \right) ,\\
        G_{4} &= \text{vol}(\text{Mink}_{2})\wedge\left(-\frac{6r^{2}}{\ell^{6}}dr\wedge d\phi  
        - \mathcal{B}_{0} \sum^{4}_{i=1} 
            d(\mu^{2}_{i})\wedge \mathcal{D}\phi_{i} \right),
    \end{aligned}
    \end{equation}
where
    \begin{equation}
    \begin{aligned}
        \mathcal{B}= \mathcal{B}_{0}\left(\frac{1}{r^{2}}-\frac{1}{r^{2}_{0}}\right)d\phi,\quad f= 1 - \frac{r_{0}^{8}}{r^{8}},~~~~ r_0^2=Q\ell^{3}, \quad 
        \mathcal{B}_{0} = Q^{2}\ell^{3}.
    \end{aligned}
    \end{equation}
Like the case of $p=3$ in \eqref{SUSYSoliton} this solution interpolates between a regular zero at $r=r_0$ (when $\phi$ has period $\phi\sim \phi + \frac{8\pi}{f'(r_{0})}\left(\frac{\ell}{r_{0}}\right)^{3}$) and AdS with a constant fibre over the S$^{7}$ as $r\to \infty$. 

By a similar argument to the one starting around \eqref{eq:startargument} this can be expressed in terms of a U(1) fibration over a squashed SE$_7$ manifold which should live within a $d=11$ analogue of our  Mink$_{d}\rightarrow$ Mink$_{d-1}$ compactifications presented earlier. The relevant class of circle compactifications of  Mink$_{3}$ in $d=11$ is given by
    \begin{equation}\label{11DClass}
    \begin{aligned}
         ds^{2} &= H^{-\frac{2}{3}}\left( ds^2(\text{Mink}_2) + (d\phi + \mathcal{A})^{2} \right) + H^{\frac{1}{3}}ds^{2}(\CY{4}),\\
         G_4&= d \left( H^{-1} \text{vol}(\text{Mink}_{2})\wedge (d\phi+\mathcal{A})) \right),
    \end{aligned}
    \end{equation}
where ${\cal F}=d{\cal A}$ must be a primative (1,1)-form on CY$_4$ directions, $D\phi = d\phi + \mathcal{A}$, and the warp factor $H$ obeys
\beq
\hat{\nabla}^2H+\hat{\cal F}^2=0,
\eeq
where as elsewhere the hats indicate that these quantities are computed on the unwarp CY$_4$ manifold. This can be derived
with the class of D1-brane branes wrapping a circle in section \ref{sec:D1circle} as follows\footnote{There is actually a more general class with a Spin(7) manifold rather than a CY$_4$, but one cannot keep the orginal NS flux non trivial in \ref{sec:D1circle}.}. The first step is to S-dualize the D1-brane configuration to a purely NS configuration, the S-dual metric, flux and dilaton are respectively
    \begin{equation}
        g'_{\mu\nu} = e^{-\Phi}g_{\mu\nu}, \quad 
        H'_{3} = F_{3}, \quad F'_{3}= 0,  \quad
        \Phi' = -\Phi.
    \end{equation}
Since now the configuration contains only NS-NS fields, we can consider it to be a solution of Type IIA supergravity (note that although the bosonic configuration is the same, the Killing spinors are different, as well as the dual field theory interpretation \cite{Itzhaki:1998dd}). Once in Type IIA, the F1 configuration can be lifted to $d=11$ as an M2-brane wrapping the $d=11$ circle spanned by $\partial_{x}$ as
    \begin{equation}\label{Lift10Dto11D}
    \begin{aligned}
            ds^{2} &= e^{-\frac{2}{3}\Phi'}ds^{2}_{\text{IIA}} + e^{\frac{4}{3}\Phi'}dx^{2},\\
            G_{4} &= - H'_{3}\wedge dx,
    \end{aligned}
    \end{equation}
Using this, the metric and flux of section \ref{sec:D1circle} (with $H_3=0$ and Spin(7) $\to$ CY$_4$), leads to \eqref{11DClass} upon identifying  $ds^2(\text{Mink}_2)=-dt^2+dx^2$.

We can now use \eqref{11DClass} to generalise the AdS$_4$ soliton in a similar fashion to \eqref{eq: gensolbrane2}. The derivation is analogous to that in the previous section so we will skip the details, the resulting class of solutions takes the form
    \begin{equation}\label{11DSmeared}
    \begin{aligned}
         ds^{2} &= h_{1}^{-\frac{2}{3}}\left( ds^2(\text{Mink}_2)+ \frac{f}{\Delta}d\phi^{2}\right) + h_{1}^{\frac{1}{3}}\left(\frac{dr^{2}}{f} + r^{2}ds^{2}(\text{KE}^{\frac{7-p}{2}}) + r^2\Delta(d\psi+ \eta+{\cal B})^2\right), \\
         G_4 &= \text{vol}(\text{Mink}_{2})\wedge d\left( h^{-1}_{1} \left(d\phi + q \left(D\psi - 
         \frac{r^{7-p}_{0}}{q_{0}} d\phi \right) \right) \right) ,
    \end{aligned}
    \end{equation}
with $h_{1}$ given by \eqref{eq:hpgensol} and $f$, $q$, $\mathcal{B}$ and $\Delta$ as in \eqref{FunctionsSolitonKE}. Flux quantization of $G_{7}=\star G_{4}$ works exactly as in \eqref{BraneCharges} for $p=1$
    \begin{equation}
        N_{M_{2}} = \frac{1}{(2\pi)^{6}}\int_{\text{SE}^{7}}  G_{7} =  \frac{6\ell^{6}}{(2\pi)^{6}}\text{Vol}(\text{SE}^{7}).
    \end{equation}
This generalises the $d=11$ AdS$_4$ soliton of \cite{Anabalon:2021tua} in two ways: First the internal space can now be a generic squashed-Einstein manifold, second it adds smeared source M2 branes at $r=r_0$. It would be interesting to study the field theoretic consequences of this, but this is beyond the scope of this work.

\section*{Acknowledgements}
We thanks Carlos Nunez for collaboration on related topics. The work of NM and RS is supported by the Ram\'on y Cajal fellowship RYC2021-033794-I, and by grants from the Spanish government MCIU-22-PID2021-123021NB-I00 and principality of Asturias SV-PA-21-AYUD/2021/52177.

\appendix

\section{Type II supergravity conventions}\label{sec:convensions}
In this appendix we give details on the conversions we use throughout this paper.\\
~\\
We work in type II supergravity whose Bosonic field content is the metric $g_{MN}$ (of signature $(-,+,...,+)$), the dilaton $\Phi$, NS 3-form $H$ and RR Poly form
\beq
F_+=F_0+F_2+F_4+F_6+F_{10},~~~~~F_-= F_1+F_3+F_5+F_7+F_9
\eeq
where $F_k$ are the usual democratic fluxes  and  the upper/lower signs should be taken in IIA/IIB here and elsewhere. The RR fluxes obey the self duality constraint
\beq
\star_{10}\lambda(F_{\pm})=F_{\pm}
\eeq
where for an arbitrary $k$-form $X_k$ we have $\lambda(X_k)=  (-1)^{[\frac{k}{2}]}X_k$ for $[n]$ the integer part of $n$ and our conventions for the hodge dual are such that in  dimension $d$ and in all signatures, with respect to a vielbein $e^{\underline{M}}$ we have
\beq
\star (e^{\underline{M}_1}\wedge...\wedge e^{\underline{M}_k})=\frac{1}{(d-k)!} \epsilon^{\underline{M}_1...\underline{M}_k}_{~~~~~~~\underline{N}_{k+1}...\underline{N}_d}e^{\underline{N}_{k+1}}\wedge...\wedge e^{\underline{N}_d},
\eeq
where $\underline{M}=(0,1....9)$ represents a tangent space index.
The equations of motion and flux Bianchi identities of type II supergravity, away from the loci of sources, can be compactly expressed in this notation as
\begin{align}
&d_H F=0,~~~~ dH=0,~~~~ d(e^{-2\Phi}\star H)=\frac{1}{2}(F_{\pm},F_{\pm})_8,~~~~2R- H^2-8 e^{\Phi}\nabla^2 e^{-\Phi}=0,\nn\\[2mm]
 &{\cal E}_{MN}=0,~~~~{\cal E}_{MN}= R_{MN}+2\nabla_{M}\nabla_{N}\Phi-\frac{1}{2} H^2_{MN}-\frac{e^{2\Phi}}{4} (F_{\pm})^2_{MN}=0,\label{eq:compactEOM}
\end{align}
where $(F_{\pm},F_{\pm})_8$ is the 8-form part of $F_{\pm}\wedge \lambda(F_{\pm})$, $d_H= d-H\wedge$ and for an arbitrary k-form $X_k$ and polyform $X$  we define
\begin{align}
(X_k)_{M}&:= \iota_{dx^M} X_k,~~~~ X_k^2:= \sum_{k} \frac{1}{k!} (X_k)_{M_1\dots M_k}(X_k)^{M_1\dots M_k},~~~~\nn\\
X^2_{MN}&:=\sum_{k} \frac{1}{(k-1)!} (X_k)_{MM_1\dots M_{k-1}}(X_k)_N{}^{M_1\dots M_{k-1}}.\label{eq:sqdefs}
\end{align}
Through these and the self duality constraint of $X_k$ it follows that \eqref{eq:compactEOM} is equivalent to the following conditions in IIA
\begin{align}
&dF_0=0,~~~dF_2=H F_0,~~~dF_4=H\wedge F_2,~~~~d\star F_4+H\wedge F_4=0,~~~d\star F_2+H\wedge \star F_4=0,\nn\\[2mm]
&dH=0,~~~d(e^{-2\Phi}\star H)=-F_0 \star F_2+ \frac{1}{2}F_4\wedge F_4+F_2\wedge \star F_4,\nn\\[2mm]
&2R- H^2-8 e^{\Phi}\nabla^2 e^{-\Phi}=0,\nn\\[2mm]
&R_{MN}+2 \nabla_{M}\nabla_{N}\Phi-\frac{1}{2} H^2_{MN}-\frac{e^{2\Phi}}{2}\left((F_2)^2_{\mu\nu}+(F_4)^2_{\mu\nu}-\frac{1}{2}g_{MN}(F_0^2+F_2^2+F_4^2)\right)=0,
\end{align}
and the following in IIB
\begin{align}
&dF_1=0,~~~dF_3=H\wedge F_1,~~~dF_5=H\wedge F_3,~~~~d\star F_3+H\wedge F_5=0,~~~d\star F_1+H\wedge F_5=0,\nn\\[2mm]
&dH=0,~~~d(e^{-2\Phi}\star H)=F_1\wedge \star F_3+F_5\wedge F_3\nn\\[2mm]
&2R- H^2-8 e^{\Phi}\nabla^2 e^{-\Phi}=0,\nn\\[2mm]
&R_{MN}+2 \nabla_{M}\nabla_{N}\Phi-\frac{1}{2} H^2_{MN}-\frac{e^{2\Phi}}{2}\left((F_1)^2_{\mu\nu}+(F_3)^2_{\mu\nu}+\frac{1}{2}(F_5)^2_{\mu\nu}-\frac{1}{2}g_{MN}(F_1^2+F_3^2)\right)=0,
\end{align}
which may be more familiar to some readers.\\
~~\\
We will now quote our spinor conventions relevant for the conditions for supersymmetry preservation in type II supergravity. Our flat space $d=10$ gamma matrices are $\Gamma_{\underline{M}}$ and obey
\beq
\{\Gamma_{\underline{M}},\Gamma_{\underline{N}}\}= \eta_{\underline{M}\underline{N}},
\eeq
for $\eta=\text{Diag}(-1,1,...,1)$, while their curved space analogues are $\Gamma_{M}=e^{\underline{N}}_{~M} \Gamma_{\underline{N}}$. We take the chirality operator to be
\beq
\hat\Gamma=\Gamma^{0...9}
\eeq
and introduce an intertwiner $B$ such that
\beq
B^{-1}\Gamma_{M}B=\Gamma_{M},
\eeq
for  $B B^*=B B=1$. The Majorana conjugate of some  $d=10$ spinor $\epsilon$ is defined as
\beq
\epsilon^c= B \epsilon^{*}.
\eeq
The conditions unbroken supersymmetry are expressed in terms of two non-trivial Majorana-Weyl Killing spinors $(\epsilon_1,\epsilon_2)$, meaning  that they obey
\beq
\epsilon^c_{1,2}=\epsilon_{1,2},~~~~\hat\Gamma\epsilon_1=\epsilon_1,~~~\hat\Gamma \epsilon_2=\mp \epsilon_2,
\eeq
and so each contain at most 16 independent real supercharges for a maximum total of 32.
The bosonic content of type II supergravity preserves supersymmetry if their exist non zero $\epsilon_{1,2}$ obeying the constraints
\begin{subequations}
\begin{align}
& \left(\nabla_M-\frac{1}{4}H_M\right) \epsilon_1 + \frac{e^\Phi}{16}F_{\pm} \Gamma_M\epsilon_2=0,~~~~ \left(\nabla_M+\frac{1}{4}H_M\right) \epsilon_2 \pm \frac{e^\Phi}{16} \lambda(F_{\pm}) \Gamma_M \epsilon_1=0, \label{eq:10dsusysseqs1s}\\
& \left(\nabla-\frac{1}{4}H-d\Phi\right)\epsilon_1=0,~~~~
\left(\nabla+\frac{1}{4}H-d\Phi\right)\epsilon_2=0,\label{eq:10dsusysseqs4s}
\end{align}
\end{subequations}
where 
\beq
\lambda(C_k)=(-)^{[\frac{k}{2}]}C_k,
\eeq
and forms should be understood as acting on spinors through the Clifford map. Specifically a form and its components contracted with an antisymmetric product of gamma matrices are equivalent objects
\beq
C_k=\frac{1}{k!}(C_k)_{\underline{M}_1...\underline{M}_k}e^{\underline{M}_1...\underline{M}_k}\equiv \slashed{C}_k=\frac{1}{k!}(C_k)_{\underline{M}_1...\underline{M}_k}\Gamma^{\underline{M}_1...\underline{M}_k}\label{eq:Cliffordmap}.
\eeq
This equivalence holds essentially because the vielbein $e^{\underline{M}}$ provides a representation of the gamma matrices, such that
\beq
\Gamma^{\underline{M}} \slashed{C}_k\equiv(e^{\underline{M}}\wedge+ \iota_{e^{\underline{M}}}) C_k,~~~~ \slashed{C}_k\Gamma^{\underline{M}}\equiv(-1)^k(e^{\underline{M}}\wedge- \iota_{e^{\underline{M}}}) C_k.\label{eq:veilbeinasgammas}
\eeq
The the spin covariant derivative is defined as
\beq
\nabla_M\epsilon= \partial_M\epsilon+\frac{1}{4}(\omega_M)_{\underline{A}\underline{B}}\Gamma^{\underline{A}\underline{B}}\epsilon,
\eeq
for $\omega_{\underline{A}\underline{B}}$ the usual anti-symmetric torsion free spin connection 1-form obeying $d e^{\underline{A}}+\omega^{\underline{A}}_{~\underline{B}}\wedge e^{\underline{B}}=0$. 

Finally we note that our conventions imply that
\beq
\hat\Gamma F_{\pm}=\star\lambda(F_{\pm}),
\eeq
where $\hat\Gamma$ should be viewed as acting on $F$ through the Clifford map.

\section{Derivation of Mink\texorpdfstring{$_2$}{2} Bi-linear conditions}\label{sec:mink2susyderivation}
In this appendix we derive necessary and sufficient conditions for warped Mink$_2$ solutions of type II supergravity to preserve Minimal supersymmetry in terms of bi-linears.\\
~\\
A Mink$_2$ solution may always be decomposed in the form
\begin{align}
ds^2&=e^{2A}ds^2(\text{Mink}_2)+ ds^2(\text{M}_8),~~~~H= e^{2A}H_1\wedge \text{vol}(\text{Mink}_2)+ H_3,\nn\\[2mm]
F_{\pm}&= f^{(8)}_{\pm}+ e^{2A}\text{vol}(\text{Mink}_2)\wedge \star_8\lambda(f^{(8)}_{\pm}),
\end{align}
where the dilaton $\Phi$ and $(e^{A},H,H_1,f^{(8)}_{\pm})$ have support on only M$_8$, which has no dependence on Mink$_2$. We will take $\mu$ to be coordinates on Mink$_2$ and $a$ coordinates on M$_8$.

We decompose our $d=10$ gamma matrices in a form compatible with the $2+8$ split of the metric as 
\beq
\Gamma_{\mu}=e^{A}\alpha^{(2)}_{\mu}\otimes \hat\gamma^{(8)},~~~~~\Gamma_{a}=\mathbb{I}\otimes \gamma^{(8)}_a.
\eeq
for $\alpha^{(2)}_{\mu}$ a real basis of gamma matrices on Mink$_2$ with chirality matrix $\hat\alpha^{(2)}=(\alpha^{(2)})^{01}$ and $\gamma^{(8)}_a$ a basis on M$_8$ with chirality matrix $\hat\gamma^{(8)}=(\gamma^{(8)})^{1...8}$. Our $d=10$ conventions then imply that the $d=10$ chirality matrix and intertwiner decompose as 
\beq
\hat \Gamma=\hat\alpha^{(2)}\otimes \hat\gamma^{(8)},~~~B=\mathbb{I}\otimes B^{(8)},
\eeq
for $B^{(8)}B^{(8)*}=1$ and $(B^{(8)})^{-1}\gamma^{(8)}_a B^{(8)}=\gamma^{(8)*}_a$. 

A peculiarity about the embedding of Mink$_2$ solution into $d=10$ is that it is possible to define Majorana-Weyl spinors on both Mink$_2$ and M$_8$. This means that while it is possible to define Majorana spinors $\zeta^{(2)}_{\pm}$ of even and odd chirality on Mink$_2$ (that are real in our conventions), minimal supersymmetry will only include one of these, making it chiral. We will choose, without loss of generality, to include only $\zeta^{(2)}_+$ , which we will refer to as ${\cal N}=(1,0)$ supersymmetry (see \cite{Rosa:2013lwa} for related work on ${\cal N}=(2,0)$), such that the $d=10$ spinors decompose as  
\beq
\epsilon_1=  \zeta^{(2)}_+\otimes  \chi^{(8)}_{1+},~~~~~\epsilon_2=  \zeta^{(2)}_+\otimes  \chi^{(8)}_{2\mp},\label{eq:mink2spinordecomposition}
\eeq
where and $(\chi^{(8)}_{1+},\chi^{(8)}_{1\mp})$ are  Majorana-Weyl spinors on M$_8$.

We would like to derive geometric conditions that imply \eqref{eq:10dsusysseqs1s}-\eqref{eq:10dsusysseqs4s} for Mink$_2$ solution in type II supergravity. Fortunately geometric conditions for generic solutions were already found in \cite{Tomasiello:2011eb}, so our task is to refine these to the case on Mink$_2$.  The conditions of \cite{Tomasiello:2011eb} are phrased in terms of the following forms 
\begin{align}
K&=\frac{1}{2}(K_1+K_2),~~~~ \tilde{K}=\frac{1}{2}(K_1-K_2),~~~~ K_i=\frac{1}{32}\overline{\epsilon}_i\Gamma_{\underline{M}}\epsilon_i e^{\underline{M}}\nn\\[2mm]
\Psi_{\pm}&= \frac{1}{32}\sum_{n=0}^{10}\frac{1}{n!}\overline{\epsilon}_2\Gamma_{\underline{M}_{n}....\underline{M}_1}\epsilon_1 e^{\underline{M}_1...\underline{M}_n},\label{eq:forms10d}
\end{align}
where $\overline{\epsilon}=(\Gamma_{0}\epsilon)^{\dag}$ and $\slashed{\Psi}_{\pm}=\epsilon_1\otimes \overline{\epsilon}_2$.
Supersymmetry requires that the differential 
\beq
d\tilde{K}= \iota_{K} H,~~~~ d_H(e^{-\Phi}\Psi_{\pm})=- (\iota_{K}+\tilde{K}\wedge )F_{\pm},~~~~ \nabla_{(M}K_{M)}=0,\label{eq:10dsusy1}
\eeq
and ``pairing'' (for $(X,Y)_{k}$  the $k$-form part of $X\wedge \lambda(Y)$)
\begin{subequations}
\begin{align}
\left( e_{1+}\Psi_{\pm} e_{2+},~\Gamma^{MN}\bigg[\pm \Psi_{\pm} d_H\left(e^{-\Phi}\Psi_{\pm} e_{+2}\right)+\frac{e^{\Phi}}{2}(\star d(e^{-2\Phi}\star e_{+2}))\Psi_{\pm}-F_{\pm}\bigg]\right)_{10}=0,\label{eq:10dsusy2}\\[2mm]
\left( e_{1+}\Psi_{\pm} e_{2+},~\bigg[ d_H\left(e^{-\Phi} e_{+1}\Psi_{\pm} \right)-\frac{e^{\Phi}}{2}(\star d(e^{-2\Phi}\star e_{+1}))\Psi_{\pm}-F_{\pm}\bigg]\Gamma^{MN}\right)_{10}=0,\label{eq:10dsusy3}
\end{align}
\end{subequations}
constraints all hold, where $e_{1,2+}$ are  null 1-forms obeying
\beq
e_{1+} \cdot K_{1}=e_{2+} \cdot K_{2} =\frac{1}{2},\label{eq:1-formnullnorms}
\eeq
but which are otherwise free - Note that terms like $\Psi_{\pm} e_{+2}$ can be understood as the forms that are equivalent to $\slashed{\Psi}_{\pm} \slashed{e}_{+2}$ under the  Clifford map. 

For the case at hand, as the spinors decompose as a tensor product of spinors on Mink$_2$ and M$_8$, the forms in \eqref{eq:forms10d} must decompose in terms of wedge products of forms on these respective whose components are bi-linears of $\zeta_+$ and $(\chi_{1+},\chi_{2\mp})$ respectively. On Mink$_2$ the only such form we can define is
\beq
v= \bar{\zeta}_+\alpha^{(2)}_{\mu}\zeta_+ dx^{\mu},
\eeq
which is necessarily closed, null and such that $\nabla^{(2)}_{(\mu}v_{\nu)}=0$ and $\iota_v\text{vol}(\text{Mink}_2)=-v$, while on M$_8$ we have 
\beq
\slashed{\Psi}^{(8)}_{\mp}=e^{-A}\chi^{(8)}_{1+}\otimes \chi^{(8)}_{2\mp}\equiv\Psi^{(8)}_{\mp}=e^{-A}\frac{1}{16}\sum_{n=0}^8\frac{1}{n!}\chi^{(8)\dag}_{2\mp}\gamma_{\underline{a_n}...\underline{a_1}}\chi^{(8)}_{1+}e^{\underline{a}_1...\underline{a}_n}\label{eq:psi8poly}
\eeq 
for $e^{\underline{a}}$ a vielbein on M$_8$. We find that
\begin{align}
K&=\frac{1}{64}(||\chi_{1+}||^2+||\chi_{2\mp}||^2) e^{A} v,\nn\\[2mm]
\tilde{K}&=\frac{1}{64}(||\chi_{1+}||^2-||\chi_{2\mp}||^2) e^{A} v,\nn\\[2mm]
\Psi_{\pm}&=\mp \frac{1}{2}e^{2A}v\wedge \Psi^{(8)}_{\mp},\label{eq:10toedecompo}
\end{align}
Given this $\nabla_{(M}K_{M)}=0$ demands we impose
\beq
d(e^{-A}(||\chi_{1+}||^2+||\chi_{2\mp}||^2))=0,\label{eq:spinor norms}
\eeq
making $K$ a null Killing vector, while $d\tilde{K}= \iota_{K} H$ that
\beq
d(e^{A}(||\chi_{1+}||^2-||\chi_{2\mp}||^2))= e^{A}(||\chi_{1+}||^2+||\chi_{2\mp}||^2)H_1.
\eeq
Then given that 
\beq
(K+\iota_{\tilde{K}})F_{\pm}=\frac{1}{64}e^A v\wedge \left((||\chi_{1+}||^2-||\chi_{2\mp}||^2)f^{(8)}_{\pm}-(||\chi_{1+}||^2+||\chi_{2\mp}||^2)\star_8\lambda(f^{(8)}_{\pm})\right),
\eeq
the remaining condition in \eqref{eq:10dsusy1} imposes 
\beq
d_{H_3}(e^{2A}\Psi^{(8)}_{\mp})=\pm\frac{ e^{A}}{32}\left(-(||\chi_{1+}||^2-||\chi_{2\mp}||^2)f^{(8)}_{\pm}+(||\chi_{1+}||^2+||\chi_{2\mp}||^2)\star_8\lambda(f^{(8)}_{\pm})\right)
\eeq
To deal with the pairing constraints \eqref{eq:10dsusy2}-\eqref{eq:10dsusy3} we note that the negative chirality spinor on Mink$_2$ furnishes us with a second 1-form $u$ defined as
\beq
u= \bar{\zeta}_-\alpha^{(2)}_{\mu}\zeta_- dx^{\mu},
\eeq
which like $V$ is also closed, null and such that $\nabla^{(2)}_{(\mu}u_{\nu)}=0$ and and $\iota_u\text{vol}(\text{Mink}_2)=u$, but additionally $\zeta_-$ can be normalised such that
\beq
u.v=-1.
\eeq
This means we can satisfy \eqref{eq:1-formnullnorms} by choosing
\beq
e_{i+}=- \frac{8 e^{A}}{||\chi_i||^2},~~~i=1,2.
\eeq
Reducing the $d=10$ pairing conditions to constraints on M$_8$ is now a very similar computation to appendix B of \cite{Macpherson:2021lbr}. After a lengthy computation we find that \eqref{eq:10dsusy2}-\eqref{eq:10dsusy3} impose only 2 further constraints  following from the parts with an index along each of Mink$_2$ and M$_8$, namely
\begin{align}
e^A(\Psi^{(8)}_{\mp},f^{(8)}_{\pm})_7 =\mp \frac{1}{16}e^{-\Phi}\star_8\left( 2(||\chi_1||^2+||\chi_2||^2) dA- (||\chi_1||^2-||\chi_2||^2) H_1\right),\nn\\[2mm]
e^A(\Psi^{(8)}_{\mp},\star_8 \lambda(f^{(8)}_{\pm}))_7 =\pm \frac{1}{16}e^{-\Phi}\star_8\left( 2 (||\chi_1||^2-||\chi_2||^2) dA- (||\chi_1||^2+||\chi_2||^2) H_1\right)\label{eq:8dpairings},
\end{align}
the parts of \eqref{eq:10dsusy2}-\eqref{eq:10dsusy3} along only  Mink$_2$ or M$_8$ are implied. We have now derived a set of necessary and sufficient geometric conditions for supersymmetry.

One can solve \eqref{eq:spinor norms} as 
\beq
||\chi_{1+}||^2+||\chi_{1\mp}||^2= 2 c_0e^{A},
\eeq
for $c_0$ a positive constant, then we can without loss of generality take
\beq
||\chi_{1+}||^2-||\chi_{1\mp}||^2= 2 c_0 e^{A}\cos\beta,
\eeq
this leads to the conditions \eqref{eq:BPSM81}-\eqref{eq:BPSM84} quoted in the main text.

\section{Deriving conditions for a restricted Mink\texorpdfstring{$_1$}{1} class}\label{sec:compactifyingMink2}
In this appendix we generalise the Mink$_2$ supersymmetry conditions of appendix \ref{sec:mink2susyderivation} by replacing
\beq
e^{2A}ds^2(\text{Mink}_2)\rightarrow e^{2A}(-dt^2+D\phi^2),~~~~~ D\phi=d\phi+{\cal A},~~~~~{\cal F}= d{\cal A}
\eeq
 for $\partial_t$ and $\partial_{\phi}$ isometries of the metric that we demand the fluxes also respect. We will additionally make the ansatz that the spinor decomposition of \eqref{eq:mink2spinordecomposition} is not modified. This is ultimately equivalent to deriving necessary and sufficient conditions for minimally supersymmetric stationary solutions of type II supergravity under two non general assumptions: 1) That  $K^{\mu}\partial_{\mu}$ as defined in  \eqref{eq:forms10d} is an null Killing vector rather than time-like which is possible more generally. 2) That this null Killing vector is spanned by a time-like and space-like Killing vector which the spinors on the remaining $d=8$ internal space are uncharged under.

We consider solutions that in IIA/IIB take the form  
\begin{align}
ds^2&=e^{2A}(-dt^2+ D\phi^2)+ ds^2(\text{M}_8),~~~~H= e^{2A}H_1\wedge dt\wedge D\phi+ e^{A}H_2^1\wedge dt+ e^{A}H_1^2\wedge D\phi+ H_3,\nn\\[2mm]
F_{\pm}&= f^{(8)}_{\pm}+e^A D\phi\wedge f^{(8)}_{\mp}+ e^{2A}dt\wedge D\phi\wedge \star_8\lambda(f^{(8)}_{\pm})+ e^A dt\wedge \star_8 \lambda(f^{(8)}_{\mp}) ,
\end{align}
where
\beq
D\phi=d\phi+{\cal A},~~~~~{\cal F}= d{\cal A},
\eeq
and $(e^{A},f^{(8)}_{\pm},f^{(8)}_{\mp},H_1,H_2^1,H_2^2,H_3)$and the dilaton $\Phi$ have support on only M$_8$, so that  $\partial_{\phi}$ and $\partial_{t}$ are isometries of the full ansatz. As we assume that \eqref{eq:mink2spinordecomposition} still holds the $d=10$ bi-linears still take the form of \eqref{eq:10toedecompo} modulo the modification that v is no longer closed, in fact now
\beq
v= dt+D\phi~~~~\Rightarrow ~~~dv={\cal F}.\label{eq:nonclosureofv}
\eeq
We now set about inserting this ansatz into the necessary and sufficient conditions for $d=10$ supersymmetry, first we consider the differential constraints \eqref{eq:10dsusy1}.

We still have that  $\nabla_{(M}K_{N)}=0$ leads to the constraint on the norms of the internal spinors $(\chi_{1+},\chi_{2\mp})$  \eqref{eq:spinor norms}, which this time we immediately solve using \eqref{eq:norms8d}.

Due to \eqref{eq:nonclosureofv} and the generalised $H$, we now find that $d\tilde{K}=\iota_{K}H$ gives rise to 
\beq
d(e^{2A}\cos\beta)= e^{2A}H_1,~~~~ e^{A}(H^1_2-H^{2}_2)=-e^{2A}\cos\beta {\cal F},\label{eq:mink2compactintermediatecond}
\eeq
which follow from that parts of $d\tilde{K}=\iota_{K}H$ that are parallel and orthogonal to $v$ respectively.

We now have that
\begin{align}
(\tilde{K}\wedge +\iota_{K})F_{\pm}=&-\frac{c_0}{32}e^{A}\bigg[e^{A}v\wedge \big(-\cos\beta f^{(8)}_{\pm}+\star_8\lambda (f^{(8)}_{\pm})\big)\nn\\[2mm]
-&(1+e^{2A}\cos\beta dt\wedge D\phi)\wedge (f^{(8)}_{\mp}-\star_8\lambda(f^{(8)}_{\mp}))\bigg],
\end{align}
so that the final condition in \eqref{eq:10toedecompo} now splits into parts parallel to $v$, $Dt\wedge D\phi$ and with legs only along the M$_8$ directions as respectively
\begin{align}
d_{H_3}(e^{2A}\Psi^{(8)}_{\mp})&=\pm\frac{ e^{2A}c_0}{16}\left(-\cos\beta f^{(8)}_{\pm}+\star_8\lambda(f^{(8)}_{\pm})\right),\nn\\[2mm]
f^{(8)}_{\mp}-\star_8 \lambda(f^{(8)}_{\mp})&= \pm \frac{16 e^{A-\Phi}}{c_0}{\cal F}\wedge\Psi^{(8)}_{\mp},\nn\\[2mm]
e^{-\Phi}(H^{1}_2-H^2_2)\wedge \Psi^{(8)}_{\mp}&=\pm\frac{e^{A}c_0}{16}\cos\beta(f^{(8)}_{\mp}-\star_8 \lambda(f^{(8)}_{\mp}))
\end{align}
where the first of these reproduces the condition we found for Mink$_2$ and the last of these is implied by the second condition in \eqref{eq:mink2compactintermediatecond}. The remaining condition informs us that $f_{\mp}^{(8)}$ decomposes into self dual and anti self dual parts (under the combined action of $\star_8\lambda$) as
\beq
f^{(8)}_{\mp}=\pm \frac{16}{c} e^{A-\Phi}{\cal F}\wedge \Psi^{(8)}_{\mp}+ g^{(8)}_{\mp},~~~~  g^{(8)}_{\mp}= \star_8 \lambda( g^{(8)}
_{\mp}),
\eeq
and that ${\cal F}$ must obey in IIA and IIB respectively
\begin{align}
&{\cal F}\wedge\iota_V\star_8\Phi_3=0~~~~\Rightarrow~~~~\star_8 {\cal F}=-{\cal F}\wedge \Phi_3\wedge V\nn\\[2mm]
\sin\alpha {\cal F}\wedge \text{Im}\Omega^{(8)}&=0,~~~~\star_8{\cal F}= - {\cal F}\wedge (\frac{1}{2}J^{(8)}\wedge J^{(8)}+\cos\alpha \text{Re}\Omega^{(8)}),
\end{align}
 where we refer to the G$_2$ and SU(4)-structure forms of section \ref{eq:Mink2Gstructures}. 

We note that the differential constraints have split into conditions that we already had for Mink$_2$ plus additional conditions involving the new flux components and field strength ${\cal F}$, there is no mixing between the two sets of conditions. Remarkably this patten persists with the pairing constraints which we find give rise in general to
\begin{align}
&(\Psi^{(8)}_{\mp},f^{(8)}_{\pm})_7 =\mp \frac{c_0}{8}e^{-\Phi}\star_8( 2 dA- \cos\beta H_1),\nn\\[2mm]
&(\Psi^{(8)}_{\mp},\star_8 \lambda(f^{(8)}_{\pm}))_7 =\pm \frac{c_0}{8}e^{-\Phi}\star_8( 2 \cos\beta dA-  H_1),\nn\\[2mm]
&(\Psi^{(8)}_{\mp},g^{(8)}_{\mp})_8=0,\nn\\[2mm]
&(\Psi^{(8)}_{\mp},g^{(8)}_{\mp})_6=\frac{c_0 e^{-\Phi}}{32}\bigg(\mp \cos\beta\star_8(H_1^2+H_2^2)  + X^{\pm}_{4}\wedge (H_1^2+H_2^2)\bigg),
\end{align}
where $X^{\pm}_4$ appear in IIA/IIB and can be expressed in terms of the relevant G-structure on M$_8$ as
\begin{align}
X_{4}^+&=\star_8\Phi_3+\cos\beta  V\wedge \Phi_3,\\[2mm]
X_{4}^-&=-\sin\alpha  \text{Im}(e^{-i \alpha}\Omega^{(8)})+\cos\beta(\frac{1}{2}J^{(8)}\wedge J^{(8)}+\cos\alpha \text{Re}(e^{-i \alpha}\Omega^{(8)})).
\end{align}
There are further condition that must be imposed in IIB which further constrains the self dual flux $g_4$, namely
\beq
\sin\beta \text{Im}(e^{-i \alpha}\Omega^{(8)})\wedge g_4=\frac{1}{6} e^{-\Phi} (H_1^2+H_2^2)\wedge J^{(8)}\wedge J^{(8)}\wedge J^{(8)}=0,\label{eq:pairinghard3}
\eeq
must hold in general, and additionally when $\sin\alpha=0$ one must also impose
\beq
\sin\beta Z_2\wedge J^{(8)}\wedge g_4+\frac{1}{2}\cos\alpha e^{-\Phi}Z_2\wedge\text{Im}(e^{-i \alpha}\Omega^{(8)})\wedge (H_1^2+H_2^2)=0\label{eq:pairinghard4},
\eeq
where we are using short hand for 3 complex (6 real) equations for which  $Z_2$ is set equal to each of
\beq
E^1\wedge E^2- \bar{E}^3\wedge \bar{E}^{4},~~~E^1\wedge E^3- \bar{E}^4\wedge \bar{E}^{2},~~~E^1\wedge E^4- \bar{E}^2\wedge \bar{E}^{3}
\eeq
for $E^{1,2,3,4}$ any canonical holomorphic frame for the SU(4)-structure in terms of which 
\begin{align}
J^{(8)}&= \frac{i}{2}(E^1\wedge \bar{E}^1+E^2\wedge \bar{E}^2+E^3\wedge \bar{E}^3+E^4\wedge \bar{E}^4),\nn\\[2mm]
\Omega^{(8)}&= E^1\wedge E^2\wedge E^3\wedge E^4.
\end{align}
This completes our derivation of the supersymmetry conditions, note that in the main text we fix
\beq
\frac{1}{2}(H_1^2+H_2^2+\cos\beta {\cal F})=H_2.
\eeq

\section{Deriving Mink\texorpdfstring{$_{D+1}$}{(D+1)} supersymmetry conditions from Mink\texorpdfstring{$_D$}{D}}\label{MinkDtoDm1derivation}
In this appendix we derive the G-structure conditions for minimally supersymmetric warped Mink$_{D+1}$ solution from those of Mink$_{D}$, for $D=2,3,4,5$.

\subsection{Mink\texorpdfstring{$_3$}{3} from Mink\texorpdfstring{$_2$}{2}}\label{sec:Mink3fromMink2}
Warped Mink$_3$ solutions in type II supergravity take the form
\begin{align}
ds^2&=e^{2A}ds^2(\text{Mink}_3)+ ds^2(\text{M}_7),~~~~H= e^{3A}H_0 \text{vol}(\text{Mink}_3)+ H_3,\nn\\[2mm]
F_{\pm}&= f^{(7)}_{\pm}+ e^{3A}\text{vol}(\text{Mink}_3)\wedge \star_7\lambda(f^{(7)}_{\pm}),\label{eq:Mink3bosonicfields}
\end{align}
where $(e^{2A},H_0,H_3,f^{(7)}_{\pm})$ and the dilaton have support on M$_7$.  Locally these are a special case of the Mink$_2$ solutions of \eqref{eq:Mink3bosonicfields} for which one fixes
\beq
ds^2(\text{M}_8)= e^{2A}d\phi^2+ ds^2(\text{M}_7),~~~~f^{(8)}_{\pm}=f^{(7)}_{\pm},~~~~H_1=e^{A}H_0d\phi\label{eq:decompostionMink8to7}
\eeq
for $\partial_{\phi}$ is an isometry direction - the domain of which determines whether we actually have Mink$_3$ or Mink$_2\times$S$^1$ or some orbifold there of globally.

We can derive geometric conditions for supersymmetry preservation for Mink$_3$ solutions from those of Mink$_2$ by first making the following decomposition of the $d=8$ gamma matrices
\begin{align}
\gamma^{(8)}_8&=\mathbb{I}\otimes\sigma_3,~~~~\gamma^{(8)}_{a}=\gamma^{(7)}_a\otimes \sigma_2,~~~a=1,...7,\nn\\[2mm]
B^{(8)}&= B^{(7)}\otimes \mathbb{I},~~~~\hat\gamma^{(8)}=\mathbb{I}\otimes \sigma_1,
\end{align}
where $i\gamma^{(7)}_{1...7}=1$,  $B^{(7)}B^{(7)*}=1$,
  $(B^{(7)})^{-1}\gamma^{(7)}_aB^{(7)}=-\gamma^{(7)*}_a$ and $\sigma_{1,2,3}$ the Pauli matrices. A Majorana-Wely spinor in $d=8$ has the same number of independent components as a Majorana spinor in $d=7$. As such we can decompose the latter, appearing in the decomposition of the $d=10$ spinors in \eqref{eq:deq10mink2}, in terms of the former as
\beq
\chi_{1+}^{(8)}=\chi^{(7)}_1\otimes \theta_{+},~~~~\chi_{2\mp}^{(8)}=\chi_2^{(7)}\otimes \theta_{\mp},~~~~\theta_{\pm}=\frac{1}{2}\left(\begin{array}{c}1\\ \pm 1 \end{array}\right),\label{eq:8to7dspinoes}
\eeq
where $(\chi^{(7)}_1,\chi^{(7)}_2)$ are $d=7$ Majorana spinors in  with respect to $B^{(7)}$. They must obey the constraints
\beq
|\chi^{(7)}_{1}|^2= 2e^A c_0 \cos^2\left(\frac{\beta}{2}\right),~~~~|\chi^{(7)}_{2}|^2= 2e^A c_0 \sin^2\left(\frac{\beta}{2}\right),\label{eq:norms7d}
\eeq
so that \eqref{eq:norms8d} holds. Inserting these definitions on the spinor into the definition of $\Psi^{(8)}_{\mp}$ in \eqref{eq:psi8poly} we find that they simply decompose as
\beq
\Psi^{(8)}_{\mp}=\frac{1}{2}(\Psi^{(7)}_{\mp}+\Psi^{(7)}_{\pm}\wedge W),\label{eq:psi8to7}
\eeq
where $W=e^8$ is the vielbein direction
\beq
W= e^{A} d\phi,
\eeq
and $\Psi^{(7)}_{\mp}$ are bilinears of odd/even form degree on the orthogonal M$_7$ defined as
\beq
\Psi^{(7)}_{+}+i \Psi^{(7)}_-=e^{-A}\frac{1}{8}\sum_{n=0}^7\frac{1}{n!}\chi^{(7)\dag}_{2}\gamma_{\underline{a_n}...\underline{a_1}}\chi^{(7)}_{1}e^{\underline{a}_1...\underline{a}_n}\label{eq:psi7poly},
\eeq
for $\Psi^{(7)}_{\pm}$ real bilinears of even/odd form degree. Inserting \eqref{eq:psi8to7} into the conditions for supersymmetry \eqref{eq:BPSM81}-\eqref{eq:BPSM84} and assuming the decomposition \eqref{eq:decompostionMink8to7} we find that they yield
\begin{align}
&d(e^{2A}\cos\beta)=0,~~~~ H_0=0,\\[2mm]
&d_{H_3}(e^{2A-\Phi}\Psi^{(7)}_{\mp})=\mp \frac{c_0 e^{2A}}{8}\cos\beta f^{(7)}_{\pm},\nn\\[2mm]
&d_{H_3}(e^{3A-\Phi}\Psi^{(7)}_{\pm})= \frac{e^{3A}c_0}{8}\star_7\lambda (f^{(7)}_{\pm}),\nn\\[2mm]
&(\Psi^{(7)}_{\mp},f^{(7)}_{\pm})_7=0\label{eq:Mink3susyapp},
\end{align}
which reproduce the necessary and sufficient conditions for supersymmetric Mink$_3$ solutions in appendix B of \cite{Macpherson:2021lbr} (with the inverse AdS$_3$ radius $m=0$). To map this reference to \eqref{eq:Mink3susyapp} one must fix their constants $(c_+,c_-)$ as
\beq
c_+=2 c_0,~~~~c_-=2e^{2A}\cos\beta c_0.
\eeq
There are also an additional 2 conditions following from \eqref{eq:BPSM83}-\eqref{eq:BPSM84}
\begin{align}
&(\Psi^{(7)}_{\pm},f^{(7)}_{\pm})_6 = \frac{c_0}{2}e^{-\Phi}\star_7  dA,\nn\\[2mm]
&(\Psi^{(7)}_{\mp},\star_7 \lambda(f^{(7)}_{\pm}))_6 =\pm \frac{c_0}{2}\cos\beta e^{-\Phi}\star_7   dA,
\end{align}
which are however implied by \eqref{eq:Mink3susyapp}. We thus find that \eqref{eq:Mink3susyapp} are necessary and sufficient conditions for Mink$_3$ as expected.

\subsection{Mink\texorpdfstring{$_4$}{4} from Mink\texorpdfstring{$_3$}{3}}\label{sec:Mink4fromMink3}
The embedding of Mink$_4$ into type II supergravity takes the form 
\begin{align}
ds^2&=e^{2A}ds^2(\text{Mink}_4)+ ds^2(\text{M}_6),~~~~H=  H_3,\nn\\[2mm]
F_{\pm}&= f^{(6)}_{\pm}+ e^{4A}\text{vol}(\text{Mink}_4)\wedge \star_6\lambda(f^{(6)}_{\pm}),\label{eq:Mink4bosonicfields}
\end{align}
where $(e^{2A},H_3,f^{(6)}_{\pm})$ and the dilaton have support on M$_6$. These are special cases of the Mink$_3$ solutions of \eqref{eq:Mink3bosonicfields} with $d=7$ fields tuned as
\beq
ds^2(\text{M}_7)= e^{2A}d\phi^2+ ds^2(\text{M}_6),~~~~f^{(7)}_{\pm}=f^{(6)}_{\pm},~~~~H_0=0\label{eq:decompostionMink7to6},
\eeq
for $\partial_{\phi}$ an isometry. Our aim this time is to use \eqref{eq:Mink3susyapp} and \eqref{eq:decompostionMink7to6} to arrive at necessary and sufficient conditions for Mink$_4$ solutions to preserve supersymmetry. As such we decompose the $d=7$ gamma matrices as
\begin{align}
\gamma^{(7)}_7=\hat\gamma^{(6)},~~~~\gamma^{(7)}_a=\gamma^{(6)}_a,~~~a=1,...,6,~~~~~B^{(7)}=B^{(6)},
\end{align}
where  $B^{(6)}B^{(6)*}=1$,
  $(B^{(6)})^{-1}\gamma^{(6)}_aB^{(6)}=-\gamma^{(6)*}_a$. 
We decompose our Majorana spinors in $d=7$ in terms of chiral spinors in $d=6$ $(\chi^{(6)}_{1+},\chi^{(6)}_{2\mp})$ as
\beq
\chi^{(7)}_1=\frac{1}{\sqrt{2}}(\chi^{(6)}_{1+}+\chi^{(6)}_{1-}),~~~~\chi^{(7)}_2=\frac{1}{\sqrt{2}}(\chi^{(6)}_{2\mp}+\chi^{(6)}_{2\pm}),\label{eq:spinors7to6}
\eeq
where $\chi^{(6)}_{1-}=(\chi^{(6)}_{1+})^c=B^{(6)}\chi^{(6)*}_{+}$ and similarly for $\chi^{(6)}_{2\pm}$ while \eqref{eq:norms7d} implies that
\beq
|\chi^{(6)}_{1+}|^2= 2e^A c_0 \cos^2\left(\frac{\beta}{2}\right),~~~~|\chi^{(6)}_{2\mp}|^2= 2e^A c_0 \sin^2\left(\frac{\beta}{2}\right)\label{eq:norms6d}.
\eeq
Inserting \eqref{eq:spinors7to6} into the definition of the $d=7$ bilinears \eqref{eq:psi7poly}, we find that they decompose in IIA/IIB as
\beq
\Psi^{(7)}_{\mp}=-(\text{Re}\Psi^{(6)}_{\mp}+\text{Re}\Psi^{(6)}_{\pm}\wedge W),~~~~~\Psi^{(7)}_{\pm}=(\text{Im}\Psi^{(6)}_{\pm}-\text{Im}\Psi^{(6)}_{\mp}\wedge W),\label{eq:psi7to6}
\eeq
where $W=e^7=e^{A}d\phi$ and $\Psi^{(6)}_{\pm}$ are $d=6$ bilinears orthogonal to it defined as
\beq
\Psi^{(6)}_{\pm}=e^{-A}\frac{1}{8}\sum_{n=0}^6\frac{1}{n!}\chi^{(6)\dag}_{2\mp}\gamma_{\underline{a_n}...\underline{a_1}}\chi^{(6)}_{1+}e^{\underline{a}_1...\underline{a}_n}\label{eq:psi6poly}.
\eeq
We can now insert these into the supersymmetry conditions for Mink$_3$ solution \eqref{eq:Mink3susyapp} to extract analogous conditions for Mink$_4$. The differential constraints are implied by
\begin{align}
&d(e^{2A}\cos\beta)=0,\nn\\[2mm]
&d_{H_3}(e^{3A-\Phi}\Psi^{(6)}_{\pm})=0,\nn\\[2mm]
&d_{H_3}(e^{2A-\Phi}\text{Re}\Psi^{(6)}_{\mp})=\mp\frac{c_0}{8}e^{2A}\cos\beta f^{(6)}_{\pm} \nn,\\[2mm]
&d_{H_3}(e^{4A-\Phi}\text{Im}\Psi^{(6)}_{\mp})=\mp \frac{c_0}{8} e^{4A}\star_6\lambda (f^{(6)}_{\pm})\label{eq:Mink4susyapp},
\end{align}
which reproduce the necessary conditions for supersymmetric Mink$_4$ solutions first appearing in \cite{Grana:2005sn}, but in the conventions of \cite{Tomasiello:2011eb}. The pairing constraint in \eqref{eq:Mink3susyapp} gives rise to a single additional condition
\beq
(\Psi^{(6)}_{\pm},f^{(6)}_{\pm})_6=0,
\eeq
which, as shown in \cite{Tomasiello:2011eb},  is implied by \eqref{eq:Mink4susyapp}.

\subsection{Mink\texorpdfstring{$_5$}{5} from Mink\texorpdfstring{$_4$}{4}}\label{sec:Mink5fromMink4}
Mink$_5$ solutions of type II supergravity take the form 
\begin{align}
ds^2&=e^{2A}ds^2(\text{Mink}_5)+ ds^2(\text{M}_5),~~~~H=  H_3,\nn\\[2mm]
F_{\pm}&= f^{(5)}_{\pm}+ e^{5A}\text{vol}(\text{Mink}_5)\wedge \star_5\lambda(f^{(5)}_{\pm}),\label{eq:Mink5bosonicfields}
\end{align}
with $(e^{2A},H_3,f^{(5)}_{\pm})$ and the dilaton independent of the Mink$_5$ directions, they can be embedded into Mink$_4$ solutions by fixing
\beq
ds^2(\text{M}_6)= e^{2A}d\phi^2+ ds^2(\text{M}_5),~~~~f^{(6)}_{\pm}=f^{(5)}_{\pm},\label{eq:decompostionMink6to5},
\eeq
in \eqref{eq:decompostionMink7to6}. We decompose the 6 dimensional gamma matrices as
\begin{align}
\gamma_6^{(6)}&=\mathbb{I}\otimes \sigma_3,~~~~\gamma_6^{(a)}=\gamma_a^{(5)}\otimes\sigma_2,\nn\\[2mm]
B^{(6)}&=B^{(5)}\otimes \sigma_2,~~~~\hat\gamma^{(6)}=\mathbb{I}\otimes \sigma_1,
\end{align}
for $\sigma_{1,2,3}$ the Pauli matrices, $B^{(5)}B^{(5)*}=-1$ and $(B^{(5)})^{-1}\gamma^{(5)}_aB^{(5)}=\gamma_a^{(5)*}$, and 6 dimensional chiral spinors as in IIA/IIB as
\beq
\chi^{(6)}_{1+}=e^{i\theta_{\mp}}\chi^{(5)}_1\otimes \theta_{+} ,~~~~\chi^{(6)}_{2\mp}=\chi^{(5)}_2\otimes \theta_{\mp},~~~~e^{i\theta_{-}}=i,~~~~e^{i\theta_{+}}=1,
\eeq
where $\theta_{\pm}$ are eigenvectors of $\sigma_1$ as in \eqref{eq:8to7dspinoes} and $(\chi^{(5)}_1,\chi^{(5)}_2)$ are arbitrary $d=5$ spinors obeying
\beq
|\chi^{(5)}_{1}|^2= 2e^A c_0 \cos^2\left(\frac{\beta}{2}\right),~~~~|\chi^{(5)}_{2}|^2= 2e^A c_0 \sin^2\left(\frac{\beta}{2}\right)\label{eq:norms5d}.
\eeq
Upon plugging this spinor decomposition into \eqref{eq:psi6poly} we find 
\beq
\Psi^{(6)}_{\mp}=\frac{1}{2}(\Psi^{(5)}_{\mp}\pm i \Psi^{(5)}_{\pm}\wedge W),~~~~\Psi^{(6)}_{\pm}=\frac{1}{2}(\tilde{\Psi}^{(5)}_{\pm}\mp i \tilde{\Psi}^{(5)}_{\mp}\wedge W),\label{eq:6to5}
\eeq
for $W=e^6=e^{A}d\phi$ and 
\begin{align}
\Psi^{(5)}_{+}+\Psi^{(5)}_{-}&=e^{-A}\frac{1}{4}\sum_{n=0}^5\frac{1}{n!}\chi^{(5)\dag}_{2}\gamma_{\underline{a_n}...\underline{a_1}}\chi^{(5)}_{1}e^{\underline{a}_1...\underline{a}_n}\nn\\[2mm]
\tilde{\Psi}^{(5)}_{+}+\tilde{\Psi}^{(5)}_{-}&=e^{-A}\frac{1}{4}\sum_{n=0}^5\frac{1}{n!}\chi^{(5)c\dag}_{2}\gamma_{\underline{a_n}...\underline{a_1}}\chi^{(5)}_{1}e^{\underline{a}_1...\underline{a}_n}\label{eq:psi5poly}.
\end{align}
Substituting for \eqref{eq:6to5} in \eqref{eq:Mink4susyapp} we find the following necessary and sufficient conditions for Mink$_5$ solutions
\begin{align}
&d(e^{2A}\cos\beta)=0,\nn\\[2mm]
&d_{H_3}(e^{3A-\Phi}\tilde{\Psi}^{(5)}_{\pm})=0,\nn\\[2mm]
&d_{H_3}(e^{4A-\Phi}\tilde{\Psi}^{(5)}_{\mp})=0,\nn\\[2mm]
&d_{H_3}(e^{3A-\Phi}\text{Im}\Psi^{(5)}_{\pm})=0,\nn\\[2mm]
&d_{H_3}(e^{4A-\Phi}\text{Im}\Psi^{(5)}_{\mp})=0,\nn\\[2mm]
&d_{H_3}(e^{2A-\Phi}\text{Re}\Psi^{(5)}_{\mp})=\mp\frac{c_0}{4}e^{2A}\cos\beta f_{\pm}^{(5)} ,\nn\\[2mm]
&d_{H_3}(e^{5A-\Phi}\text{Im}\Psi^{(5)}_{\pm})=\mp \frac{c_0}{4}\star\lambda(f^{(5)}_{\pm}).\label{eq:mink5bpsapps}
\end{align}

\subsection{Mink\texorpdfstring{$_6$}{6} from Mink\texorpdfstring{$_5$}{5}}\label{sec:Mink6fromMink5}
The final type of Mink$_D$ solution that will interest us in this work are Mink$_6$ solutions, because in type II supergravity these have the smallest internal space that can support more than simply an identity-structure. They are decomposable as
\begin{align}
ds^2&=e^{2A}ds^2(\text{Mink}_6)+ ds^2(\text{M}_4),~~~~H=  H_3,\nn\\[2mm]
F_{\pm}&= f^{(4)}_{\pm}+ e^{6A}\text{vol}(\text{Mink}_6)\wedge \star_4\lambda(f^{(4)}_{\pm}),\label{eq:Mink6bosonicfields}
\end{align}
where $(e^{2A},H_3,f^{(4)}_{\pm})$ and the dilaton have support on M$_4$, and can be embedded in Mink$_5$ as
\beq
ds^2(\text{M}_5)= e^{2A}d\phi^2+ ds^2(\text{M}_4),~~~~f^{(5)}_{\pm}=f^{(4)}_{\pm},~~~~H_0=0\label{eq:decompostionMink5to6},
\eeq
for $\partial_{\phi}$ an isometry. We decompose the $d=5$ gamma matrices in terms of $d=4$ ones as
\beq
\gamma^{(5)}_5=\hat\gamma^{(4)},~~~~\gamma^{(5)}_a=\gamma^{(4)}_a,~~~a=1,...,4
\eeq
for $B^{(5)}=B^{(4)}$ such that $B^{(4)}B^{(4)*}=-1$ and $(B^{(4)})^{-1}\gamma_{a}^{(4)}B^{(4)}=\gamma_{a}^{(4)*}$. 

We can now extract necessary and sufficient conditions for supersymmetric Mink$_6$ from those of Mink$_5$ by suitably defining the spinor - however some extra care is required with respect to the cases considered earlier in this appendix. The issue is that the Majorana conjugate for both $d=6$  Lorentzian spinors and $d=4$ Euclidean spinors is chirality preserving, while Mink$_6$ supports spinors of both chiralities. As such minimal supersymmetry is chiral for Mink$_6$ (like for Mink$_2$) \textit{i.e.} ${\cal N}=(1,0)$, while the generic supersymmetric embeddings of Mink$_6$ into solutions with a Mink$_5$ factor will preserve at least ${\cal N}=(1,1)$, which is non minimal and also only compatible with identity-structure. Thus to realise ${\cal N}=(1,0)$ Mink$_6$ solutions we should decompose the $d=5$ spinors of the previous section in terms of $d=4$ spinors of fixed chirality $(\chi^{(4)}_{1+},\chi^{(4)}_{2\mp})$, we take in IIA/IIB
\beq
\chi^{(5)}_{1}=\mp \chi^{(4)}_{1+},~~~~\chi^{(5)}_{2}=\chi^{(4)}_{2\mp},
\eeq
such that
\beq
|\chi^{(4)}_{1+}|^2= 2e^A c_0 \cos^2\left(\frac{\beta}{2}\right),~~~~|\chi^{(4)}_{2\mp}|^2= 2e^A c_0 \sin^2\left(\frac{\beta}{2}\right)\label{eq:norms4d}.
\eeq
Then upon inserting our decomposition of the $d=5$ spinors into \eqref{eq:psi5poly}, we find that the $d=5$ biliears decomposes as
\beq
\Psi^{(5)}_{\mp}+\Psi^{(5)}_{\pm}=\Psi^{(4)}_{\mp}\wedge(W\mp 1) ,~~~~\tilde{\Psi}^{(5)}_{\mp}+\tilde{\Psi}^{(5)}_{\pm}=\tilde{\Psi}^{(4)}_{\mp}\wedge(W\mp 1),\label{eq:5to4polyforms}
\eeq
where $W=e^5=e^{A}d\phi$ and
\begin{align}
\Psi^{(4)}_{\mp}&=e^{-A}\frac{1}{4}\sum_{n=0}^4\frac{1}{n!}\chi^{(4)\dag}_{2\mp}\gamma_{\underline{a_n}...\underline{a_1}}\chi^{(4)}_{1+}e^{\underline{a}_1...\underline{a}_n}\nn\\[2mm]
\tilde{\Psi}^{(4)}_{\mp}&=e^{-A}\frac{1}{4}\sum_{n=0}^4\frac{1}{n!}\chi^{(4)c\dag}_{2\mp}\gamma_{\underline{a_n}...\underline{a_1}}\chi^{(4)}_{1+}e^{\underline{a}_1...\underline{a}_n}\label{eq:psi4poly}.
\end{align}
Thus, upon substituting for \eqref{eq:5to4polyforms} in \eqref{eq:mink5bpsapps} we derive necessary and sufficient conditions for ${\cal N}=(1,0)$ supersymmetric Mink$_6$  solutions, namely
\begin{align}
&d(e^{2A}\cos\beta)=0,\nn\\[2mm]
&d_{H_3}(e^{4A-\Phi}\tilde{\Psi}^{(4)}_{\mp})=0,\nn\\[2mm]
&d_{H_3}(e^{4A-\Phi}\text{Im}\Psi^{(4)}_{\mp})=0,\nn\\[2mm]
&d_{H_3}(e^{2A-\Phi}\text{Re}\Psi^{(4)}_{\mp})=\frac{1}{4}c_0 e^{2A}\cos\beta f^{(4)}_{\pm},\nn\\[2mm]
&d_{H_3}(e^{6A-\Phi}\text{Re}\Psi^{(4)}_{\mp})= -\frac{1}{4}c_0 e^{6A}\star_4 \lambda(f^{(4)}_{\pm}).
\end{align}

\section{G-structures and parametrising bilinears}\label{sec:gstructuresapp}
Thought this work we will be interested in Mink$_D$ solutions in type II supergravity that preserve at least minimal supersymmetry. Minimal supersymmetry implies that their internal $d=10-D$ dimensional manifold M$_{d}$  supports a specific G-structure determined by the dimension of the space and the chiral and Majorana (or not) nature of the internal spinors it can support. In this appendix we will give details on these G-structures, the forms that define them and how they relate to the internal spinors introduced in appendices \ref{sec:mink2susyderivation} and \ref{MinkDtoDm1derivation}. We in large part follow the conventions of the very informative appendix B of \cite{Gauntlett:2003cy}.\\
~\\
The G-structures that will be of interest to us in this work are SU(n)-structures for $n=2,3,4$, Spin(7)-structures and G$_2$-structures. Each of these has a canonical dimension associated to it, but can also be embedded into higher dimensions. Of these G-structures, SU(4) has the smallest canonical dimension $d=4$, an internal space with $d<4$ can only support an identity-structure.

SU(n)-structures are canonically $2n$-dimension and are characterised by a real (1,1)-form $J^{(2n)}$ and a $(n,0)$-form $\Omega^{(2n)}$ which obey
\beq
J^{(2n)}\wedge\Omega^{(2n)}=0,~~~~\text{vol}(\text{M}_{2n})=\frac{1}{n!}(J^{(2n)})^n=\frac{1}{2^n}(i)^{-n(n+2)}\Omega^{(2n)}\wedge\overline{\Omega}^{(2n)}.
\eeq
They can always be expressed in terms of a canonical complex frame $E^a$ as
\begin{align}
J^{(2n)}&=\frac{i}{2}\sum_{a=1}^nE^a\wedge \overline{E}^a,~~~~\Omega^{(2n)}=E^1\wedge ...\wedge E^n,\label{eq:SU(n)}
\end{align}
for $E^1=e^1+i e^2,~E^2=e^3+i e^4$ and so on. Note that one can always decompose an SU(n+1)-structure in terms of an SU(n)-structure and a holomorphic 1-form $Z$ that is orthogonal to it as
\beq
J^{(2(n+1))}=J^{(2n)}+\frac{i}{2}Z\wedge \overline{Z},~~~~\Omega^{(2(n+1))}=\Omega^{(2n)}\wedge Z.\label{eq:sutosu}
\eeq
Spin(7)-structures are canonically 8 dimensional and are characterised by a real self dual 4-form $\psi_4$ obeying
\beq
\psi_4\wedge\psi_4=14 \text{vol}(\text{M}_8).
\eeq
This can always be decomposed in terms of an SU(4)-structure as
\beq
\psi_4=\frac{1}{2}J^{(8)}\wedge J^{(8)}+\text{Re}\Omega^{(8)},\label{eq:spin73formb}
\eeq
and so like wise have a canonical decomposition in terms of their vielbein as in \eqref{eq:SU(n)}. Finally G$_2$-structures have canonical dimension 7 and come equipped with an associative 3-form $\Phi_3$ obeying
\beq
\Phi_3\wedge \star_7\Phi_3=7 \text{vol}(\text{M}_7).
\eeq
This can always be decomposed in terms of an SU(3)-structure and an additional vielbein direction $U$ that lies orthogonal to it as
\beq
\Phi_3=J^{(6)}\wedge U-\text{Im}\Omega^{(6)},~~~\star_7\Phi_3=\frac{1}{2}J^{(6)}\wedge J^{(6)}+\text{Re}\Omega^{(6)}\wedge U\label{eq:G23form},
\eeq
where in the frame of \eqref{eq:SU(n)} we would have $U=e^7$. Next we will explain how the spinors defined on each of our internal spaces give to the rise to these structures.

\subsection{Bilinears on M\texorpdfstring{$_8$}{8}}
We will begin with $d=8$, in this case we have  Majorana-Weyl spinors $(\chi^{(8)}_{1+},\chi^{(8)}_{2\mp})$ in  IIA/IIB which can in general be decomposed in terms of unit norm Majorana-Weyl spinors $\chi^{(8)}_{\pm}$ as in \eqref{eq:8dspinordecompIIA}. In general give $\chi^{(8)}_{\pm}$ rise to the components of the Spin(7) and G$_2$ structure forms as well as a vielbein component $V$ that is orthogonal to $\Phi_3$ as
\beq
V_a=\chi^{(8)\dag}_{-}\gamma^{(8)}_a\chi^{(8)}_{+},~~~(\Phi_3)_{a_1a_2a_3}=\chi^{(8)\dag}_{-}\gamma^{(8)}_{a_1a_2a_3}\chi^{(8)}_{+},~~~(\psi_4)_{a_1...a_4}=-\chi^{(8)\dag}_{+}\gamma^{(8)}_{a_1...a_4}\chi^{(8)}_{+},\label{eq:8dformsdef}.
\eeq
As explained in \cite{Gauntlett:2003cy}, moving to the canonical frame is equivalent to taking the unit norm spinors to obey the projections
\beq
\gamma^{(8)}_{1234}\chi^{(8)}_{\pm}=\gamma^{(8)}_{1256}\chi^{(8)}_{\pm}=\gamma^{(8)}_{1357}\chi^{(8)}_{\pm}=-\chi^{(8)}_{\pm},~~~~\hat\gamma^{(8)}\chi^{(8)}_{\pm}=\pm \chi^{(8)}_{\pm},\label{eq:8dprojectors}
\eeq 
in terms of which $V=e^8$ and $(\Phi_3,\psi_4)$ take the canonical from described above. With respect to \eqref{eq:8dprojectors} it is then a simple matter to show that  $\chi^{(8)}_{\pm}$ give rise to two bilinears that will be of relevance to us, namely
so if we define
\begin{align}
\slashed{\Psi}^{(8)}_{\text{G}_2}&=\chi^{(8)}_{+}\otimes \chi^{(8)\dag}_{-}~~~\Rightarrow~~~\Psi^{(8)}_{\text{G}_2}=\frac{1}{16}\bigg(V- \Phi_3+\star_8 \Phi_3-\iota_{V}\text{vol}(\text{M}_8)\bigg),\nn\\[2mm]
\slashed{\Psi}^{(8)}_{\text{Spin(7)}}&=\chi^{(8)}_{+}\otimes \chi^{(8)\dag}_{+}~~~\Rightarrow~~~\Psi^{(8)}_{\text{Spin(7)}}=\frac{1}{16}\bigg(1-\psi_4+\text{vol}(\text{M}_8)\bigg),
\end{align}
which are defined in terms of the forms that make up a $d=8$ G$_2$ and Spin(7)-structure respectively.
Given \eqref{eq:8dspinordecompIIA}, the definition of the 8d bilinear \eqref{eq:psi8poly}  on quickly establishes using the above that in type IIA we have
\beq
\Psi^{(8)}_-=c_0\sin\beta\Psi^{(8)}_{\text{G}_2}.
\eeq
In IIB the computation is a little trickier, but is made easier if one first observes that 
\beq
\slashed{\Psi}^{(8)}_+=e^{-A}\chi^{(8)}_{1+}\otimes \chi^{(8)\dag}_{2+}=c_0\sin\beta\bigg[\cos\alpha \chi^{(8)}_{+}\otimes \chi^{(8)\dag}_{+}-\sin\alpha\chi^{(8)}_{+}\otimes \chi^{(8)\dag}_{-}.\slashed{U}\bigg].
\eeq
We then find after using \eqref{eq:veilbeinasgammas}  that
\beq
\Psi^{(8)}_+=\sin\beta c_0 \left(\cos\alpha  \Psi^8_{\text{Spin(7)}}+ \sin\alpha (U\wedge-\iota_U) \Psi^{(8)}_{\text{G}_2}\right)=  \frac{c_0}{16}\sin\beta \text{Re}\bigg[e^{i\alpha}e^{-i J^{(8)}}-e^{-i \alpha}\Omega^{(8)}\bigg],
\eeq
where we assume the decompositions of \eqref{eq:spin73formb} and \eqref{eq:G23form}, and that
\beq
J^{(8)}=J^{(6)}+ U\wedge V,~~~~\Omega^{(8)}= \Omega^{(6)}\wedge (U+i V)
\eeq
as we are free to do without loss of generality. Given what we have derived this far it is possible to show that any 2-form ${\cal F}$ obeying the constraints \eqref{eq:calFMink2toMink1} is such that
\beq
\slashed{\cal F}\chi^{(8)}_{1+}=\slashed{\cal F}\chi^{(8)}_{2\mp}=0,
\eeq
which it is easiest to show in the frame where \eqref{eq:8dprojectors} hold.

\subsection{Bilinears on M\texorpdfstring{$_7$}{7}}
In $d=7$ we have two Majorana spinors $(\chi^{(7)}_1,\chi^{(7)}_2)$ which can be decomposed in terms of a single $d=7$   unit norm Majorana spinor $\chi^{(8)}$ and a vielbein component $U$ as
\beq
\chi^{(7)}_1=\sqrt{2e^{A}c_0}\cos\left(\frac{\beta}{2}\right)\chi^{(7)},~~~~\chi^{(7)}_2=\sqrt{2e^{A}c_0}\sin\left(\frac{\beta}{2}\right)\left(\cos\alpha -\sin\alpha i \slashed{U}\right)\chi^{(7)}.
\eeq  
This single spinor gives rise to a G$_2$-structure 3-form as
\beq
(\Phi_3)_{a_1a_2a_3}=\chi^{(7)\dag}\gamma_{a_1a_2a_3}\chi^{(7)},
\eeq
and the canonical frame discussed earlier is equivalent to $\chi^{(7)}$ obeying the productions
\beq
\gamma^{(7)}_{1234}\chi^{(7)}=\gamma^{(7)}_{1256}\chi^{(7)}=\gamma^{(7)}_{1357}\chi^{(7)}=-\chi^{(7)}.\label{eq:d=7projections}
\eeq
As we are now in odd dimensions the Clifford map for bilinears works as
\begin{align}
\slashed{\Psi}^{(7)}_+&=i \slashed{\Psi}^{(7)}_-=e^{-A}\chi^{(7)}_1\otimes \chi_2^{{(7)}\dag},\nn\\[2mm]
\Psi^{(7)}_+&=\frac{1}{8}e^{-A}\sum_{n ~\text{even}}\chi^{(7)\dag}_2\gamma_{\underline{a}_n...\underline{a}_1}\chi^{(7)}_1 e^{\underline{a}_1...\underline{a}_n},\nn\\[2mm]
\Psi^{(7)}_-&=\star_7\lambda(\Psi^{(7)}_+)=\frac{1}{8}e^{-A}\sum_{n ~\text{odd}}\chi^{(7)\dag}_2\gamma_{\underline{a}_n...\underline{a}_1}\chi^{(7)}_1 e^{\underline{a}_1...\underline{a}_n}.
\end{align}
In terms of \eqref{eq:d=7projections} it is straight forward to establish that $\chi^{(7)}$ gives rise to the bilinears
\begin{align}
\slashed{\Psi}^{(7)}_{\text{G}_2+}&=i \slashed{\Psi}^{(7)}_{\text{G}_2-}=\chi^{(7)}_1\otimes \chi_2^{{(7)}\dag}~~~\Rightarrow\nn\\[2mm]
\slashed{\Psi}^{(7)}_{\text{G}_2+}&=\frac{1}{8}\left(1-\star_7\Phi_3\right),~~~~\slashed{\Psi}^{(7)}_{\text{G}_2-}=\frac{1}{8}\left(-\Phi_3+\text{vol}(\text{M}_7)\right),
\end{align}
which define a G$_2$-structure. Then given that
\beq
e^{-A}\chi^{(7)}_1\otimes \chi_2^{{(7)}\dag}= c_0 \sin\beta\left(\cos\alpha\chi^{(7)}\otimes \chi^{{(7)}\dag}+i\sin\alpha \chi^{(7)}\otimes \chi^{{(7)}\dag} \slashed{U}\right) ,
\eeq
we find using  using \eqref{eq:veilbeinasgammas} in both IIA and IIB that
\beq
\Psi^{(7)}_{\pm}=c_0\sin\beta \left(\cos\alpha \Psi^{(7)}_{\text{G}_2\pm}+\sin\alpha(U\wedge-\iota_U) \Psi^{(7)}_{\text{G}_2\mp}\right).
\eeq
From here one need only decompose $\Phi_3$ in terms of $U$ as in \eqref{eq:G23form} to reach the form of \eqref{eq:d7biliears} in the main text. 

Finally we note that using the canonical frame in which \eqref{eq:d=7projections} hold, it is a simple matter to establish that imposing \eqref{eq:primativecondmink33} is equivalent to imposing
\beq
\slashed{\cal F}\chi^{(7)}_{1}=0,~~~\slashed{\cal F}\chi^{(7)}_{2}=0.
\eeq

\subsection{Bilinears on M\texorpdfstring{$_6$}{6}}
For the case of $d=6$ the internal space supports two chiral spinors  $(\chi^{(6)}_{1+},\chi^{(6)}_{2\mp})$ that we can in general decompose in terms of some of unit norm $\chi^{(6)}_+$ and a unit norm holomorphic 1-form $Z$ as
\begin{align}
\chi^{(6)}_{1+}&= \sqrt{2 c_0 e^{A}}\cos\left(\frac{\beta}{2}\right)\chi_+,\nn\\[2mm]
\chi^{(6)}_{2-}&= \sqrt{2 c_0 e^{A}}\sin\left(\frac{\beta}{2}\right)e^{i\alpha}\left(\kappa_{\|}\chi^c_++\kappa_{\perp}\frac{1}{2}\overline{Z}\chi_+\right),\nn\\[2mm]
\chi^{(6)}_{2+}&= \sqrt{2 c_0 e^{A}}\sin\left(\frac{\beta}{2}\right)e^{-i\alpha}\left(\kappa_{\|}\chi^{(6)}_+-\kappa_{\perp}\frac{1}{2}Z\chi^c_+\right),\label{eq:6dspinordecomp}
\end{align}
for $\kappa_{\|}^2+\kappa_{\perp}^2=1$. The spinor $\chi_+$ defines an SU(3)-structure with its associated $(1,1)$ and $(3,0)$ forms defined through
\beq
J^{(6)}_{a_1a_2}=-i \chi^{(6)\dag}_+\gamma^{(6)}_{a_1a_2}\chi^{(6)}_+,~~~~\Omega^{(6)}_{a_1a_2a_3}=-\chi^{(6)\dag}_+\gamma^{(6)}_{a_1a_2a_3}\chi^{(6)}_+.
\eeq
In a canonical frame we can assume $\chi^{(6)}$ obeys the projections
\beq
\gamma^{(6)}_{1234}\chi^{(6)}_+=-\chi^{(6)}_+,~~~~\gamma^{(6)}_{1256}\chi^{(6)}_+=-\chi^{(6)}_+,
\eeq
and $\gamma^{(6)}_{135}\chi^{(6)}_+=-\chi^{(6)c}_+$ without loss of generality, we can also choose to align $Z=e^5+i e^6$ if we wish. It is then a simple matter to establish that for $\chi^{(6)}_-=(\chi^{(6)}_+)^c$
\beq
\slashed{\Psi}^{(6)}_{\text{SU(3)}\pm}=\chi^{(6)}_+\otimes \chi^{(6)}_{\pm},~~~\Rightarrow~~~ \Psi^{(6)}_{\text{SU(3)}+}=\frac{1}{8} e^{-i J^{(6)}},~~~~\Psi^{(6)}_{\text{SU(3)}-}=\frac{1}{8} \Omega^{(6)}.
\eeq
Then by substituting \eqref{eq:6dspinordecomp} into \eqref{eq:psi6poly} and decomposing $(J^{(6)},\Omega^{(6)})$ as in \eqref{eq:sutosu} we find in both IIA and IIB that the bilinears are given by \eqref{eq:SU(3)XSU(3)bilinears} in the main text.

Using the cannonical frame of this section it is quick to show that \eqref{eq:Mink4conds} implies that
\beq
\slashed{\cal F}\chi^{(6)}_{+}=0,~~~~\slashed{\cal F}\chi^{(6)}_{\mp}=0,
\eeq
as one would expect for primitive (1,1)-forms.
\subsection{Bilinears on M\texorpdfstring{$_5$}{5}}
A single  unit norm spinor $\chi^{(5)}$ in 5 euclidean dimensions defines an SU(2)-structure, with associated forms defined via
\beq
V_a=\chi^{(5)\dag}\gamma^{(5)}_a\chi^{(5)},~~~J^{(4)}_{a_1a_2}=-i\chi^{(5)\dag}\gamma^{(5)}_{a_1a_2}\chi^{(5)},~~~\Omega^{(4)}_{a_1a_2a_3}=-\chi^{(5)\dag}\gamma^{(5)}_{a_1a_2a_3}\chi^{(5)},
\eeq
meaning that the metric decomposes as
\beq
ds^2(\text{M}_5)=  ds^2(\text{M}_{\text{SU(2)}})+V^2,
\eeq
with $(J^{(4)},\Omega^{(4)})$ with legs on the vielbein directions spanning $\text{M}_{\text{SU(2)}}$ only. 
 Minimal supersymmetry for in type II supergravity for a $d=5$ dimensional internal space means their are two spinors $(\chi^{(5)}_1,\chi^{(5)}_2)$ it supports in general. We can decompose these in terms of $\chi^{(5)}$ and a unit norm holomorphic 1-form $Z$ that is orthogonal to $V$ as
\beq
\chi^{(6)}_1= \sqrt{2c_0 e^{A}}\cos\left(\frac{\beta}{2}\right)\chi^{(6)},~~~~\chi^{(6)}_2= \sqrt{2c_0 e^{A}}\sin\left(\frac{\beta}{2}\right)\left(a\chi^{(6)}+b \chi^{(6)c}+\frac{c}{2}\overline{Z}\chi^{(6)}\right),\label{eq:5dspinorsdecomp}
\eeq
where $|a|^2+|b|^2+c^2=1$.
In a canonical frame we can assume that $\chi^{(5)}$ obeys the projector 
\beq
\gamma^{(5)}_5\chi^{(5)}=\chi^{(5)},\label{eq:5dframecond}
\eeq
as well as $\gamma^{(5)}_{12}\chi^{(5)}=i \chi^{(5)}$ and  $\gamma^{(5)}_{13}\chi^{(5)}=-\chi^{(5)c}$, which aligns $V=e^5$ and we are free to choose $Z=e^3+ie^4$. In terms of this one easily establishes that
\begin{align}
\slashed{\Psi}^{(5)}_{\text{SU(2)}+}&=\slashed{\Psi}^{(5)}_{\text{SU(2)}-}=\chi^{(5)}\otimes \chi^{(5)\dag}~~~~\tilde{\slashed{\Psi}}^{(5)}_{\text{SU(2)}+}=\tilde{\slashed{\Psi}}^{(5)}_{\text{SU(2)}-}=\chi^{(5)}\otimes \chi^{(5)c\dag},~~~\Rightarrow
\nn\\[2mm]
\Psi^{(5)}_{\text{SU(2)}}&=\Psi^{(5)}_{\text{SU(2)}+}+\Psi^{(5)}_{\text{SU(2)}-}= \frac{1}{4}(1+V)\wedge e^{-i J^{(4)}},~~~\tilde{\Psi}^{(5)}_{\text{SU(2)}}=\tilde{\Psi}^{(5)}_{\text{SU(2)}+}+\tilde{\Psi}^{(5)}_{\text{SU(2)}-}= \frac{1}{4}(1+V)\wedge \Omega^{(4)}.\nn
\end{align}
Then by using \eqref{eq:5dspinorsdecomp} in \eqref{eq:psi5poly} we find \eqref{eq:5dbilinears} in the main text,
 where we have decomposed
\beq
J^{(4)}= \frac{i}{2}\left(W\wedge \overline{W}+Z\wedge \overline{Z}\right),~~~~\Omega^{(4)}=W\wedge Z.\label{eq:SU(2)decomp}
\eeq

As was true in the proceeding sections a 2-form ${\cal F}$ subject to \eqref{eq:thisexampleofcalFconds} which also constrains the G-structure is equivalent to imposing
\beq
\slashed{\cal F}\chi^{(5)}_1=0,~~~~\slashed{\cal F}\chi^{(5)}_2=0,
\eeq
which is easiest to show in the frame for which the conditions discussed around \eqref{eq:5dframecond} hold.

\subsection{Bilinears on M\texorpdfstring{$_4$}{4}}

Finally we must consider the case of $d=4$ that supports chiral spinors $(\chi^{(4)}_{1+},\chi^{(4)}_{1\mp})$ that may be decomposed in terms of two unit norm chiral spinors $(\chi^{(4)}_+,\chi^{(4)}_-)$ and a unit norm holomorphic 1-form $Z$  as
\begin{align}
\chi^{(4)}_{1+}&=\sqrt{2c_0 e^{A}}\cos\left(\frac{\beta}{2}\right)\chi^{(4)}_+,\nn\\[2mm]
\chi^{(4)}_{1-}&=\sqrt{2c_0 e^{A}}\sin\left(\frac{\beta}{2}\right)\chi^{(4)}_-,\nn\\[2mm]
\chi^{(4)}_{1+}&=\sqrt{2c_0 e^{A}}\sin\left(\frac{\beta}{2}\right)\left(a\chi^{(4)}_++ b \chi_+^c\right),\label{eq:deq4spinordecomp}
\end{align}
where $|a|^2+|b|^2=1$. The spinors $(\chi^{(4)}_+,\chi^{(4)}_-)$ give rise to2  unit norm  holomorphic 1-forms $(W,Z)$ that span M$_4$ and SU(2)-structure forms as
\beq
W_a=\chi^{\dag}_-\gamma_a^{(4)}\chi_+,~~~Z_a=\chi^{c\dag}_-\gamma_a^{(4)}\chi_+,~~~J^{(4)}_{a_1a_2}=-i\chi_+^{\dag}\gamma_{a_1a_2}\chi_+,~~~\Omega^{(4)}_{a_1a_2}=-i\chi_+^{c\dag}\gamma_{a_1a_2}\chi_+.
\eeq
In a canonical frame we can take $\chi^{(4)}_{\pm}$ to obey
\beq
\gamma^{(4)}_{12}\chi^{(4)}_{\pm}=\pm i \chi^{(4)}_{\pm},~~~~\gamma^{(4)}_{13}\chi^{(4)}_{\pm}=\mp i \chi^{(4)c}_{\pm},
\eeq
without loss of generality. In this frame it does not take long to establish that 
\begin{align}
\slashed{\Psi}^{(4)}_{\text{SU}(2)}&=\chi^{(4)}_+\otimes \chi^{(4)\dag}_+,~~~~\tilde{\slashed{\Psi}}^{(4)}_{\text{SU}(2)}=\chi^{(4)}_+\otimes \chi^{(4)c\dag}_+,~~~\Rightarrow\nn\\[2mm]
\Psi^{(4)}_{\text{SU}(2)}&=\frac{1}{4}e^{-i  J^{(4)}},~~~~\tilde{\Psi}^{(4)}_{\text{SU}(2)}=\frac{1}{4}\Omega^{(4)}
\end{align}
giving rise to an SU(2)-structure and
\begin{align}
\slashed{\Psi}^{(4)}_{\mathbb{I}}&=\chi^{(4)}_+\otimes \chi^{(4)\dag}_+,~~~~\tilde{\slashed{\Psi}}^{(4)}_{\mathbb{I}}=\chi^{(4)}_+\otimes \chi^{(4)c\dag}_+,~~~\Rightarrow\nn\\[2mm]
\Psi^{(4)}_{\mathbb{I}}&=\frac{1}{4}W\wedge e^{\frac{1}{2}Z\wedge \overline{Z}},~~~~\tilde{\Psi}^{(4)}_{\mathbb{I}}=-\frac{1}{4}Z\wedge e^{\frac{1}{2}W\wedge \overline{W}}
\end{align}
giving rise to an identity structure. Then by plugging \eqref{eq:deq4spinordecomp} into \eqref{eq:5to4polyforms} we find \eqref{eq:4dbilinears} in the main text.

Finally using the canonical frame of this section it is simply to show that a primitive (1,1)-form will always satisfy
\beq
\slashed{\cal F}\chi^{(4)}_{+}=0,
\eeq
from which it follows that in the limit of \eqref{eq:prima} we have $\slashed{\cal F}\chi^{(4)}_{1+}=\slashed{\cal F}\chi^{(4)}_{2+}=0$.

\end{document}